%% file: Melzani_reconnection_ion_electron_relat.tex
\documentclass{aa}

\usepackage{amsmath,amssymb,textcomp}
\usepackage{graphicx}
\usepackage[varg]{txfonts}
\usepackage{import}% To use .pdf_tex from Inkscape in figures.
\usepackage{color}

\usepackage{natbib}
\usepackage{hyperref}
\hypersetup{pdfborder={0 0 0}}% Remove frames around links.
\hypersetup{
     colorlinks = true, 
     linkcolor = blue,
     anchorcolor = black,
     citecolor = blue, 
     filecolor = black, 
     urlcolor = blue
}
\newcommand{\tot}{\mathrm{tot}}
\newcommand{\dif}{\mathrm{d}}
\newcommand{\idem}{~~\textacutedbl}
\newcommand{\cs}{\mathrm{cs}}
\newcommand{\bg}{\mathrm{bg}}
\newcommand{\out}{\mathrm{out}}
\newcommand{\ins}{\mathrm{in}}
\newcommand{\lec}{\mathrm{e}}
\renewcommand{\ion}{\mathrm{i}}
\newcommand{\therm}{\mathrm{th}}
\newcommand{\pe}{\mathrm{pe}}
\newcommand{\ce}{\mathrm{ce}}
\newcommand{\ci}{\mathrm{ci}}
\newcommand{\wci}{\omega_\mathrm{ci}}
\newcommand{\wce}{\omega_\mathrm{ce}}
\newcommand{\wpe}{\omega_\mathrm{pe}}

\def\dintcl{\int\limits_{\mathrm{cell}}\kern-0.7em\int\kern-1.4em\bigcirc\kern.7em}
\renewcommand{\b}{\textbf}
\definecolor{orange}{rgb}{0.6,0.1,0}
\definecolor{darkblue}{rgb}{0.,0.,0.5}

\renewcommand{\note}[1]{}
\begin{document}

\title{Relativistic magnetic reconnection in collisionless ion-electron plasmas explored with particle-in-cell simulations}

\author{Micka\"el Melzani\inst{1}         \and
        Rolf Walder\inst{1}               \and
        Doris Folini\inst{1}              \and 
        Christophe Winisdoerffer\inst{1}  \and 
        Jean M. Favre\inst{2}
        }

\institute{\'{E}cole Normale Sup\'{e}rieure, Lyon, CRAL, UMR CNRS 5574, 
           Universit\'{e} de Lyon, France\\
           \hspace{0.3cm} E-mail: mickael.melzani@ens-lyon.fr
           \and
           CSCS Lugano, Switzerland
}

\offprints{M. Melzani}

\date{Received ... ; accepted ...}

\authorrunning{M. Melzani et al.}
\titlerunning{Relativistic reconnection in ion-electron plasmas}

%%%%%%%%%%%%%%%%%%%%%%%%%%%%%%%%%%%%%%%%%%%%%%%%%%%%%%%%%%%%%%%%%%%%%%%%%%%%%%%%%%%%%%%%%%%%%%%%%%%%%%%%%%%%%%%%%%%%%%%%%%%%%%%%%%%%%%%%%%%%%%%%
%%%%%%%%%%%%%%%%%%%%%%%%%%%%%%%%%%%%%%%%%%%%%%%%%%%%%%%%%%%%%%%%%%%%%%%%%%%%%%%%%%%%%%%%%%%%%%%%%%%%%%%%%%%%%%%%%%%%%%%%%%%%%%%%%%%%%%%%%%%%%%%%
%%%%%%%%%%%%%%%%%%%%%%%%%%%%%                             Abstract                                                %%%%%%%%%%%%%%%%%%%%%%%%%%%%%%
%%%%%%%%%%%%%%%%%%%%%%%%%%%%%%%%%%%%%%%%%%%%%%%%%%%%%%%%%%%%%%%%%%%%%%%%%%%%%%%%%%%%%%%%%%%%%%%%%%%%%%%%%%%%%%%%%%%%%%%%%%%%%%%%%%%%%%%%%%%%%%%%
%%%%%%%%%%%%%%%%%%%%%%%%%%%%%%%%%%%%%%%%%%%%%%%%%%%%%%%%%%%%%%%%%%%%%%%%%%%%%%%%%%%%%%%%%%%%%%%%%%%%%%%%%%%%%%%%%%%%%%%%%%%%%%%%%%%%%%%%%%%%%%%%

\abstract{%
Magnetic reconnection is a leading mechanism for magnetic energy conversion and high-energy non-thermal particle production 
in a variety of high-energy astrophysical objects, including ones with relativistic ion-electron plasmas
(e.g., microquasars or AGNs) -- a regime where first principle studies are scarce.
We present 2D particle-in-cell (PIC) simulations of low $\beta$ ion-electron plasmas under relativistic conditions, 
i.e., with inflow magnetic energy exceeding the plasma rest-mass energy.\newline
We identify outstanding properties: 
(i) For relativistic inflow magnetizations (here $10 \leq \sigma_\lec \leq 360$), 
the reconnection outflows are dominated by thermal agitation instead of bulk kinetic energy.
(ii) At large inflow electron magnetization ($\sigma_\lec \geq 80$), 
the reconnection electric field is sustained more by bulk inertia than by thermal inertia.
It challenges the thermal-inertia-paradigm and its implications.
(iii) The inflows feature sharp transitions at the entrance of the diffusion zones. 
These are not shocks but results from particle ballistic motions, all bouncing at the same location,
provided that the thermal velocity in the inflow is far smaller than the inflow $E\times B$ bulk velocity.
(iv) Island centers are magnetically isolated from the rest of the flow, 
and can present a density depletion at their center.
(v) The reconnection rates are slightly larger than in non-relativistic studies.
They are best normalized by the inflow relativistic Alfv\'en speed projected in the outflow direction, 
which then leads to rates in a close range (0.14--0.25) thus allowing for an easy estimation of the reconnection electric field.
%
% We also underline similarities with known non-relativistic setups or relativistic pair plasma studies,
% e.g., relativistic outflows being dominated by thermal agitation instead of bulk kinetic energy, all the more that the inflow is relativistic;
% the overall flow structure with or without a guide field; reconnection rates decreasing with increasing guide field strength.
%
% We also put together a formalism to work with relativistic collisionless plasmas based on averaging Vlasov equation and keeping lab-frame quantities,
% that remains valid for arbitrary particle distribution functions.
}%


\maketitle

%%%%%%%%%%%%%%%%%%%%%%%%%%%%%%%%%%%%%%%%%%%%%%%%%%%%%%%%%%%%%%%%%%%%%%%%%%%%%%%%%%%%%%%%%%%%%%%%%%%%%%%%%%%%%%%%%%%%%%%%%%%%%%%%%%%%%%%%%%%%%%%%
%%%%%%%%%%%%%%%%%%%%%%%%%%%%%%%%%%%%%%%%%%%%%%%%%%%%%%%%%%%%%%%%%%%%%%%%%%%%%%%%%%%%%%%%%%%%%%%%%%%%%%%%%%%%%%%%%%%%%%%%%%%%%%%%%%%%%%%%%%%%%%%%
\section{Introduction}\label{sec:intro}
%%%%%%%%%%%%%%%%%%%%%%%%%%%%%%%%%%%%%%%%%%%%%%%%%%%%%%%%%%%%%%%%%%%%%%%%%%%%%%%%%%%%%%%%%%%%%%%%%%%%%%%%%%%%%%%%%%%%%%%%%%%%%%%%%%%%%%%%%%%%%%%%
%%%%%%%%%%%%%%%%%%%%%%%%%%%%%%%%%%%%%%%%%%%%%%%%%%%%%%%%%%%%%%%%%%%%%%%%%%%%%%%%%%%%%%%%%%%%%%%%%%%%%%%%%%%%%%%%%%%%%%%%%%%%%%%%%%%%%%%%%%%%%%%%

Magnetic reconnection has been the focus of extended studies since 
its first introduction by \citet{Giovanelli1947,Giovanelli1948}
to explain the sudden release of energy in solar flares.
The term itself was coined by \citet{Dungey1958}.
It is now the key ingredient for theories of coronal heating, solar flares and jets, and coronal mass ejections in the Sun
\citep{Priest1987}, of magnetic storms and substorms in the Earth magnetosphere \citep{Paschmann2013}, 
and for the behavior of fusion plasmas with,
e.g., the sawtooth oscillation in tokamaks \citep{Biskamp2000}.
Space physics proofs that magnetic reconnection can quickly convert magnetic energy
into kinetic energies (bulk flow, heat, non-thermal particles), with fast variability and high efficiency.
Such attributes made it most attractive for high-energy astrophysics to explain,
for example, radiation \citep{Romanova1992} and flares \citep{Giannios2009} in active galactic nuclei (AGN) jets
or in gamma-ray bursts \citep{Lyutikov2006c,Lazar2009},
the heating of AGN and microquasar coronae and associated flares \citep{Matteo1998,Merloni2001,Goodman2008,Reis2013},
the flat radio spectra from galactic nuclei and AGNs \citep{Birk2001},
the heating of the lobes of giant radio galaxies \citep{Kronberg2004},
the $\sigma$-paradox and particle acceleration at pulsar wind termination shocks \citep{Kirk2003,Petri2007b,Sironi2011b},
GeV-TeV flares from the Crab nebulae \citep{Cerutti2012,Cerutti2012b,Cerutti2013},
transient outflow production in microquasars and quasars \citep{deGouveia2005,deGouveia2010,Kowal2011,McKinney2012,Dexter2013},
gamma-ray burst outflows and non-thermal emissions \citep{Drenkhahn2002,McKinney2012b},
X-ray flashes \citep{Drenkhahn2002},
soft gamma-ray repeaters \citep{Lyutikov2006b},
flares in double pulsar systems \citep{Lyutikov2013},
or energy extraction in the ergosphere of black holes \citep{Koide2008}.
As pointed out by \citet{Uzdensky2006}, magnetic reconnection is of dynamical importance 
in any environment where magnetic fields dominate the energy budget, so that 
the energy transfer can have dynamical and observable consequences, 
and where the rates of reconnection are fast, which is known to hold 
in collisionless plasmas \citep{Birn2001} or in collisional but turbulent plasmas \citep{Lazarian1999,Lazarian2011}.

Many of the above mentioned environments are collisionless \citep{Ji2011}, 
so that fast reconnection must be triggered and sustained by non-ideal terms 
others than collisional ones, which implies kinetic processes 
on scales of the order of the electron inertial length or Larmor radius,
with particles largely out of equilibrium and possibly comprising high-energy tails. 
These non-ideal terms can be linked to particle inertia and wave-particle resonant interactions, 
or to finite Larmor radius effects in magnetic field gradients.
Simulation studies thus require full kinetic codes such as Vlasov solvers 
or particle-in-cell algorithms.

% Studies of non-relativistic reconnection are many, and feature fast reconnection rates,
% but no consensus is yet reached on the exact mechanisms responsible for this. 
% The Hall mediated reconnection paradigm \citep{Birn2001,Shay2007} was recently challenged by kinetic simulations
% presenting fast rates without making use of the Hall physics \citep{Bessho2005,Daughton2007}.
% 3D?
% acceleration?

Most of the above environments are also relativistic,
either because of relativistic velocities (bulk flows or currents), 
or because the thermal kinetic energy and/or the magnetic energy density exceeds
the rest-mass energy of the particles.
The latter translates into the magnetization of the inflow, 
$\sigma_{\ins,s} = B_\ins^2/(\mu_0 n_\ins m_s c^2)$ with $s$ denoting ions or electrons, 
being larger than unity.
This magnetic energy can be transferred to the particles, and as it is larger than the particles rest mass, 
relativistic particles are expected. 
The relation $h_{0,\out,s}\Gamma_{\out,s}  = 1 + \sigma_{\ins,s}$,
with $h_{0,\out,s}$ the enthalpy and $\Gamma_{\out,s}$ the bulk Lorentz factor of the reconnection outflow
(see Sect.~\ref{sec:outflows}, Eq.~\ref{equ:outflow_energy_bis}), indeed shows that either relativistic temperatures ($h_{0,\out,s}>1$) 
or relativistic bulk velocities ($\Gamma_{\out,s}>1$) are obtained for the outflows.
The relevant magnetization is thus not that of the plasma, which is low because of the ion mass, 
but that of each species taken individually.

Studies of relativistic reconnection are more scarce than their non-relativistic counterparts
\citep[for the latter, see the reviews by][]{Birn2007,Treumann2013},
and they mainly deal with pair plasmas: 
for relativistic pair plasmas, they include 2D MHD simulations \citep{Watanabe2006,Zenitani2011,Takamoto2013,Baty2013},
two-fluid simulations \citep{Zenitani2009b,Zenitani2009},
test-particle simulations \citep{Bulanov1976,Romanova1992,Larrabee2003,Cerutti2012}\note{Larrabee's paper 
include a retroaction of the particles on the fields, in an iterative way.},
1D PIC simulations \citep{Petri2007b},
2D PIC simulations \citep{Jaroschek2008,Sironi2011b,Bessho2012,Cerutti2012b,Cerutti2013,Zenitani2001,Zenitani2005,Zenitani2008b,Zenitani2007},
and 3D PIC simulations \citep{Zenitani2008b,Zenitani2008c,Sironi2011b,Kagan2012,Cerutti2013b,Sironi2014}.
Relativistic reconnection in ion-electron plasmas is less studied. We find 
a test-particle simulation \citep{Romanova1992},
a resolution of the diffusion equation \citep{Birk2001},
and a discussion by \citet{Sakai2002} in a 2D PIC simulations of laser fusion beams.
% where reconnection occurs when the filaments of the filamentation instability merge.

The focus of the present work is on relativistic reconnection -- as compared to non-relativistic studies -- 
and on ion-electron plasmas -- as compared to pair plasmas. 
Our goal is to carve out aspects that are particular to this regime, 
to shed light on the underlying physical causes, 
and to ultimately put our findings in the, admittedly speculative, larger astrophysical context of 
microquasar and AGN disk coronae and magnetospheres,
and of other possible environments with ion-electron relativistic plasmas.
Part of our results are also of interest for pair plasmas and for non-relativistic cases.

In Sect.~\ref{sec:pb_setup} we describe the simulation setup and parameters. 
% We define the non-relativistic and relativistic Alfv\'en speeds, as well as various magnetization parameters.
Section~\ref{sec:results} presents the results of simulations with antiparallel asymptotic magnetic fields.
We investigate the structure of the two-scale diffusion region in Sect.~\ref{sec:inflow}, 
and explain why we see sharp transitions at the entrance of this region.
Next, we turn to the relativistic Ohm's law.
In non-relativistic reconnection, non-ideal terms are dominated by thermal inertia, i.e., by the divergence of off-diagonal 
elements of the pressure tensor. There are, however, PIC studies (see references of Sect.~\ref{Sec:Ohms_law}) 
suggesting that for relativistic reconnection, thermal inertia can be dominated by bulk inertia.
In Sect.~\ref{Sec:Ohms_law} we show that this is the case in our simulations with large inflow magnetization.
We demonstrate in Sect.~\ref{sec:outflow_analytical_estimate_harder} that this is to be expected
on the basis of an analytical model.
Concerning the reconnection outflows, mass and energy conservation imply that relativistic inflow magnetization 
results in relativistic temperatures and/or relativistic bulk velocities in the outflows, but say nothing on the 
balance between the two. 
In Sect.~\ref{sec:outflows} we show that in our simulations, thermal energy largely dominates over bulk kinetic energy.
We demonstrate analytically in Sect.~\ref{sec:outflow_analytical_estimate_harder} that this is to be expected for large inflow 
magnetization, under the assumption that thermal inertia significantly contributes in Ohm's law.
This is an important question that has observational consequences.
In Sect.~\ref{sec:islands_no_guide_field} we detail the structure of the magnetic islands 
and of their central density dips and isolated centers.
Section~\ref{sec:rec_electric_field} studies the reconnection electric field.
The relevant normalization is non-trivial for relativistic setups,
and we propose to use the relativistic Alfv\'en speed in the inflow, which leads to rates in a close range.
Section~\ref{sec:results_guide_field} highlights differences resulting from the presence of a guide magnetic field.
% \modifbis{We describe the overall structure in Sect.~\ref{sec:guide_field_overall_structure},
% stressing that the relation $E<B$ holds everywhere, and that non-idealness and particle acceleration is allowed at places where $\b{E}\cdot\b{B}\neq 0$.
% Section~\ref{sec:rec_electric_field_guide_field} shows that the reconnection rates decrease with increasing guide fields.}
We summarize and conclude our work in Sect.~\ref{sec:ccl}, and discuss applications to astrophysical objects.

%%%%%%%%%%%%%%%%%%%%%%%%%%%%%%%%%%%%%%%%%%%%%%%%%%%%%%%%%%%%%%%%%%%%%%%%%%%%%%%%%%%%%%%%%%%%%%%%%%%%%%%%%%%%%%%%%%%%%%%%%%%%%%%%%%%%%%%%%%%%%%%%
%%%%%%%%%%%%%%%%%%%%%%%%%%%%%%%%%%%%%%%%%%%%%%%%%%%%%%%%%%%%%%%%%%%%%%%%%%%%%%%%%%%%%%%%%%%%%%%%%%%%%%%%%%%%%%%%%%%%%%%%%%%%%%%%%%%%%%%%%%%%%%%%
\section{Problem setup}\label{sec:pb_setup}
%%%%%%%%%%%%%%%%%%%%%%%%%%%%%%%%%%%%%%%%%%%%%%%%%%%%%%%%%%%%%%%%%%%%%%%%%%%%%%%%%%%%%%%%%%%%%%%%%%%%%%%%%%%%%%%%%%%%%%%%%%%%%%%%%%%%%%%%%%%%%%%%
%%%%%%%%%%%%%%%%%%%%%%%%%%%%%%%%%%%%%%%%%%%%%%%%%%%%%%%%%%%%%%%%%%%%%%%%%%%%%%%%%%%%%%%%%%%%%%%%%%%%%%%%%%%%%%%%%%%%%%%%%%%%%%%%%%%%%%%%%%%%%%%%

%%%%%%%%%%%%%%%%%%%%%%%%%%%%%%%
%%%%%%%%%%%%%%%%%%%%%%%%%%%%%%%
\subsection{Description of the relativistic Harris equilibrium}
%%%%%%%%%%%%%%%%%%%%%%%%%%%%%%%
%%%%%%%%%%%%%%%%%%%%%%%%%%%%%%%

\begin{table}[tbp]
\caption{\label{tab:param_physical_tearing}Parameters of the current sheet.
They hold for a mass ratio of 25, and are independent of the background plasma parameters. 
The free variables are $\wce/\wpe$ and $L/d_\ion$.
The electron and ion temperatures are the same, normalized as
$\Theta_s = T_s/(m_s c^2)$. 
The ions and electrons counterstream with opposite velocities $\pm U_\lec\hat{\b{y}}$
(given here in units of $c$) and associated Lorentz factors $\Gamma_\lec$.
The sheet half-width in units of ion inertial lengths is $L/d_\ion$, 
while in units of the thermal Larmor radii (at current sheet center) it is $L/r_\mathrm{ce}$.
}
\centering
\begin{tabular}{c|c||c|c|c|c|c}
 $\wce/\wpe$ & $L/d_\ion$ & $\Gamma_\lec U_\lec$ & $\Theta_\lec$ & $\Theta_\ion$ & $r_\ce/d_\lec$ & $L/r_\ce$ \\
\hline
 1 & 0.5 & 0.20 & 0.25 & 0.01  & 0.7 & 3.8 \\
 3 & 0.5 & 0.53 & 2.40 & 0.096 & 1.6 & 1.6 \\
 6 & 1   & 0.70 & 10   & 0.4   & 3.3 & 1.5 \\
\end{tabular}
\end{table}

\begin{table*}[tbp]
\caption{\label{tab:param_magnetization}Physical input parameters of the simulations 
and resulting magnetizations of the \textit{background} plasma.
The enthalpy of the background plasma is $h_{0,\bg,s}$. 
Its cold magnetization $\sigma^\mathrm{cold}_s(B)$ is defined by Eq.~\ref{equ:sigma_s_cold},
$\sigma^\mathrm{hot}_s$ by Eq.~\ref{equ:sigma_s}, and 
$\sigma_{\ion+\lec}(B)$ by Eq.~\ref{equ:sigma_tot}.
In all cases, we assume $\Gamma_\ins \sim 1$.
The background plasma $\beta_s = n_sT_s/(B^2/2\mu_0) = 2\Theta_s/\sigma_{s}^\mathrm{cold}(B)$ includes the guide field 
($\sigma_{s}^\mathrm{cold}(B_\mathrm{tot})=\sigma_{s}^\mathrm{cold}(B_0)+\sigma_{s}^\mathrm{cold}(B_\mathrm{G})$).
The Alfv\'en speeds, defined in Sect.~\ref{sec:Alfven_velocities}, 
do not take into account the temperature, and are given in units of $c$.
For the relativistic Alfv\'en speed, when there is a guide field we display the $\hat{\b{z}}$-projection:
${V^\mathrm{R}_\mathrm{A,in}}\cos\theta$, with $\theta = \arctan B_\mathrm{G}/B_0$.
}
\centering
\begin{tabular}{c|c|c||c|c|c|c|c|c||c|c}
 $\wce/\wpe$ & $n_\mathrm{bg}/n_\cs(0)$ & $B_\mathrm{G}/B_0$ & & $T_{\bg,s}$ (K) & $\beta_s$ & $h_{0,\bg,s}$ & $\sigma^\mathrm{cold}_s(B_\mathrm{rec})$ & $\sigma^\mathrm{hot}_s(B_\mathrm{rec})$ & $\sigma_{\ion+\lec}(B_\mathrm{rec})$ & ${V^\mathrm{NR}_\mathrm{A,in}}$, ${V^\mathrm{R}_\mathrm{A,in}}$ \\
\hline
 1 & 0.1 & 0   &  ion & $1.5\times10^7$ & $5\times10^{-4}$   & 1     & 0.4 & 0.4  & 0.38 & 0.63, 0.53 \\
   &     &     &  lec & \idem           & \idem              & 1.006 & 10  & 9.94 &      & \\
\hline
 3 & 0.31& 0   &  ion & $2\times10^8$   & $2.5\times10^{-3}$ & 1     & 1.16 & 1.16 & 1.11& 1.08, 0.73 \\
   &     &     &  lec & \idem           & \idem              & 1.086 & 29   & 27   &     & \\
\hline
 3 & 0.1 & 0   &  ion & $2\times10^8$   & $7.5\times10^{-4}$ & 1     & 3.6 & 3.6 & 3.26 & 1.90, 0.88 \\
   &     &     &  lec & $3\times10^9$   & $1.1\times10^{-2}$ & 2.57  & 90  & 35  &      & \\
\hline
 3 & 0.1 & 0   &  ion & $1.5\times10^7$ & $5.6\times10^{-5}$ & 1     & 3.6 & 3.6 & 3.46 & 1.90, 0.88 \\
   &     &     &  lec & \idem           & \idem              & 1.006 & 90  & 89  &      & \\
\hline
 3 & 0.1 & 0   &  ion & $2\times10^8$   & $7.5\times10^{-4}$ & 1     & 3.6 & 3.6 & 3.45 & 1.90, 0.88 \\
   &     &     &  lec & \idem           & \idem              & 1.086 & 90  & 83  &      &\\
\hline
 3 & 0.1 & 0.5 &  ion & $1.5\times10^7$ & $4.5\times10^{-5}$ & 1     & 3.6 & 3.6 & 3.46 & 1.90, 0.81 \\
   &     &     &  lec & \idem           & \idem              & 1.006 & 90  & 89  &      & \\
\hline
 3 & 0.1 & 1   &  ion & $1.5\times10^7$ & $2.8\times10^{-5}$ & 1     & 3.6 & 3.6 & 3.46 & 1.90, 0.66 \\
   &     &     &  lec & \idem           & \idem              & 1.006 & 90  & 89  &      & \\
\hline
 6 & 0.1 & 0   &  ion & $8\times10^8$   & $7.5\times10^{-4}$ & 1.014 & 14.4 & 14.2 & 13.5& 3.80, 0.97 \\
   &     &     &  lec & \idem           & \idem              & 1.37  & 360  & 260  &     & \\
\end{tabular}
\end{table*}

We use the explicit particle-in-cell code \texttt{Apar-T}, presented and tested in \citet{Melzani2013}.
Broadly speaking, it is a parallel electromagnetic relativistic 
three dimensional PIC code with a staggered grid, where the fields are
integrated via Faraday and Maxwell-Amp\`{e}re equations, currents computed by a charge conserving volume weighting (CIC),
and fields interpolated accordingly.

The simulations start from a Harris equilibrium,
which is a solution of the Vlasov-Maxwell system. 
The magnetic field is 
$\b{B}_\mathrm{rec} = \hat{\b{z}}\, B_0 \tanh\left( {x}/{L} \right)$
(see Figs.~\ref{fig:overview_lec_number} 
or~\ref{fig:xcf_wcewpe=3_NT=6000_2D_pseudocolor_illustration} for axis orientation), and is
sustained by a population of electrons and ions of equal number density 
$n_\mathrm{cs}(x) = n_\mathrm{cs}(0)/\mathrm{cosh}^2 (x/L)$ (cs stands for current sheet), flowing
with bulk velocities $U_\lec$ and $U_\ion=-U_\lec$ in the $\pm y$ directions. 
We denote the associated Lorentz factors by $\Gamma_\lec$ and $\Gamma_\ion$.
Each species follows a Maxwell-J\"uttner distribution (Eq.~\ref{equ:Jutt_Harris}) of normalized temperature 
$\Theta_s = 1/\mu_s = T_s/m_sc^2$.

We derived the equilibrium relations for relativistic temperatures and current drift speeds, as well as for 
arbitrary ion to electron mass ratios and temperature ratios, in \citet{Melzani2013}.
Details specific to the present application can be found in Appendix~\ref{app:relat_Harris_again}.
The equilibrium depends on the ratio $\wce/\wpe$, 
with $\wpe=(n_\cs(0) e^2/(\epsilon_0m_\lec))^{1/2}$
the electron plasma pulsation defined with the lab-frame number density $n_\cs(0)$ (not including background particles)
and $\wce = eB_0/m_\lec$ the electron cyclotron pulsation in the asymptotic magnetic field (not including the guide field, $e>0$).
Our simulations are, however, not loaded exactly with the equilibrium values,
but with a temperature and current speed uniformly 
in excess of $\sim 10\%$ in order to shorten the otherwise rather long stable phase.

We also set a background plasma of number density (for electrons or for ions) 
$n_\mathrm{bg}$ and of temperature $T_{\bg,\lec}$ for electrons and $T_{\bg,\ion}$ for ions.
Finally, a guide magnetic field is sometimes considered, i.e., a uniform component $\b{B}_\mathrm{G} = B_\mathrm{G}\hat{\b{y}}$.
Adding the background plasma or the guide field does not change the Harris equilibrium.

%%%%%%%%%%%%%%%%%%%%%%%%%%%%%%%
%%%%%%%%%%%%%%%%%%%%%%%%%%%%%%%
\subsection{Magnetization and energy fluxes}
%%%%%%%%%%%%%%%%%%%%%%%%%%%%%%%
%%%%%%%%%%%%%%%%%%%%%%%%%%%%%%%

There are several ways to characterize the magnetization of the configuration. 
The ratio $\wce/\wpe$ has no direct physical meaning and is mostly used as a simulation label.\note{with 
$\wce$ defined in the inflow and $\wpe$ defined at the center of the current sheet:
 \begin{equation}\label{equ:wcewpe_us}
  \left(\frac{\wce}{\wpe}\right)^2 = \frac{e^2B_0^2/m_\lec^2}{n_\cs(0)e^2/(\epsilon_0m_\lec)} = \frac{B_0^2}{\mu_0n_\cs(0)m_\lec c^2}.
 \end{equation}}

The magnetization $\sigma^\mathrm{hot}_s$ of the background plasma species $s$ is the ratio of the energy flux 
in the reconnecting magnetic field to that in the particles (rest-mass, thermal, bulk).
The electromagnetic energy flux is the Poynting flux. Far from the current sheet, it reads
\begin{equation}\label{equ:energy_flux_fields}
 \frac{\b{E}\wedge\b{B}}{\mu_0} = \frac{\b{E}\wedge\b{B}_\mathrm{rec}}{\mu_0} + \frac{\b{E}\wedge\b{B}_\mathrm{G}}{\mu_0} = \frac{B_\mathrm{rec}^2}{\mu_0}v_{E_y\times B_\mathrm{rec}}\hat{\b{x}} + \frac{B_\mathrm{G}^2}{\mu_0}v_{E_y\times B_\mathrm{rec}}\hat{\b{x}},
\end{equation}
where $\b{B} = B_\mathrm{rec}\hat{\b{z}} + B_\mathrm{G}\hat{\b{y}}$ and $v_{E_y\times B_\mathrm{rec}} = E_y/B_\mathrm{rec}$. 
This splitting of the energy flux in two contributions, 
one from the magnetic field that will reconnect, the other from the guide field that will mostly be compressed,
is possible only if the electric field is normal to $\b{B}$, 
which is indeed the case in the ideal outer area because of the tendency of the plasma to screen parallel electric fields.
For the particles, the energy flux of species $s$ is (Eq.~\ref{equ:energy_relat_lab_all_species}):
\begin{equation}\label{equ:energy_flux_particles}
  n_{\mathrm{lab},s}\langle v\gamma m_sc^2\rangle_s =  n_{\mathrm{lab},s} \Gamma_s h_{0,s} \bar{v}_s m_sc^2,
\end{equation}
with $n_{\mathrm{lab},s}$ the particle number density in the lab frame ($=\Gamma_s$ times that in the comobile frame),
$\langle\cdot\rangle_s$ denoting an average over momentum of the distribution function,
$m_s$ the particle mass, $\bar{v}_s$ their bulk velocity, $\Gamma_s$ the associated Lorentz factor, and 
$h_{0,s}$ their comobile enthalpy (drawn in Fig.~\ref{fig_kappa_32} for a thermal distribution).
All in all, the magnetization of species $s$ is\note{Another possibility is to include the energy in the guide field, 
which may be justified because the guide field does not reconnect:
\begin{equation}\label{equ:sigma_s_with_B_guide}
\begin{aligned}
 \sigma^\mathrm{tot}_s &= \frac{E\times B_\mathrm{rec}/\mu_0}{n_{\mathrm{lab},s}\langle v\gamma m_sc^2\rangle_s + E\times B_\mathrm{G}/\mu_0} = \frac{B_\mathrm{rec}^2/\mu_0}{n_\mathrm{lab}m_sc^2\Gamma_s h_{0,s} + B_\mathrm{G}^2/\mu_0} \\
    &= \frac{\sigma^\mathrm{cold}_s(B_\mathrm{rec})}{\Gamma_s h_{0,s} + \sigma^\mathrm{cold}_s(B_\mathrm{G})},~~~~\sigma^\mathrm{cold}_s(B) = \frac{B^2}{\mu_0n_\mathrm{lab}m_sc^2}.
\end{aligned}
\end{equation}}: 
\begin{equation}\label{equ:sigma_s}
\begin{aligned}
 \sigma^\mathrm{hot}_s(B_\mathrm{rec}) &= \frac{E\times B_\mathrm{rec}/\mu_0}{n_{\mathrm{lab},s}\langle v\gamma m_sc^2\rangle_s} = \frac{B_\mathrm{rec}^2/\mu_0}{n_{\mathrm{lab},s}m_sc^2\Gamma_s h_{0,s}} \\
    &= \frac{\sigma^\mathrm{cold}_s(B_\mathrm{rec})}{\Gamma_s h_{0,s}},
\end{aligned}
\end{equation}
with $\sigma^\mathrm{cold}_s$ the magnetization of the plasma without taking into account temperature effects and relativistic 
bulk motion:
\begin{equation}\label{equ:sigma_s_cold}
 \sigma^\mathrm{cold}_s(B) = \frac{B^2}{\mu_0n_{\mathrm{lab},s}m_sc^2}.
\end{equation}
If $\sigma^\mathrm{cold}_s(B_\mathrm{rec})$ exceeds unity, then it is possible to pass to the particles an amount of energy 
from the reconnecting field that exceeds their rest-mass, 
i.e., it is possible to obtain relativistic particles.
We do not include the guide field $B_\mathrm{G}\hat{\b{y}}$ in the definition of the magnetization because it is mostly compressed and 
does not transfer energy to the particles.

Finally, the total magnetization of the plasma is:
\begin{equation}\label{equ:sigma_tot}
 \sigma_{\ion+\lec}(B_\mathrm{rec}) = \frac{B_\mathrm{rec}^2/\mu_0}{\sum_sn_{\mathrm{lab},s}m_sc^2\Gamma_s h_{0,s}}
    = \frac{\sigma^\mathrm{cold}_\ion(B_\mathrm{rec})}{\sum_s\Gamma_s h_{0,s}(m_s/m_\ion)}.
\end{equation}
In the inflow part of our simulations, 
we have for the range of background temperatures considered here: $h_{0,\ion} \sim 1$, $h_{0,\lec}<2.6$, and $m_\ion=25m_\lec$,
so that the particle energy flux is largely dominated by the rest-mass energy flux of the ions, 
which is $n_\mathrm{lab,i}\bar{v}_\ion m_\ion c^2$, 
and has no temperature dependence. 
\note{We thus have 
\begin{equation}\label{equ:sigma_Zenitani_for_us}
 \sigma_\mathrm{i+e}^\mathrm{tot} \simeq \frac{B_\mathrm{rec}^2}{\mu_0n_\mathrm{lab} m_\ion c^2\Gamma_\mathrm{in}}.
\end{equation}}
We thus have $\sigma_\mathrm{i+e} \sim \sigma^\mathrm{cold}_\ion$,
and $\sigma_\mathrm{i+e}$ is not a good representative of the electron physics and of the possibility that they be relativistically magnetized.
% The link between Eq.~\ref{equ:sigma_Zenitani_for_us} and Eq.~\ref{equ:wcewpe_us} 
% is $(\wce/\wpe)^2 = (m_\ion/m_\lec)[n_\mathrm{in}/n_\cs(0)]\Gamma_\mathrm{in} \times \sigma_\mathrm{i+e}^\mathrm{tot}$.
The inflow magnetizations in our simulations are presented in Table~\ref{tab:param_magnetization}.
\note{We note that this coincides with the definition of
\citet{Zenitani2009b} for a pair plasma,
\begin{equation}\label{equ:sigma_Zenitani}
 \sigma_\mathrm{i+e}^\mathrm{tot} = \left( \frac{B^2}{(2h_0n_0mc^2)\mu_0\Gamma^2} \right)_\mathrm{in} = \left( \frac{|\b{E}\wedge \b{B}|/\mu_0}{2n_\mathrm{lab}\langle v\gamma mc^2\rangle} \right)_\mathrm{in},
\end{equation}
where the factor 2 is here because the energy flux is twice that for a single species.
}

%%%%%%%%%%%%%%%%%%%%%%%%%%%%%%%
%%%%%%%%%%%%%%%%%%%%%%%%%%%%%%%
\subsection{Alfv\'en velocities}
%%%%%%%%%%%%%%%%%%%%%%%%%%%%%%%
%%%%%%%%%%%%%%%%%%%%%%%%%%%%%%%
\label{sec:Alfven_velocities}

We give the definitions of the Alfv\'en speeds that will be used to discuss the normalization of 
the reconnection electric field in Sect.~\ref{sec:rec_electric_field}.
They are reported in Table~\ref{tab:param_magnetization}.

The Alfv\'en velocity in the inflow plasma, far from the current sheet,
% and with no temperature correction, which is justified except in the hottest case $T_{\bg,\lec}=3\times10^9$\,K), 
is expressed in the comobile plasma frame and is, respectively in the non-relativistic and relativistic cases:
\begin{subequations}
\begin{align}
 V_\mathrm{A,in}^\mathrm{NR}  &= \frac{B_\tot}{\sqrt{\mu_0 n_{0,\bg} (m_\lec+m_\ion)}}, \label{equ:inflow_NR_Alfven_speed} \\
 V_\mathrm{A,in}^\mathrm{R}   &= c\,\left(\frac{\sigma_{\ion+\lec}(B_\tot)}{1+\sigma_{\ion+\lec}(B_\tot)}\right)^{1/2} \simeq \frac{V_\mathrm{A,in}^\mathrm{NR}}{\sqrt{\left(V_\mathrm{A,in}^\mathrm{NR}\right)^2/c^2+1}}, \label{equ:inflow_R_Alfven_speed}
\end{align}
\end{subequations}
where $\sigma_{\ion+\lec}(B_\tot)$ is to be expressed in the comobile frame 
(Eq.~\ref{equ:sigma_tot} with $n_{\mathrm{lab},s}=n_{0,\bg}$ the comobile density and $\Gamma_s=1$),
and where $B_\tot = (B_0^2 + B_\mathrm{G}^2)^{1/2}$.
For the relativistic expression~\ref{equ:inflow_R_Alfven_speed}, the first equality is general and 
derived from the relativistic ideal MHD description \citep{Gedalin1993},
while the second holds only because $m_\ion \gg m_\lec$ and the total enthalpy is dominated by the ion contribution.
When there is a guide magnetic field, we show in Sect.~\ref{sec:rec_electric_field_guide_field} that it is relevant 
to project the Alfv\'en speed into the direction of the reconnecting magnetic field ($\hat{\b{z}}$), i.e., to consider 
$V_\mathrm{A,in}^\mathrm{R}\cos\theta$ with $\tan\theta = B_\mathrm{G}/B_0$.

A hybrid Alfv\'en speed is often defined in the literature, 
as depending on the asymptotic 
magnetic field (without the guide field) and on the 
comobile density at the center of the current sheet:
\begin{equation}\label{equ:hybrid_NR_Alfven_speed}
 V_\mathrm{A,0}^\mathrm{NR} = \frac{B_0}{\sqrt{\mu_0 n_{0,\cs}(0)(m_\lec+m_\ion)}} = \left(\frac{m_\lec}{m_\lec+m_\ion}\right)^{1/2} \frac{\wce}{\omega_{0,\pe}}\,c,
\end{equation}
where a subscript 0 indicates a comobile quantity.
Its relativistic generalization is denoted by $V_\mathrm{A,0}^\mathrm{R}$,
and is obtained with Eq.~\ref{equ:inflow_R_Alfven_speed} but with parameters of the plasma at the center of the current sheet in 
the magnetization.
% \begin{equation}\label{equ:hybrid_R_Alfven_speed}
%  V_\mathrm{A,0}^\mathrm{R} = \frac{V_\mathrm{A,0}^\mathrm{NR}}{\sqrt{\left(V_\mathrm{A,0}^\mathrm{NR}\right)^2+h_0(T)c^2}}.
% \end{equation}

%%%%%%%%%%%%%%%%%%%%%%%%%%%%%%%
%%%%%%%%%%%%%%%%%%%%%%%%%%%%%%%
\subsection{Simulation parameters and resolution tests}
%%%%%%%%%%%%%%%%%%%%%%%%%%%%%%%
%%%%%%%%%%%%%%%%%%%%%%%%%%%%%%%

\begin{figure}[tbp]
\centering
  \includegraphics[width=\columnwidth]{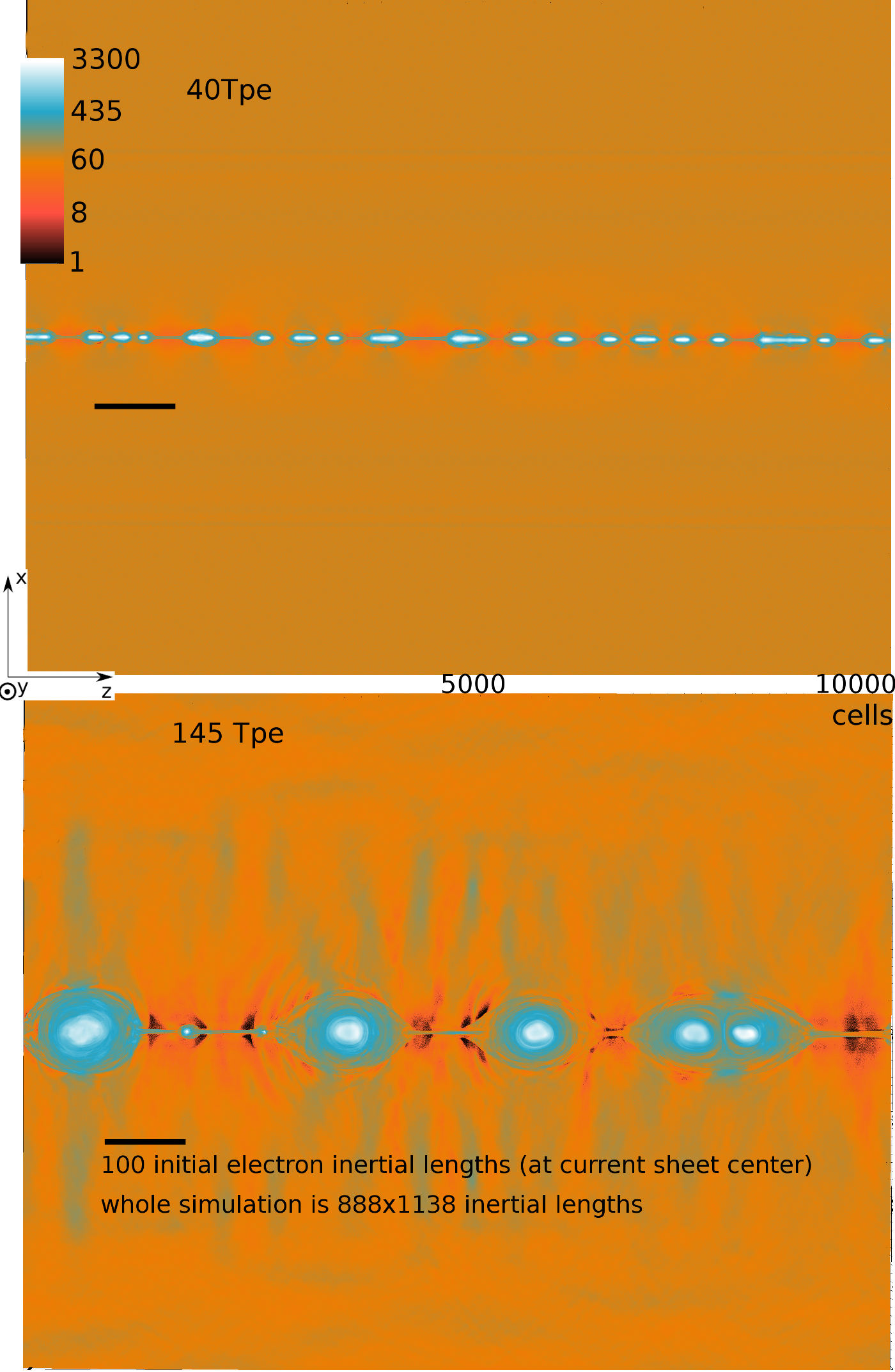}
  \caption{\label{fig:overview_lec_number}Electron number density in the whole simulation domain, at two different times.
  From run $\wce/\wpe=3$, $\sigma_\lec^\mathrm{hot}=89$, $T_\bg=1.5\times10^7$\,K.
  Units are the number of electron superparticles per cell, lengths are in cell numbers.
  Here $40T_\pe = 30\omega_\ci^{-1} = 750\wce^{-1}$.
  }
\end{figure}

The physical parameters of the main simulations are given in Tables~\ref{tab:param_physical_tearing} and~\ref{tab:param_magnetization}.
We consider a mass ratio $m_\ion/m_\lec = 25$, except for one simulation with pairs. 
The background plasma number density is
$n_\bg=0.1n_\cs(0)$ or $0.3n_\cs(0)$. 
Its temperature is varied between $T_\bg = 1.5\times10^7\,$K ($2.5\times10^{-3}m_\lec c^2$)
and $3\times10^9\,$K ($0.5m_\lec c^2$).
The magnetization depends on the ratio $\wce/\wpe = 1$, 3, or 6,
leading to inflow magnetizations $\sigma^\mathrm{hot}_s$ between 10 and 260 for electrons, or 0.4 and 14 for ions. 
The current sheet is either of initial half-width $L = 0.5d_\ion = 2.5d_\lec$, or $L=1d_\ion$ in the $\wce/\wpe = 6$ case,
with $d_\lec$, $d_\ion$ the inertial lengths defined at current sheet center. 
We stress that for relativistic temperatures the sheet width in terms of Larmor radii will not be the same for simulations with 
different $\wce/\wpe$ (Eq.~\ref{equ:rce_over_de}); see $L/r_\mathrm{ce}$ in Table~\ref{tab:param_physical_tearing}.

The numerical resolution is set by the number of cells $n_x$ per electron inertial length $d_\lec$,
by the number of timesteps $n_t$ per electron plasma period $2\pi/\wpe$, 
and by the number of computer particles (the so-called superparticles) per cell $\rho_{\mathrm{sp}}$.
The quantities $d_\lec$, $\omega_\pe$, and $\rho_{\mathrm{sp}}$ are defined at $t=0$ at the center of 
the current sheet, where the particle density is highest.
For the simulations of Table~\ref{tab:param_magnetization}, we take $n_x=9$ and $n_t=150$
($250$ for $\wce/\wpe=6$).
We checked with a simulation with twice this resolution ($\wce/\wpe=3$, $n_x=18$, $n_t=250$) 
that all of the presented results are not affected\footnote{Throughout this section, 
``all the results'' means the time evolution of the reconnection rate 
and of the width of the diffusion zone for electrons and ions, 
the distribution functions of high-energy particles, 
the temperature curves in the diffusion zone,
the energy content in the outflows,
as well as the relative weight of the terms in Ohm's law.}.
% \modif{(still need to check this one)}

Concerning the number of superparticles per cell, the simulations of Table~\ref{tab:param_magnetization}
use $\rho_{\mathrm{sp}}=1820$ ($1090$ for $n_\bg/n_\mathrm{cs}(0)=0.3$).
This corresponds, for the case $n_\bg/n_\mathrm{cs}(0)=0.1$, to 1650 electron and ion superparticles per cell 
for the plasma of the current sheet, 
and to 170 for the background plasma. The density profile of the current sheet plasma is set by changing the number 
of superparticles per cell when going away from the center.
We stressed in \citet{Melzani2013,Melzani2013b} that because of their low numbers of superparticles 
per cell when compared to real plasmas, PIC simulations present high levels of collisionality. 
One should thus ensure that collisionless kinetic processes remain faster than collisional effects (e.g., for thermalization),
essentially by taking a large enough number $\Lambda^\mathrm{PIC}$ of superparticles per Debye sphere.
For example, with $\Theta_\lec=2.4$
the electron Debye length is 20 cells large, and we have initially at the center of the current sheet: 
$\Lambda^\mathrm{PIC} \sim 364\times20\times20 = 7.3\times10^5$ superparticles.
For a background plasma with $T_\bg=2\times10^8$\,K: $\Lambda^\mathrm{PIC}=133$.
\note{Note: we use $\tilde{\lambda}_D = n_xv_\mathrm{th}/c = n_x\Theta^{1/2}$ for the non-relativistic case, 
and for the UR case we take $\tilde{\lambda}_D = \sqrt{2}\times n_x\Theta^{1/2}$. 
(cf \url{http://iopscience.iop.org/0741-3335/55/6/065006/article})
For $n_x=9$ and $\Theta=2.4$, it gives $19.7$ cells.}
We performed a simulation identical to that with $\wce/\wpe=3$, $n_\bg/n_\cs(0)=0.1$, $T_\bg=2\times10^8$\,K,
but with half the superparticles, and found no change in the results.
It shows that the main simulations use a large enough $\rho_\mathrm{sp}$.

Boundaries are periodic along $z$ and $y$. At the top and bottom $x$ boundaries we use reflective boundaries, i.e., 
we place a perfectly conducting wall that reflects waves and particles.
The number of cells for the standard simulations is $4100\times6144$. 
The length along $y$ is of no dynamical importance, and the dimensions correspond
to a 2D simulation with $455$ initial electron inertial lengths along $x$ and $683$ along $z$, 
with typically $4\times10^9$ superparticles.
It takes $70\,T_\mathrm{pe}$ for light waves to start from the current sheet, 
reflect at the $\pm x$ boundaries, and come back to the sheet.
This corresponds to $(18,\,52,\,106)\wci^{-1}$, or $(450,\,1300,\,2650)\wce^{-1}$
for, respectively, simulations with $\wce/\wpe=1,\,3,\,6$.
The light travel time in the $z$ direction is larger.
Except for run $\wce/\wpe=1$, all the analyses presented here are for smaller times and are thus not affected by boundaries.
To check this, we performed a larger simulation with $8000\times10240$ cells (i.e., 
$888\times1138$ initial electron inertial length) for the case $\wce/\wpe=3$, $n_\bg/n_\cs(0)=0.1$, $T_\bg=2\times10^8$\,K, with 
the same $n_t$, $n_x$, $\rho_\mathrm{sp}$.
The corresponding light-crossing time is now $136\,T_\mathrm{pe} = 101\wci^{-1} = 2535\wce^{-1}$.
All the results are the same, which shows that we do not suffer from boundary effects.

%%%%%%%%%%%%%%%%%%%%%%%%%%%%%%%%%%%%%%%%%%%%%%%%%%%%%%%%%%%%%%%%%%%%%%%%%%%%%%%%%%%%%%%%%%%%%%%%%%%%%%%%%%%%%%%%%%%%%%%%%%%%%%%%%%%%%%%%%%%%%%%%
%%%%%%%%%%%%%%%%%%%%%%%%%%%%%%%%%%%%%%%%%%%%%%%%%%%%%%%%%%%%%%%%%%%%%%%%%%%%%%%%%%%%%%%%%%%%%%%%%%%%%%%%%%%%%%%%%%%%%%%%%%%%%%%%%%%%%%%%%%%%%%%%
\section{Results with no guide field}
%%%%%%%%%%%%%%%%%%%%%%%%%%%%%%%%%%%%%%%%%%%%%%%%%%%%%%%%%%%%%%%%%%%%%%%%%%%%%%%%%%%%%%%%%%%%%%%%%%%%%%%%%%%%%%%%%%%%%%%%%%%%%%%%%%%%%%%%%%%%%%%%
%%%%%%%%%%%%%%%%%%%%%%%%%%%%%%%%%%%%%%%%%%%%%%%%%%%%%%%%%%%%%%%%%%%%%%%%%%%%%%%%%%%%%%%%%%%%%%%%%%%%%%%%%%%%%%%%%%%%%%%%%%%%%%%%%%%%%%%%%%%%%%%%
\label{sec:results}

This Section explores results for simulations with no guide field, where the magnetic field above and 
below the current sheet is antiparallel.
% Section~\ref{sec:overall_evolution} describes the overall evolution of the reconnection event,
% Sect.~\ref{sec:inflow} explores the structure of the diffusion region and the inflows, 
% Sect.~\ref{Sec:Ohms_law} details Ohm's law,
% Sect.~\ref{sec:outflows} concerns the reconnection outflows, 
% Sect.~\ref{sec:islands_no_guide_field} the islands,
% and Sec~\ref{sec:rec_electric_field} the reconnection electric field.

%%%%%%%%%%%%%%%%%%%%%%%%%%%%%%%
%%%%%%%%%%%%%%%%%%%%%%%%%%%%%%%
\subsection{Overall structure and evolution}
%%%%%%%%%%%%%%%%%%%%%%%%%%%%%%%
%%%%%%%%%%%%%%%%%%%%%%%%%%%%%%%
\label{sec:overall_evolution}

\begin{figure}[tbp]
\centering
  \includegraphics[width=\columnwidth]{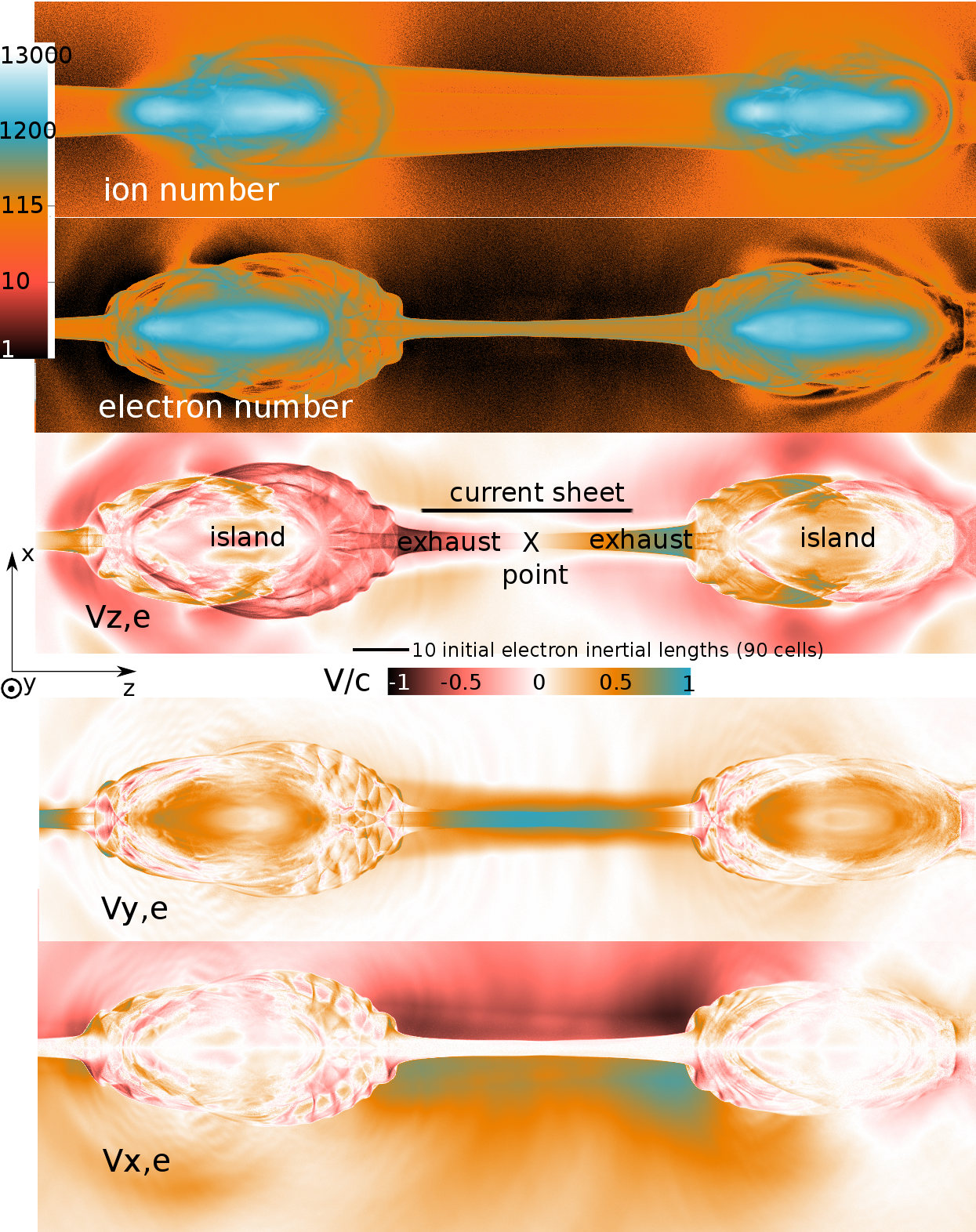}
  \caption{\label{fig:xcf_wcewpe=3_NT=6000_2D_pseudocolor_illustration}
  Zoom around a X-point showing various fluid quantities.
  From run $\wce/\wpe=3$, $\sigma_\lec^\mathrm{hot}=89$, $T_\bg=1.5\times10^7$\,K, at $t=40T_\pe=30\omega_\ci^{-1}=750\omega_\ce^{-1}$% (iteration 6000) 
  and $z\sim3024$ cells. Units for particle densities are particle numbers per cell.
  The temperatures at the same time are shown in Fig.~\ref{fig:xcf_wcewpe=3_NT=6000_2D_pseudocolor_temperature_NT=6000},
  and cuts along $x$ and $z$ in Figs.~\ref{fig_wcewpe=1_NT=21300_cutalongX_partNumber_velocity} (upper-right),
  \ref{fig:wcewpe=3_NT=6001_Ohms_law},
  \ref{fig:wcewpe=3_cut_Z_NT6000_temperature_only}, and \ref{fig_wcewpe=3_NT=6000_cutalongZ_summary} (right).
  }
\end{figure}

\begin{figure}[tb]
 \centering
 \def\svgwidth{\columnwidth}
 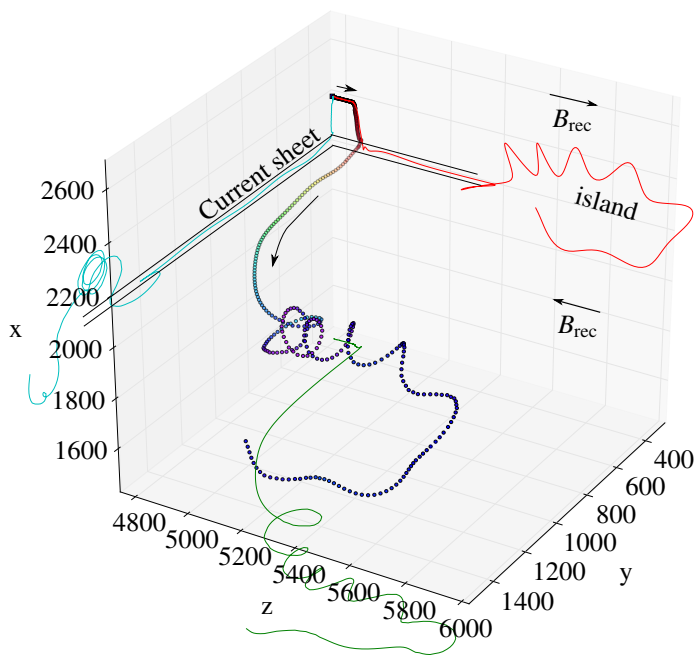
 \caption{\label{fig:traj_3D_no_guide_filed}Typical trajectory for a particle, here an electron.
 Note the Speiser-like oscillatory motion inside the current sheet.
 Axis scales are given in cell numbers, with 9 cells representing one initial electron inertial length $d_\lec$. Dot colors are the particle
 Lorentz factor. Solid lines are projections onto the $x$-$y$, $y$-$z$, and $z$-$x$ planes.
 The $y$ direction is not described in the simulations, and is here reconstructed on the basis of an invariance of the 
 electromagnetic fields along $y$.}
\end{figure}

We first summarize the general picture, rising important points that are detailed in the next subsections.
The initial kinetic equilibrium is unstable to the collisionless tearing mode, 
which in all presented simulations is triggered by the noise level and by the slightly out of equilibrium initial state.
We have studied in detail the linear phase of this instability with PIC simulations \citep{Melzani2013} for pair plasmas,
and found growth rates within 5\% of the analytical derivations of \citet{Petri2007} made on the basis of 
a linearization of the Vlasov-Maxwell system.
Physically, the instability is driven by the particles freely bouncing in the layer of magnetic field reversal \citep{Coppi1966}, 
with a mechanism similar to a filamentation instability: perturbations in $B_x$ and $B_z$ lead to a bunching of the particles,
which in turn increase the magnetic field perturbation.
It leads to the formation of alternating X- and O-points, here with no privileged location because 
we impose no localized initial perturbation (Fig.~\ref{fig:overview_lec_number}).

With the appearance of X- and O-points, the magnetic flux variations across fixed contours
induce an out-of-plane electric field $E_y\hat{\b{y}}\propto-\hat{\b{y}}$, that amplifies the initial current along $-y$,
that increases the magnetic field in order to cancel the former magnetic flux variations and prevent 
reconnection. However, non-ideal processes forbid an ideal plasma response, and allow the triggering of reconnection and the 
existence of a finite electric field at the current sheet center, where ideal Ohm's law would otherwise read $E_y=0$.
This electric field $E_y$ is at the heart of the reconnection process, as it is responsible for 
the transfer of energy from the magnetic field to particles in the diffusion region.
We detail Ohm's law and the contribution of each non-ideal terms in Sect.~\ref{Sec:Ohms_law}.

Plasma and magnetic fields decouple in the non-ideal region, and flux tubes can ``reconnect'', producing new flux tubes 
strongly bent that accelerate the plasma outward in the $\pm\hat{\b{z}}$ directions, thus producing the exhaust outflows.
This depletion of particles and/or the spreading of the electric field $E_y$ 
outside the current sheet create an inflow from the ideal zone toward the 
current sheet: particles $E\times B$ drift at a speed $\b{v}_\mathrm{in} = \b{E}\times \b{B}/B^2 = E_y/B_z \hat{\b{x}}$.
The incoming particles are then accelerated along $\hat{\b{y}}$ by $E_y$ once they enter the non-ideal region
where they are unmagnetized (because there $E>B$). 
The structure of this central region is investigated in Sect.~\ref{sec:inflow}.

The exhaust outflows, which in the MHD view are driven by the magnetic field tension force, 
are produced by particles accelerated by $E_y$ along $\hat{\b{y}}$ 
and then slowly rotating %(because $E_y<0$, from $-\hat{\b{y}}$ to $\pm\hat{\b{z}}$ for ions) 
due to the increasingly strong magnetic field $B_x$ as one goes away from the X-point (Fig.~\ref{fig:traj_3D_no_guide_filed}).
As they do so, particles still gain energy as long as $q_sE_y\hat{\b{y}}\cdot\b{v} > 0$. 
The energy content of these outflows comprises a Poynting flux, bulk kinetic energy, and thermal energy,
with respective weights that depend on the background plasma parameters, as studied in Sect.~\ref{sec:outflows}.
The balance between inflow and outflow can lead, depending on the conditions, to a steady state Sweet-Parker-like configuration.

In the configuration of the simulations the initial perturbation is not localized in space, 
so that several X-points appear, with 
islands in-between that collect the flux of particles and of reconnected magnetic field.
The islands are trapped between two exhausts, and
the bulk energy of the outflows is converted into heat by random scatterings in the complex electric and magnetic structures 
at the island entrance and inside the islands (Fig.~\ref{fig:xcf_wcewpe=3_NT=6000_2D_pseudocolor_temperature_NT=6000}),
with however particle distributions not necessarily thermal (see Sect.~\ref{sec:islands_no_guide_field}).
The islands grow, and since they are threaded by parallel currents (along $-\hat{\b{y}}$), attract each others
via the Lorentz force and merge, thus growing even more.
As time goes on, the island number dwindles and the space in-between them increases, 
forming elongated current sheets composed of a X-point surrounded by two 
elongated exhausts (see Fig.~\ref{fig:xcf_wcewpe=3_NT=6000_2D_pseudocolor_illustration}).
We stop the simulations when only two or three islands remain.

%%%%%%%%%%%%%%%%%%%%%%%%%%%%%%%
%%%%%%%%%%%%%%%%%%%%%%%%%%%%%%%
\subsection{Inflow: two-scale diffusion region and sharp transitions}
%%%%%%%%%%%%%%%%%%%%%%%%%%%%%%%
%%%%%%%%%%%%%%%%%%%%%%%%%%%%%%%
\label{sec:inflow}

% \begin{figure}[tbp]
% \centering
%   \includegraphics[width=\columnwidth]{images/fig_wcewpe=1_NT=21300_cutalongX_partNumber_velocity_2.pdf}
%   \includegraphics[width=\columnwidth]{images/fig_wcewpe=3_NT=0000000006000_cutX_various_slides_pdftex.pdf}
%   \includegraphics[width=\columnwidth]{images/wcewpe3_Te_bg2e8_ok_cut_along_x_z2452_NT6000_various.pdf}
%   \includegraphics[width=\columnwidth]{images/wcewpe3_TbgHot_cut_along_x_z2420_NT6000_various.pdf}
%   \caption{\label{fig_wcewpe=1_NT=21300_cutalongX_partNumber_velocity}Cut along $x$ through the X-point for, from top to bottom, 
%   run wcewpe1: $t=142T_\pe=35\omega_\ci^{-1}=875\omega_\ce^{-1}$ (iteration 21300), $z=4824$;
%   run wcewpe3: $t=40 T_\pe=30\omega_\ci^{-1}=750\omega_\ce^{-1}$ (iteration 6000),  $z=3024$;
%   run wcewpe3: $t=40 T_\pe=30\omega_\ci^{-1}=750\omega_\ce^{-1}$ (iteration 6000),  $z=2452$;
%   run wcewpe3: $t=40 T_\pe=30\omega_\ci^{-1}=750\omega_\ce^{-1}$ (iteration 6000),  $z=2420$.}
% \end{figure}

\begin{figure*}[tbp]
 \centering
 \def\svgwidth{\textwidth}
 \begin{tiny}
 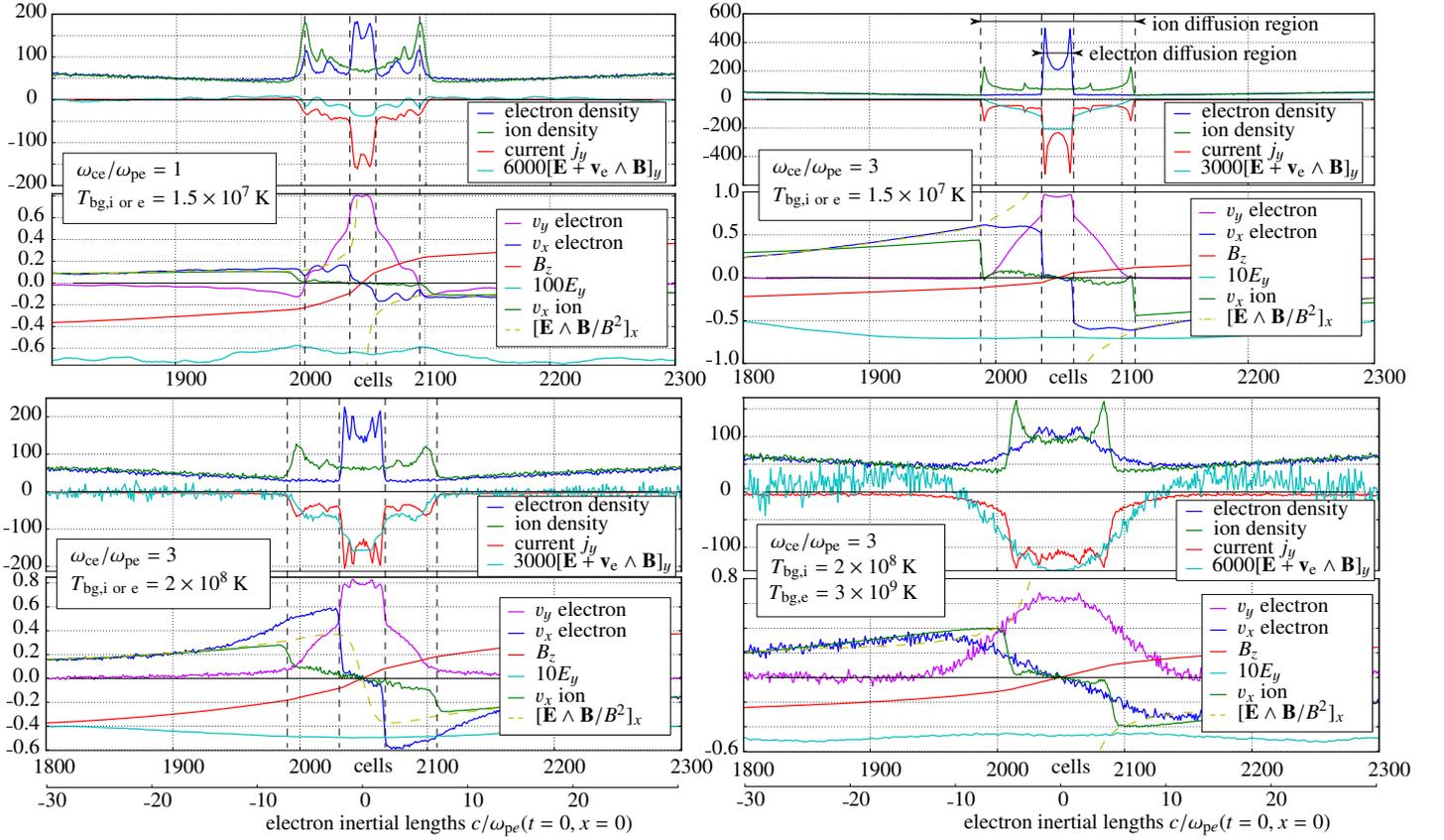
 \end{tiny}
 \caption{\label{fig_wcewpe=1_NT=21300_cutalongX_partNumber_velocity}Cut along $x$ through the X-point for four simulations,
  as indicated in the insets. The times are 
  $t=40 T_\pe=30\omega_\ci^{-1}=750\omega_\ce^{-1}$ %(iteration 6000) 
  for the cases $\wce/\wpe=3$,
  and $t=142T_\pe=35\omega_\ci^{-1}=875\omega_\ce^{-1}$ %(iteration 21300) 
  for the case $\wce/\wpe=1$. 
  }
\end{figure*}

% From top to bottom, 
%   run wcewpe1: $t=142T_\pe=35\omega_\ci^{-1}=875\omega_\ce^{-1}$ (iteration 21300), $z=4824$;
%   run wcewpe3: $t=40 T_\pe=30\omega_\ci^{-1}=750\omega_\ce^{-1}$ (iteration 6000),  $z=3024$;
%   run wcewpe3: $t=40 T_\pe=30\omega_\ci^{-1}=750\omega_\ce^{-1}$ (iteration 6000),  $z=2452$;
%   run wcewpe3: $t=40 T_\pe=30\omega_\ci^{-1}=750\omega_\ce^{-1}$ (iteration 6000),  $z=2420$.

We examine the inflow of plasma into the diffusion region. 
In the literature, for antiparallel reconnection (i.e., no guide field),
its width is found to scale with the particles inertial length, a result that we show 
to hold also for relativistic reconnection in Sect.~\ref{sec:diff_width_2}.
The originality of our results is the formation of a very sharp transition at the entrance of the 
diffusion regions, that we explore in Sect.~\ref{sec:sharpness_boundaries}.

%%%%%%%%%%%%
\subsubsection{Width of the diffusion region}\label{sec:diff_width_2}
%%%%%%%%%%%%
The diffusion region is, by definition, the area where impeding mechanisms 
(which can be collisions, inertia and collective interactions, or finite Larmor effects)
prevent the particles from responding in an ideal way to the electric fields induced by magnetic flux variations. 
The magnetic field and the plasma are then not coupled anymore, 
the former can freely diffuse, and reconnection can start or be sustained.
Defining the diffusion region is thus a matter of finding the area where the non-ideal processes 
dominate over ideal behavior.

Defining unambiguous criteria to identify this region is a complex subject \citep{Ishizawa2005,Klimas2010},
especially in the presence of a guide field \citep{Hesse2002,Hesse2004,Liu2014},
in 3D simulations \citep{Pritchett2013}, or in asymmetric configurations \citep{Zenitani2011c}.
In our case, as we show in Sect.~\ref{sec:sharpness_boundaries}, 
there is a sharp increase in particle density when the inflow plasma reach the central part,
where particles are retained by bouncing motions around the reversing magnetic field. 
It is associated with a sharp drop in inflow velocity, a rise in temperature, and 
a violation of the frozen-in relation $\b{E}+\bar{\b{v}}_s\wedge\b{B}=0$.
We identify the diffusion region with this area of increased density.

\begin{figure}[tbp]
 \centering
 \def\svgwidth{\columnwidth}
 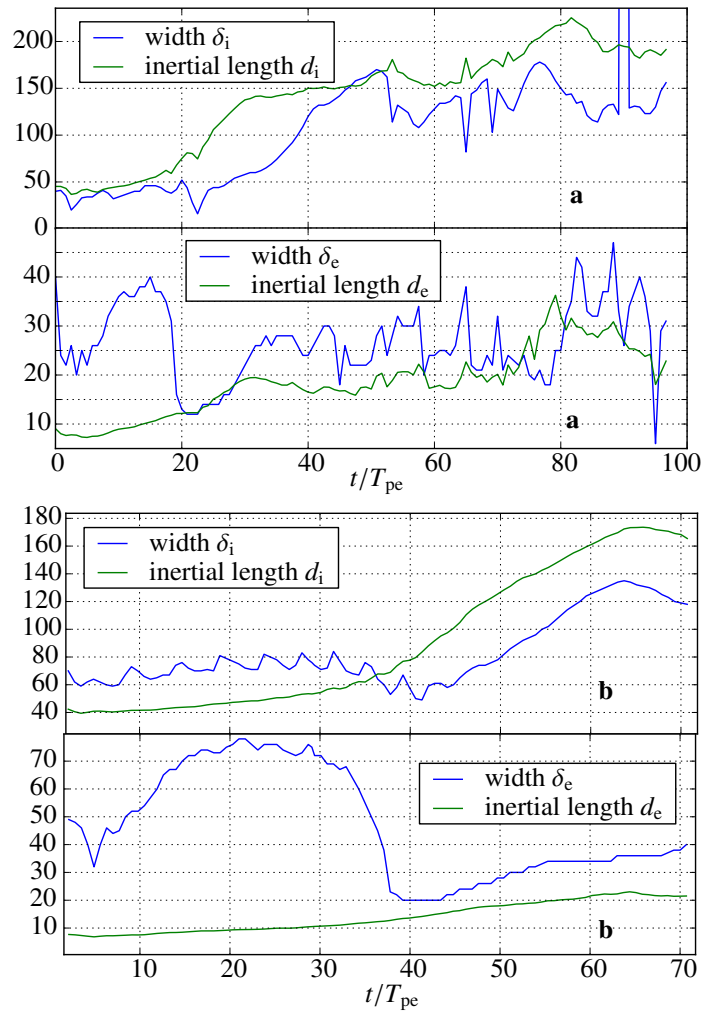
  \caption{\label{fig:scale_diff_region_nice_withothers}Width of the diffusion region $\delta_s$, 
           and inertial length $d_s$ measured at the center of the diffusion region, 
           for ions and electrons.
           Panels $a$ are for the simulation with $\wce/\wpe=3$, $\sigma_\lec^\mathrm{hot}=89$, 
           $T_\bg = 1.5\times10^7$\,K and $B_\mathrm{G}=0$; panels $b$ for $\wce/\wpe=6$, $\sigma_\lec^\mathrm{hot}=260$.
           The phase between 0 and $18T_\pe$ for panels $a$ (0 and $33T_\pe$ for panels $b$) corresponds to 
           times where the tearing instability has not yet started. The decrease in $\delta_\lec$ between 18 and $20T_\pe$ for $a$ 
           (33 and $37T_\pe$ for $b$) is the linear and non-linear growth phase of the instability.}
\end{figure}

Because of their lighter mass and fastest response, 
electrons remain frozen to the magnetic field longer than ions. 
The ion diffusion region is thus larger than the electron one.
In all the antiparallel simulations we find the expected two-scale (ion and electron) diffusion region.
It can be seen in Fig.~\ref{fig_wcewpe=1_NT=21300_cutalongX_partNumber_velocity},
where we present a cut along $x$ through a X-point at a given time for different simulations.
The width $\delta_s$ of the diffusion region is roughly given by the inertial length $d_s$ 
of the corresponding species, defined with the 
particle density at the center of the diffusion region, i.e., 
$\delta_s \sim d_s = c/\sqrt{n_s(x=0,t)q_s^2/\epsilon_0m_s}$, 
with throughout all simulations $0.5\leq\delta_\ion/d_\ion\leq1$ 
and $1\leq\delta_\lec/d_\lec\leq1.5$ (Fig.~\ref{fig:scale_diff_region_nice_withothers}).
This is also the scaling found in PIC simulations of non-relativistic ion-electron magnetic reconnection 
\citep[e.g.,][]{Pritchett2001,Klimas2010}.

In the case of hot background electrons ($T_{\bg,\lec}=3\times10^9$\,K, 
with a corresponding background plasma $\beta_\lec=1.1\times10^{-2}$),
the transition is less sharp and the width is larger than the inertial length. 
The sharpness is discussed in Sect.~\ref{sec:sharpness_boundaries}, 
and the larger extent is expected because inflowing particles have larger speeds, 
and thus larger Larmor radii and larger bouncing motions.
The width $\delta_s$ should thus also depend on the $\beta_s$ parameter of the inflow.

%%%%%%%%%%%%%%%%%%  OLD VERSION, LONG  %%%%%%%%%%%%%%%%%%%
\note{
We now detail the structure of the diffusion region. 
We underline the difficulty of its definition in Sect.~\ref{sec:diff_width_1}.
In the literature, for antiparallel reconnection (i.e., no guide field),
its width is found to scale with the particles inertial length, a result that we show 
to hold also for relativistic reconnection in Sect.~\ref{sec:diff_width_2}.
The originality of our results is the formation of a very sharp transition at the entrance of the 
diffusion regions, that we explore in Sect.~\ref{sec:diff_width_2}.

%%%%%%%%%%%%
{Width of the diffusion region: theoretical arguments}% \label{sec:diff_width_1}
%%%%%%%%%%%%
The diffusion region is, by definition, the area where impeding mechanisms\note{such as collisions, inertia and collective interactions, or finite Larmor effects,}
prevent the particles from responding in an ideal way to the electric fields induced by magnetic flux variations. 
The magnetic field and the plasma are then not coupled anymore, 
the former can freely diffuse, and reconnection can start.
Defining the diffusion region is thus a matter of finding the area where the non-ideal processes 
dominate over ideal behavior.

A practical point of view to identify the non-ideal area is to notice that
non-idealness means that the frozen-in relation $\b{E}+\bar{\b{v}}_s\wedge\b{B}=0$ does not hold
(with $\bar{\b{v}}_s$ the mean velocity of particles of species $s$).
The diffusion region can thus be defined in simulations as the 
area where the frozen-in condition is violated.
We note that in the presence of a guide field, breaking the frozen-in condition implies the existence of 
an electric field parallel to the magnetic field, $\b{E}\cdot\b{B} \neq 0$ (see Sect.~\ref{sec:results_guide_field}).

The reasons to behave non-ideally can be
collisions, particle inertia and collective interactions, and finite Larmor effects.
They will become important whenever they occur on a length scale $d < r_{\mathrm{c}s}$, where
$r_{\mathrm{c}s}$ is the Larmor radius of the particles of species $s$, 
as then the trajectories of the particles, while they respond to the induced electric fields, 
will be too much perturbed\note{(the response to a variation is an adiabatic motion in a B field,
and occurs in an unexpected way if a process perturbs the orbit on a scale $<r_{\mathrm{c}s}$)}. 
The scale $d$ is thus, respectively for each impeding process:\note{make a list of this}
% \begin{itemize}
 the mean free path $l_\mathrm{mfp}$ between collisions;
 the oscillation length of the particles due to fluctuating fields and finite inertia 
       (which is the Debye length $\lambda_{\mathrm{D}s} = v_{th,s}/\omega_{\mathrm{p}s} = v_\mathrm{th}/c\times d_s$ 
       for electrostatic Langmuir oscillations, but the inertial length $d_s$ is often used instead);
 the gradient scale of the magnetic field $l_{\nabla B}$\note{We note 
    that the conditions for non-reconnection or idealness just invoked 
    ($d>r_{\mathrm{c}s}$ with $d=l_\mathrm{mfp}$, $d=d_s$, $d=l_{\nabla B}$), are equivalent 
    to saying that the particles are magnetized, or adiabatic, i.e., that the drift theory 
    applies and that they conserve their first magnetic invariant.
    This criteria $d>r_{\mathrm{c}s}$ is thus often justified by saying that particles should remain 
    frozen to the field in order to forbid reconnection.
    However, in an ideal plasma the particle magnetization becomes ambiguous at a magnetic null:
    $r_{\mathrm{c}s} \propto m_s/B$ has no clear limit when $m_s=0$ and $B=0$.
    On another hand, at the null $E/B=+\infty$, so that particles are definitely not 
    magnetized there, but reconnection still cannot be triggered.
    We thus prefer to justify the criteria by saying that particles must not be hindered in their 
    response to induced electric fields.}.
% \end{itemize}

Since the Larmor radius and the inertial length vary differently with the particle mass, 
we immediately see that there will be two diffusion regions, one for the ions and one for the electrons.
Since electrons are lighter than ions, they have smaller Larmor radii and will remain frozen-in longer 
than ions.
This can be estimated simply in the case of Langmuir oscillations at frequency $\omega_{\mathrm{p}s}$: 
retaining only bulk inertia in Ohm's law,
\begin{equation}
 m_s\frac{\dif\bar{\b{v}}_s}{\dif t} = q_s (\b{E}+\bar{\b{v}}_s\wedge\b{B}),
\end{equation}
we see that the inertial term 
$m_s\dif\bar{\b{v}}_s/\dif t \sim m_s\omega_{\mathrm{p}s} \bar{v}_s$ dominates over 
the ideal part $q_s \bar{\b{v}}_s\wedge\b{B} \sim m_s\omega_{\mathrm{c}s}\bar{v}_s$ 
when $\omega_{\mathrm{p}s} > \omega_{\mathrm{c}s}$. 
Assuming a linear variation of $B(x) \propto B_0 x/l_{\nabla B}$, 
this is equivalent to $x < l_{\nabla B} \sqrt{m_s n_s} / (B_0\sqrt{\epsilon_0})$. 
Assuming a particle density $n_s\sim\mathrm{cst}$ among the species, 
we see that the width $\delta_s$ of the diffusion region is (under these hypotheses)
$\delta_s \propto \sqrt{m_s}$.
It also implies $\delta_s \propto d_s$.

We stress that this estimation is at best approximate. In reality, the waves involved can be non-electrostatic,
the magnetic field gradient scale $l_{\nabla B}$ may be set up by the dynamics (it does decrease 
after the initial phase in our simulations),
and in steady state the width of the diffusion region may be fixed also by other mechanisms. 
Other estimations can be envisaged: for example the width of the Speiser layer \citep{Speiser1965}: 
$\delta_s \propto \sqrt{r_{\mathrm{c}s}l_{\nabla B}}$,
or the bouncing width $x$ such that $r_{\mathrm{c}s}(x) = x$. 
See, e.g., \citet{Ishizawa2005,Treumann2006,Klimas2010,Zenitani2011c}.
It is also worth mentioning that the presence of a guide field complicates the structure of the diffusion region, 
which can feature a Larmor radius scaled inner structure \citep{Hesse2002,Hesse2004,Liu2014}.
}

% \begin{figure}[tbp]
% \centering
%   \includegraphics[width=\columnwidth]{images/fig_wcewpe=1_NT=21300_cutalongX_temperatures.pdf}
%   \caption{\label{fig:fig_wcewpe=1_NT=21300_cutalongX_temperatures}Cut along $x$ through the X-point for run wcewpe1. 
%   $t=142T_\pe=35\omega_\ci^{-1}=875\omega_\ce^{-1}$ (iteration 21300), $z=4824$. \modif{Figure useful?}}
% \end{figure}

%%%%%%%%%%%%
\subsubsection{Sharpness of the inflow boundaries}
%%%%%%%%%%%%
\label{sec:sharpness_boundaries}

\begin{figure}
 \centering
 \def\svgwidth{\columnwidth}
 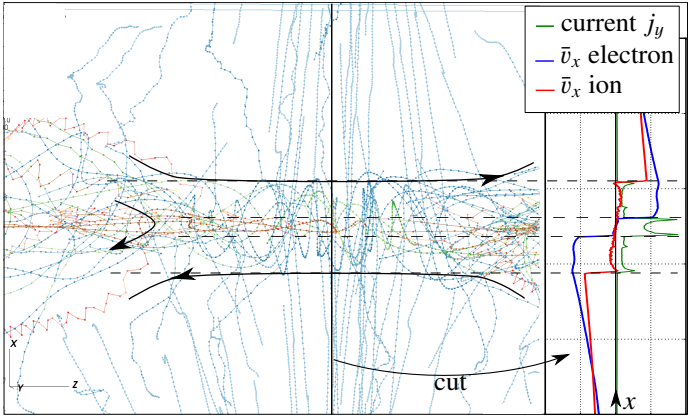
 \caption{\label{fig:visit_ballistic_motions_with_profiles}Trajectories of a sample of particles, 
  and on the right fluid velocities and current for a cut through the X-point.}
\end{figure}

We see from Fig.~\ref{fig_wcewpe=1_NT=21300_cutalongX_partNumber_velocity} that the boundary of the diffusion region, 
defined by the increase in particle number density, is very sharp in some cases 
(especially for the case $\wce/\wpe=3$ with a cold background plasma).
These sharp transitions are \textit{not} the trace of a shock between the incoming plasma and the over-dense diffusion region.
First, because the inflow bulk velocity is not supersonic, in the sense that we have 
$\bar{v}_{\ins,s} < C_\mathrm{fms}$, with $C_\mathrm{fms}$ the phase speed of waves propagating perpendicular to 
the magnetic field $\b{B}_0$, i.e., the fast magnetosonic velocity in MHD 
(Alfv\'en and slow waves do not propagate perpendicularly to $\b{B}_0$).
Second, because the width of the transition between inflow plasma and diffusion region plasma is, in the sharpest case, 
less than an electron inertial length, while we know that the thermalization of a cold inflow by collisionless kinetic processes 
occurs on a width of some inertial length, with also the formation of a precursor upstream \citep[e.g.,][]{Plotnikov2013}
that is not seen here.
Third, because there is no variations in the magnetic field across the sharp particle density variation.

Instead, we explain these sharp transitions by quasi-ballistic motions of the particles entering in the diffusion region.
Figure~\ref{fig:visit_ballistic_motions_with_profiles} shows some particle trajectories: they travel toward the diffusion region because of the electric field $E$ 
with a guiding center velocity given by $\b{v}_{E\times B} = \b{E}\times\b{B}/B^2$, they reach the magnetic field reversal, 
overshoot to the other side because of their inertia, turn around the increasingly strong magnetic field, and then 
oscillate around the $B=0$ line before being ejected toward the islands 
(because they have a large $v_y$ in the $B_x$ field, see Sect.~\ref{sec:overall_evolution} and Fig.~\ref{fig:traj_3D_no_guide_filed}
for a trajectory in 3D).
This explains well the fluid point of view of Fig.~\ref{fig_wcewpe=1_NT=21300_cutalongX_partNumber_velocity}:
averaging over particles to obtain the fluid particle number density or fluid velocity, 
the sharpness of the transitions is due to the fact that all particles of the same species turn around at roughly the same location. 

It also explains the M shape of the particle densities of Fig.~\ref{fig_wcewpe=1_NT=21300_cutalongX_partNumber_velocity}:
particles spend more time at the edge of the diffusion region, when turning back, than at the center, hence the peaks at the edges and
the depletion at the center. The fluid velocity profiles are also well explained by the particle view.
We note that a M shape is also reported in the context of the inversion electric field layer 
\citep[][and references therein]{Chen2011}.
Here we do not find any trace of the inversion layer in the electric field. 
A possible explanation may be that our electrons are relativistic.

Concerning the sharp transitions, the question is thus to know why and under which conditions all the particles of a species turn back at the same location.
They will do so if they all enter the diffusion region with the same velocity, i.e., if their thermal velocity is 
smaller than their bulk velocity: $v_{\ins,s}^\therm \ll \bar{v}_{\ins,s} \sim v_{E\times B}$.
Table~\ref{tab:sharpness_entrance_dissipation_zone} presents the ratio $v_{\ins,s}^\therm/\bar{v}_{\ins,s}$ for the different simulations 
of Fig.~\ref{fig_wcewpe=1_NT=21300_cutalongX_partNumber_velocity}.
For a given magnetization, here $\wce/\wpe=3$, we clearly see a correlation between a small 
$v_{\ins,s}^\therm/\bar{v}_{\ins,s}$ and a sharp transition.
In particular, in the case with very hot electrons ($v_{\ins,\lec}^\therm/\bar{v}_{\therm,\lec} = 1.8$) and relatively cold ions
($v_{\ins,\ion}^\therm/\bar{v}_{\ins,\ion} = 0.1$), the electron particle number and velocities present smooth variations,
while the same curves for ions do exhibit sharp transitions. 
This is in accordance with the explanation of sharp transitions by the cold nature, in term of the ratio 
$v^\therm_\ins/v_{E\times B}$, of the inflowing plasma.

The influence of the magnetization on the sharpness of the transitions is seen by comparing the two first simulations of
Table~\ref{tab:sharpness_entrance_dissipation_zone}, 
with $\wce/\wpe=1$ and 3 and same ratio $v_{\ins,s}^\therm/\bar{v}_{\ins,s}$
(plotted on the top part of Fig.~\ref{fig_wcewpe=1_NT=21300_cutalongX_partNumber_velocity}).
A smaller magnetization implies a smoother shape.
We understand this as a consequence of the fact that particles turn back on a scale given by their Larmor radii in the
magnetic field at the sheet entrance, which is smaller when $\wce/\wpe$ is higher.

As a final remark concerning resolution, we note that in the coldest cases, 
the thermal Larmor radius of the electrons are smaller than a cell length.
The resolution of the Larmor radius on the grid is, however, of no importance to integrate particle trajectories 
in constant fields,
because interpolation of field quantities to particle position then provides the same (constant) values, 
regardless of the cell size.
What matters is the temporal resolution, $\omega_{\mathrm{c}s}\Delta t < 1$ \citep[see][\S3.1, for a discussion]{Melzani2013}.
Here, we have $\omega_{\mathrm{c}s}\Delta t = 0.04$ and $0.12$ for $\wce/\wpe = 1$ or~3.
Also, a test run with a spatial and temporal resolution increased by a factor two leads to the same structures.

In summary, the sharp transitions are explained by collective bouncing motions allowed by the cold nature of the inflow:
$v_{\ins,s}^\therm \ll \bar{v}_{\ins,s}$.
Such transitions may also occur in non-relativistic reconnection, 
but then the inflow speed $\bar{v}_{\ins,s}\sim E/B$ is low and the 
background plasma should be very cold. 
They are likely to be more common in relativistic reconnection, where  $\bar{v}_{\ins,s}\sim E/B$ is larger.
This may be why, to our knowledge, they have never been reported in other simulations.

\begin{table*}[tbp]
\caption{\label{tab:sharpness_entrance_dissipation_zone}Values regarding the sharpness of the edge of the diffusion region. 
         Here $\Theta_s = T_s/(m_sc^2)$, $v_{\ins,s}^\therm=\sqrt{T_s/m_s}$, and $\bar{v}_{\ins,s}$ the fluid velocity 
         at the entrance of the diffusion region of species $s$,
         measured in Fig.~\ref{fig_wcewpe=1_NT=21300_cutalongX_partNumber_velocity}.
         Also, $\tilde{r}_{\mathrm{c}s}$ is the thermal Larmor radius in number of cells with
         $r_{\mathrm{c}s} = \langle(\gamma v_\perp)^2\rangle^{1/2} / \omega_{\mathrm{c}s}$, 
         estimated with the formula for a Maxwell-J\"uttner plasma at rest 
         as $r_{\mathrm{c}s} = (c/\wce)\sqrt{2\Theta_s\kappa_{32}(1/\Theta_s)}$ \citep{Melzani2013}. 
         The indication of sharpness is qualitative,
         see Fig.~\ref{fig_wcewpe=1_NT=21300_cutalongX_partNumber_velocity} for details.}
\centering
\begin{tabular}{c|cc||cc|cc|cc|cc}
 $\wce/\wpe$ & $T_{\bg,\lec}$ (K) & $T_{\bg,\ion}$ (K) & $\Theta_{\bg,\lec}$       & $\Theta_{\bg,\ion}$   & $v_{\ins,\lec}^\therm/\bar{v}_{\ins,\lec}$ & $v_{\ins,\ion}^\therm/\bar{v}_{\ins,\ion}$ & Sharpness: elecs. & and ions & $\tilde{r}_{\mathrm{ci},\ins}$ & $\tilde{r}_{\mathrm{ce},\ins}$ \\
 \hline
 1           & $1.5\times10^7$    & $1.5\times10^7$   & $2.5\times 10^{-3}$ & $10^{-4}$       & 0.1 & 0.02    & sharp      & sharp      & 0.6 & 3.18  \\
 3           & $1.5\times10^7$    & $1.5\times10^7$   & $2.5\times 10^{-3}$ & $10^{-4}$       & 0.1 & 0.02    & very sharp & very sharp & 0.2 & 1.1   \\
 3           & $2  \times10^8$    & $2  \times10^8$   & $3  \times 10^{-2}$ & $10^{-3}$       & 0.3 & 0.1     & sharp      & sharp      & 0.7 & 3.7   \\
 3           & $3  \times10^9$    & $2  \times10^8$   & $5  \times 10^{-1}$ & $10^{-3}$       & 1.8 & 0.1     & smooth     & sharp      & 3   & 3.7   \\
\end{tabular}
\end{table*}

%%%%%%%%%%%%%%%%%%%%%%%%%%%%%%%
%%%%%%%%%%%%%%%%%%%%%%%%%%%%%%%
\subsection{The relativistic Ohm's law}
%%%%%%%%%%%%%%%%%%%%%%%%%%%%%%%
%%%%%%%%%%%%%%%%%%%%%%%%%%%%%%%
\label{Sec:Ohms_law}

\begin{figure}[tbp]
 \centering
 \def\svgwidth{\columnwidth}
 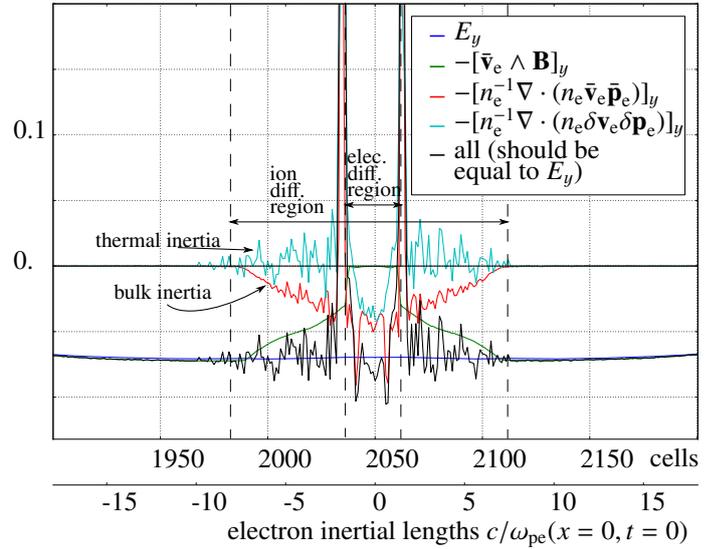
 \caption{\label{fig:wcewpe=3_NT=6001_Ohms_law}
 Different components of Ohm's law (Eq.~\ref{equ:Ohm_fluid_1}) 
 for a cut along $x$ through a X-point 
 (the same as in Figs.~\ref{fig:xcf_wcewpe=3_NT=6000_2D_pseudocolor_illustration},
 \ref{fig_wcewpe=1_NT=21300_cutalongX_partNumber_velocity} upper-right,
 \ref{fig:wcewpe=3_cut_Z_NT6000_temperature_only}, \ref{fig_wcewpe=3_NT=6000_cutalongZ_summary} right,
 and~\ref{fig:xcf_wcewpe=3_NT=6000_2D_pseudocolor_temperature_NT=6000}).
 Run $\wce/\wpe=3$, $\sigma_\lec^\mathrm{hot}=89$, $T_\bg=1.5\times10^7$\,K, $B_\mathrm{G}=0$, 
 $t=40T_\pe=30\omega_\ci^{-1}=750\omega_\ce^{-1}$.%, $z=3024$.
 }
\end{figure}

We explore the balance of terms in Ohm's law. 
The literature concerning 2D non-relativistic reconnection largely shows that for antiparallel reconnection, 
the dominant term is thermal inertia either in ion-electron \citep{Klimas2010,Shay2007,Fujimoto2009} or pair \citep{Bessho2005} plasmas,
and this fact is the key element of various analytical models \citep[e.g.,][]{Hesse2011}.
On the other hand, reconnection with a guide field is sustained by electron bulk inertia at skin-depth scales, 
and thermal inertia at Larmor radius scales \citep{Hesse2002,Hesse2004,Liu2014}.
% As we show here, the same investigations in relativistic plasmas may lead to surprises.
Existing studies with relativistic pair plasmas confirm the non-relativistic trend: 
with no guide field, \citet{Hesse2007,Bessho2012,Zenitani2009b} find a dominating contribution of thermal inertia,
while with a guide field 
\citet{Hesse2007} find a significant contribution of the time derivative of bulk momentum $\partial_t(n_\lec\bar{p}_y)$.
\note{with a guide field $B_\mathrm{G}/B_0=1.5$} \note{note the exception being \citet{Kagan2012} finding thermal inertia contributing alone with a guide field
for 3D simulations of relativistic pair plasmas}
Here we explore the situation in ion-electron relativistic plasmas for antiparallel reconnection. 
We show for the first time in the antiparallel case that bulk inertia contributes at least as much as thermal 
inertia for large inflow magnetizations. We explain that this is due to a relativistic inflow magnetization
in Sect.~\ref{sec:outflow_analytical_estimate_harder}.

Ohm's law is the fluid equation of motion for the lighter particles (the electrons),
and is a mean to evaluate the relative weight of the different non-ideal terms 
allowing reconnection with a simple \textit{fluid} picture.\note{In the ideal case, the lighter particles cancel any electric field 
    (in the frame of zero bulk velocity) induced by magnetic flux variations: $\b{E}' = \b{E}+\bar{\b{v}}\wedge \b{B} = 0$,
    and consequently forbid magnetic flux variations: 
    $\frac{\dif}{\dif t}\iint_{S(t)}\dif\b{S}\cdot\b{B} = -\oint_{C(t)} \dif\b{l}\cdot(\b{E}+\bar{\b{v}}\wedge \b{B})=0$. 
    Non-ideal effects such as collisions, particle inertia and collective interactions, or finite Larmor effects, 
    prevent this cancellation and allow magnetic reconnection.}
Understanding the weight and localization of each term is a first step toward building of
an effective resistivity for fluid models (resistive MHD, two-fluid codes, hybrid codes),
where concrete resistive parametrizations can lead to very different behaviors,
for instance changing from steady to unsteady states in \citet{Zenitani2009}, 
or triggering or not a Petsheck-like configuration depending on the localization and gradients of the resistivity 
\citep{Baty2006}.

The relativistic Ohm's law for electrons is directly derived from the equation of conservation of momentum for the electron fluid, 
Eq.~\ref{equ:fluid_2}, which is itself obtained from Vlasov equation in Appendix~\ref{sec:app_measure_relat_1}. It reads:
\begin{equation}\label{equ:Ohm_fluid_1}
\begin{aligned}
 \frac{q_\lec}{m_\lec}&(\b{E}+\bar{\b{v}}_\lec\wedge\b{B}) = \frac{1}{n_\mathrm{e}}\frac{\partial}{\partial t} (n_\mathrm{e}\bar{\b{p}}_\lec) + \frac{1}{n_\mathrm{e}}\frac{\partial}{\partial \b{x}}\cdot (n_\mathrm{e} \langle \b{p}_\lec\b{v}_\lec\rangle) \\
       &= \underbrace{\frac{1}{n_\mathrm{e}}\frac{\partial}{\partial t} (n_\mathrm{e}\bar{\b{p}}_\lec) + \frac{1}{n_\mathrm{e}}\frac{\partial}{\partial \b{x}}\cdot (n_\mathrm{e}\bar{\b{p}}_\lec\bar{\b{v}}_\lec)}_\text{bulk inertia} \\
       &~~~~~~~~~~~~~~~~~~~~~~~~~~~~~~~~~~~~~~~~~~~~+ \underbrace{\frac{1}{n_\mathrm{e}}\frac{\partial}{\partial \b{x}}\cdot (n_\mathrm{e}\langle \delta\b{p}_\lec\delta\b{v}_\lec\rangle)}_\text{thermal inertia}. 
\end{aligned}
\end{equation}
Here we use for simplicity quantities computed in the simulation (or lab) frame, e.g.,
$n_\mathrm{e}$ is the electron number density in the lab-frame (denoted by $n_\mathrm{lab,e}$ in Appendix~\ref{sec:app_measure_relat_1}). 
Bared quantities are averaged over the momentum distribution function.
We also used the definition $\delta\b{p} = \b{p}-\bar{\b{p}}$ where $\b{p}=\gamma \b{v}$ is the momentum.
Brackets $\langle\cdot\rangle_s$ denote an average in momentum space over the particles distribution function.
The left-hand side of Eq.~\ref{equ:Ohm_fluid_1} is the ideal part, and is set equal to 0 in ideal MHD. 
On the right-hand side figure terms linked to finite inertia
(i.e., particles do not respond perfectly to the electric field variations because of their inertia):
\begin{itemize}
 \item The first term is a part of bulk inertia. However, it vanishes in steady state and we will neglect it 
       (this is validated a posteriori).
 \item The second term is inertia linked to the bulk flow, and is denoted as \textit{bulk inertia}.
       It comes from the overall structure of the flow around the sheet 
       (the profiles of the mean quantities: increase of $\bar{v}$, $\bar{p}$).
 \item The third term is inertia linked to microscopic thermal motion, and is denoted as \textit{thermal inertia}.
       It comes from the divergence of the off-diagonal terms of the pressure tensor,
       and can be anticipated by a study of the temperature curves.       
\end{itemize}
We analyze Ohm's law in the direction of the reconnection electric field ($\hat{\b{y}}$ here). 
Given the invariance along $y$, the divergence terms have two contributions: 
$\sum_k\partial_k(p_kv_y) = \partial_x(p_xv_y) + \partial_z(p_zv_y)$.
A computation of the divergence requires a smoothing of the quantities, especially for the thermal inertia term which is 
very noisy.

\begin{figure}[tbp]
 \centering
 \def\svgwidth{\columnwidth}
 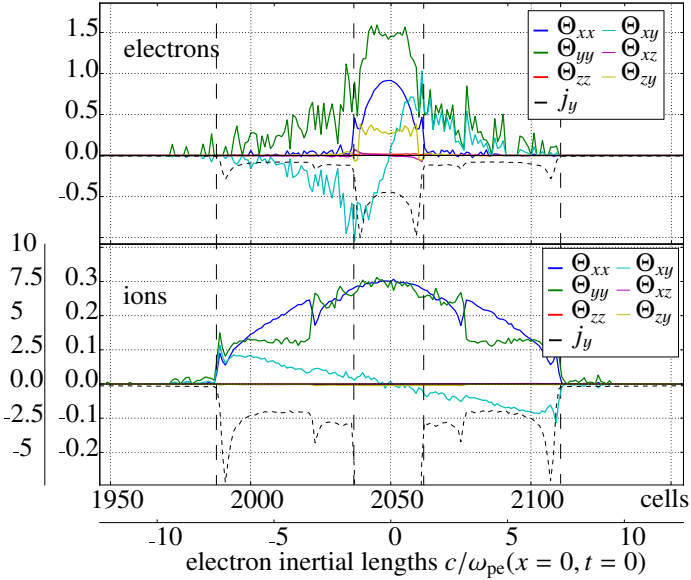
 \caption{\label{fig:wcewpe=3_cut_Z_NT6000_temperature_only}
 Temperatures for a cut along $x$ through the same X-point as in 
 Figs.~\ref{fig:xcf_wcewpe=3_NT=6000_2D_pseudocolor_illustration}, \ref{fig_wcewpe=1_NT=21300_cutalongX_partNumber_velocity} upper-right,
 \ref{fig_wcewpe=3_NT=6000_cutalongZ_summary} right, and \ref{fig:xcf_wcewpe=3_NT=6000_2D_pseudocolor_temperature_NT=6000}.
 Run with $\wce/\wpe=3$, $\sigma_\lec^\mathrm{hot}=89$, $T_\bg=1.5\times10^7$\,K.
 We define $\Theta_{mn,s} = T_{mn,s}/m_sc^2$ with $s=\ion$ or $\lec$.
 For the ion temperatures, the inner vertical axis is $\Theta_\ion$, 
 the outer one is $\Theta_\ion \times m_\ion/m_\lec$, and shows that $T_\ion > T_\lec$.}
\end{figure}

\begin{figure*}[tbp]
  \centering
  \def\svgwidth{\textwidth}
  \begin{tiny}
  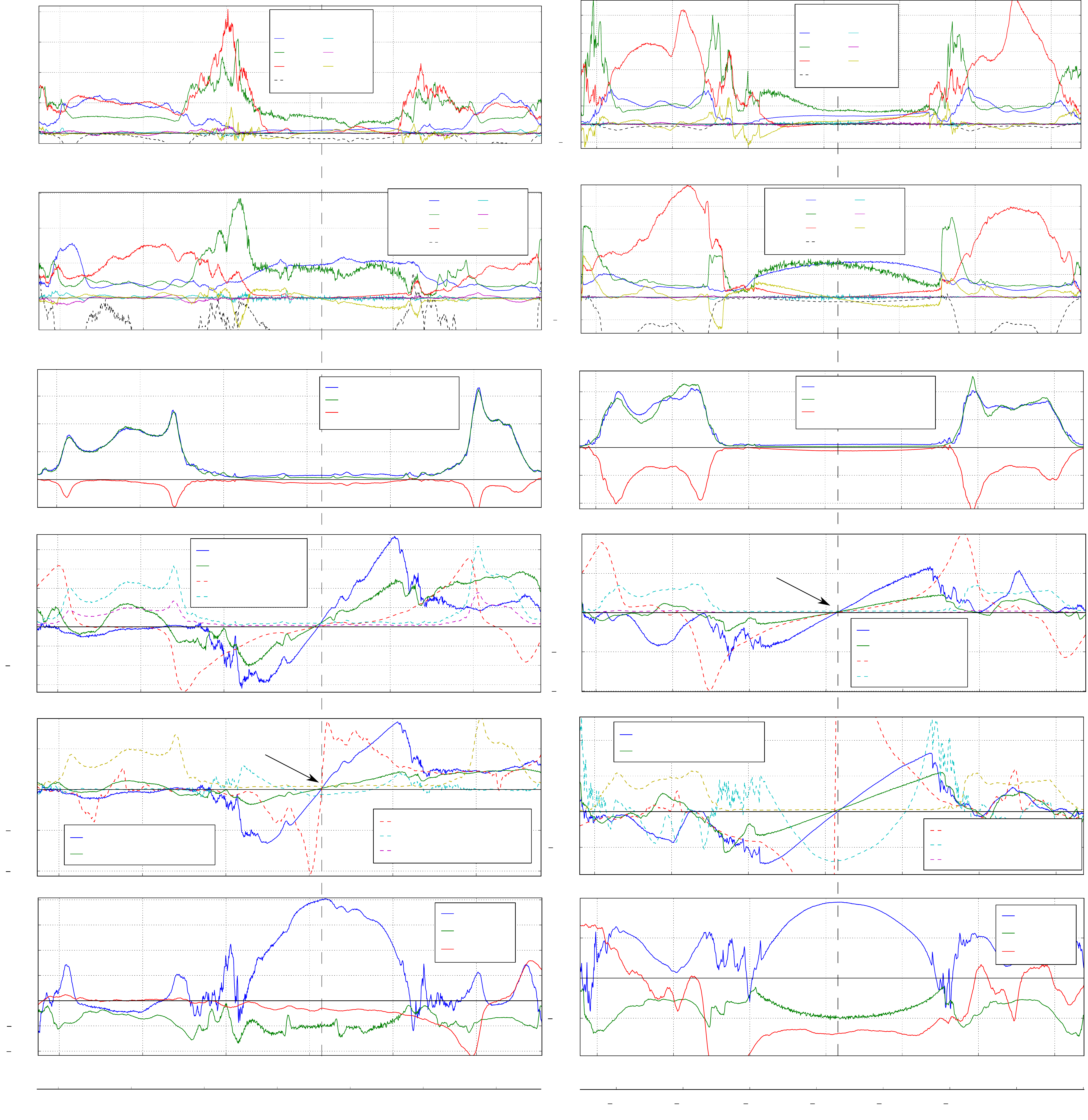
  \end{tiny}
  \caption{\label{fig_wcewpe=3_NT=6000_cutalongZ_summary}
  Cut along $z$ through the X-point.
  Left:  Run $\wce/\wpe=1$, $\sigma_\lec^\mathrm{hot}=9.9$, same X-point as in Fig.~\ref{fig_wcewpe=1_NT=21300_cutalongX_partNumber_velocity}, 
  $t=142T_\pe=35\omega_\ci^{-1}=875\omega_\ce^{-1}$. %, $z=4824$ (iteration 21300)
  Right: Run $\wce/\wpe=3$, $\sigma_\lec^\mathrm{hot}=89$, $T_\bg=1.5\times10^7$\,K, $B_\mathrm{G}=0$, same X-point and time as 
  in Figs.~\ref{fig:xcf_wcewpe=3_NT=6000_2D_pseudocolor_illustration}, 
  \ref{fig_wcewpe=1_NT=21300_cutalongX_partNumber_velocity} upper-right, 
  \ref{fig:wcewpe=3_NT=6001_Ohms_law}, 
  \ref{fig:wcewpe=3_cut_Z_NT6000_temperature_only},
  and \ref{fig:xcf_wcewpe=3_NT=6000_2D_pseudocolor_temperature_NT=6000}
  ($t=40T_\pe=30\omega_\ci^{-1}=750\omega_\ce^{-1}$). %, $z=3024$, iteration 6000
  }
\end{figure*}

We show in Fig.~\ref{fig:wcewpe=3_NT=6001_Ohms_law} the results for a cut through the X-point, 
for the simulation with $\wce/\wpe=3$ and $T_\bg=1.5\times10^7$\,K ($\sigma_\lec^\mathrm{hot}=89$). 
Ohm's law is satisfied everywhere, except at the sharp transitions at the entrance of the electron diffusion region,
where the derivatives diverge. Different areas emerge:
\begin{itemize}
 \item The electrons are ideal outside of the ion diffusion region. 
 \item In the ion diffusion region, $|\bar{\b{v}}_\lec\wedge\b{B}|$ decreases linearly.

       The bulk inertia term $\sum_k\partial_k(n_\lec\bar{p}_k\bar{v}_y)$ rises linearly to compensate. 
       The term $\partial_x(n_\lec\bar{p}_x\bar{v}_y)$ dominates over $\partial_z(n_\lec\bar{p}_z\bar{v}_y)$.
       The contribution of the former is understandable when looking at 
       $\bar{p}_x$ and $\bar{v}_y$ that increase when we get closer to the sheet in this region
       (see Fig.~\ref{fig_wcewpe=1_NT=21300_cutalongX_partNumber_velocity} for $\bar{v}_y$).

       The thermal inertia term $\sum_k\partial_k(n_\lec\delta{p}_k\delta{v}_y)$ is slightly positive, 
       and partly cancels the contribution of bulk inertia. 
       This cancellation is also reported in \citet{Fujimoto2009} and \citet{Klimas2010} for non-relativistic ion-electron plasmas, 
       and in \citet{Bessho2012} for relativistic pairs.
       Only $\partial_x(n_\lec\delta{p}_x\delta{v}_y)$ contributes, and is negative, 
       which is easily seen when looking at the temperature curves $T_{xy,\lec}$ and $T_{zy,\lec}$
       (Fig.~\ref{fig:wcewpe=3_cut_Z_NT6000_temperature_only}).
 \item In the electron diffusion region, the ${\bar{\b{v}}}_\lec\wedge\b{B}$ term vanishes 
       (because $B$ is very weak and $\bar{v}_{x}=0$).

       The bulk inertia term is constant, and due only to the term $\partial_z(n_\lec\bar{p}_z\bar{v}_y)$,
       which is expected to contribute given that $\bar{v}_y\sim\mathrm{cst}$ and $\bar{p}_z \propto z-z_\mathrm{X-point}$ in this region
       (Fig.~\ref{fig_wcewpe=3_NT=6000_cutalongZ_summary}).
       The other term, $\partial_x(n_\lec\bar{p}_x\bar{v}_y)$, vanishes because $\bar{p}_x=\mathrm{cst}=0$ in this area.

       The thermal inertia term $\sum_k\partial_k(n_\lec\delta{p}_k\delta{v}_y)$ contributes as much as the bulk inertia term.
       Only $\partial_x(n_\lec\delta{p}_x\delta{v}_y)$ contributes, and is negative, which is easily seen when looking at the 
       temperature curves $T_{xy,\lec}$ and $T_{zy,\lec}$ in Fig.~\ref{fig:wcewpe=3_cut_Z_NT6000_temperature_only}.
\end{itemize}

A cut along $x$ through other X-points in the simulation leads to the same results.
Also, a cut along $z$ through the X-point reveals that the results of the electron region 
hold throughout the center of the current sheet,
with a slow increase of the ${\bar{\b{v}}}_\lec\wedge\b{B}$ term as we get near the islands.

In summary, non-ideal terms in the ion regions are due to bulk inertia and, 
in the electron diffusion region, to an 
interestingly equal contribution of bulk and thermal inertia.
For other runs with $\wce/\wpe=3$ ($\sigma_\lec^\mathrm{hot}=27$ to 89), we also find an equal contribution from thermal and bulk inertia.
For the most magnetized run, $\wce/\wpe=6$ ($\sigma_\lec^\mathrm{hot}=260$), 
the contribution of bulk inertia exceeds that of thermal inertia by a factor 1.5 to~3.
We show in Sect.~\ref{sec:outflow_analytical_estimate_harder} with analytical estimations
that the large magnetization for electrons indeed allows bulk inertia to overreach thermal inertia,
with the former scaling as $\sigma^\mathrm{cold}_{\ins,\lec}$ 
and the latter as $\left(\sigma^\mathrm{cold}_{\ins,\lec}\right)^{1/2}$.
This effect is present in our simulations, and not in the references previously mentioned with antiparallel fields, 
because our background electron magnetization is larger. 
It is thus a new regime that challenges the thermal inertia paradigm at large electron magnetizations.
We discuss the possible consequences in Sect.~\ref{sec:astro_outlook}.

% It is also interesting to compare our work with that of \citet{Zenitani2009b}, where relativistic two-fluid simulations in pair plasmas 
% are performed, and a study of the component of Ohm's law through a line across the X-point is shown.
% They find a contribution close to what we have in the ion diffusion region, where $E_y$ is dominated by 
% bulk inertia $\partial_x(\bar{p}_x\bar{v}_y)$ and ${\bar{\b{v}}}_\lec\wedge\b{B}$, 
% indicating that up to near the entrance of the inner electron diffusion region the flow structure is similar.
% However the analogy stops inside the electron region (or near the center of their diffusion region). 
% There, Zenitani et al. find a negligible $\partial_z(\bar{p}_z\bar{v}_y)$, while it accounts for 50\% of $E_y$ in our case,
% and the contribution to $E_y$ is given by friction and viscous inertia, instead of thermal inertia 
% (absent from their study because the pressure tensor is diagonal). We note that their viscous inertia term 
% gives a profile similar to our thermal inertia term. 

%%%%%%%%%%%%%%%%%%%%%%%%%%%%%%%
%%%%%%%%%%%%%%%%%%%%%%%%%%%%%%%
\subsection{Outflow: energy content of the exhaust jets}
%%%%%%%%%%%%%%%%%%%%%%%%%%%%%%%
%%%%%%%%%%%%%%%%%%%%%%%%%%%%%%%
\label{sec:outflows}

\begin{table*}[tb]
\caption{\label{tab:energy_outflow_flux}Energy content of the outflows. 
The energy flux due to the bulk flow mean velocity is proportional to $\Gamma_{\out,s}-1$, 
 and the energy flux due to kinetic particle motions and pressure work is proportional to $\Gamma_{\out,s}(h_{0,\out,s}-1)$
 (see Eq.~\ref{equ:particles_energy_flux}). Here $B_\mathrm{G}/B_0=0$. Quantities are measured at their maximum value,
 which is reached at the end of the exhausts along $z$.}
\centering
\begin{tabular}{c|c|c|cc||c|cc|cc}
 $\wce/\wpe$ & $n_\mathrm{bg}/n_0$ & $T_{\bg,\lec}$, $T_{\bg,\ion}$ (K) & $\sigma^\mathrm{cold}_{\ins,\lec}$ & $\sigma^\mathrm{cold}_{\ins,\ion}$ & & electrons & & ions & \\
\hline
 1 & 0.1 & $1.5\times10^7$, idem           & 10 & 0.4 & $\Gamma_{\out,s}-1$               & 0.49 & 31\% & 0.02 & 20\% \\
   &     &                                 &    &     & $\Gamma_{\out,s}(h_{0,\out,s}-1)$ & 1.07 & 69\% & 0.08 & 80\% \\
\hline
 3 & 0.31& $2.0\times10^8$, idem           & 29 &  1.2 & $\Gamma_{\out,s}-1$               & 0.71 & 14\% & 0.065& 22\% \\
   &     &                                 &    &      & $\Gamma_{\out,s}(h_{0,\out,s}-1)$ & 4.34 & 86\% & 0.24 & 78\% \\
\hline
 3 & 0.1 & $3\times10^9$, $2\times10^8$    & 90 &  3.6 & $\Gamma_{\out,s}-1$               & 0.089& 1\%  & 0.056&  9\% \\
   &     &                                 &    &      & $\Gamma_{\out,s}(h_{0,\out,s}-1)$ & 15   & 99\% & 0.54 & 91\% \\
\hline
 3 & 0.1 & $1.5\times10^7$, idem           & 90 & 3.6 & $\Gamma_{\out,s}-1$               & 0.63 & 5\%  & 0.13 & 20\% \\
   &     &                                 &    &     & $\Gamma_{\out,s}(h_{0,\out,s}-1)$ & 11   & 95\% & 0.5  & 80\% \\
\hline
 3 & 0.1 & $2.0\times10^8$, idem           & 90 & 3.6  & $\Gamma_{\out,s}-1$               & 0.38 & 4\%  & 0.091& 14\% \\
   &     &                                 &    &      & $\Gamma_{\out,s}(h_{0,\out,s}-1)$ & 9.7  & 96\% & 0.56 & 86\% \\
\hline
 6 & 0.1 & $8.0\times10^8$, idem           & 360&  14 & $\Gamma_{\out,s}-1$               & 0.42 & 1\%   & 0.19 & 8\%  \\
   &     &                                 &    &     & $\Gamma_{\out,s}(h_{0,\out,s}-1)$ & 51   & 99\%  & 2.2  & 92\% \\
\end{tabular}
\end{table*}

It can easily be shown (Sect.~\ref{sec:outflow_analytical_estimate}) from analytical
considerations that the outflows from the diffusion region should
have relativistic bulk velocities and/or relativistic temperatures. In our simulation
data, the thermal part always clearly dominates over the bulk kinetic energy part,
more strongly for more relativistic cases (Sect.~\ref{sec:outflows_results_simulations}). A refined analytical
estimate explains why in Sect.~\ref{sec:outflow_analytical_estimate_harder}.

%%%%%%%%%%%%
\subsubsection{A simple analytical estimation}
%%%%%%%%%%%%
\label{sec:outflow_analytical_estimate}

As explained in Sect.~\ref{sec:overall_evolution}, bipolar outflow jets are naturally produced from each side of the X-points.
They are clearly visible in Fig.~\ref{fig:xcf_wcewpe=3_NT=6000_2D_pseudocolor_illustration}.
An estimation of the energy content of these outflows can be easily obtained in steady state,
by using the conservation of particle number and of energy. 
To do so, we consider that the diffusion region for species $s$ has a length $D_s$ (along $z$) and a width $\delta_s$ (along $x$).
We generalize the situation to cases where there is a guide field $B_\mathrm{G}$.
We denote quantities entering (leaving) this region by a subscript ``in'' (``out'', respectively).

Conservation of particle number (Eq.~\ref{equ:conservation_part_number_lab}) 
gives $n_{\ins,s} v_{\ins,s} D_s = n_{\out,s} v_{\out,s} \delta_s$. 
The inflow velocity is estimated by the $E\times B$ velocity as $v_{\ins,s} = E_y / B_0$.
This assumes that the reconnection electric field is constant inside and outside of the diffusion region, 
a fact confirmed by our simulations\note{(analytical justification in 2D:...)}, 
and that particles do $E\times B$ drift up to the very entrance of the diffusion region, 
which is true in our simulations up to a factor $\sim 2$
(Fig.~\ref{fig_wcewpe=1_NT=21300_cutalongX_partNumber_velocity}).
It also assumes that the value of the magnetic field at the entrance of the diffusion region is the asymptotic field 
$B_0$, while in our simulations $B_z$ has already dropped by a factor $\sim 3$ at this level 
(Fig.~\ref{fig_wcewpe=1_NT=21300_cutalongX_partNumber_velocity}), leading to an error in $v_{\ins,s}$ of the same order.

Regarding energy conservation (Eq.~\ref{equ:energy_relat_lab_all_species}), 
the inflow energy flux includes 
the particle energies, and the reconnecting and guide field Poynting fluxes
(Eqs.~\ref{equ:energy_flux_fields} and \ref{equ:energy_flux_particles}):
$D_sn_{\ins,s}m_sc^2\bar{p}_{\ins,s} + D_sv_{\ins,s} \cdot B_0^2/\mu_0 + D_sv_{\ins,s} \cdot B_\mathrm{G,in}^2/\mu_0$.
We assume that in the outflow the energy in the reconnected magnetic field $B_0$ is negligible compared to particle energy,
so that the energy flux is 
$\delta_s n_{\out,s}m_s c^2 \bar{p}_{\out,s} + \delta_s v_{\out,s} \cdot B_\mathrm{G,out}^2/\mu_0$.
% Here $\langle\cdot\rangle_s$ stands for an average over $\b{p}$ weighted by the distribution function $f_s(\b{x},\b{p},t)$ for species $s$,
% and $\bar{p}_{\out,s} = \langle\gamma v_z\rangle_s$ is the mean momentum.
Equating the two fluxes, combining this with the conservation of particle number, we obtain:
\begin{equation}\label{equ:outflow_energy}
%  \begin{aligned}
 \frac{\bar{p}_{\out,s}}{\bar{v}_{\out,s}} %&= \frac{\bar{p}_{s,\ins}}{\bar{v}_{s,\ins}} + \frac{B_0^2}{\mu_0 n_{s,\ins}m_s c^2} + \frac{B_\mathrm{G}^2}{\mu_0 n_{s,\ins}m_s c^2}(1-\alpha) \\
 = \frac{\bar{p}_{\ins,s}}{\bar{v}_{\ins,s}} + \sigma_{\ins,s}^\mathrm{cold}(B_0) + \sigma_{\ins,s}^\mathrm{cold}(B_\mathrm{G,in})(1-\alpha),
%  \end{aligned}
\end{equation}
with $(1-\alpha) = \left( 1-\frac{n_{\ins,s} B_\mathrm{G,out}^2}{n_{\out,s} B_\mathrm{G,in}^2} \right)$. 
The guide field is usually merely compressed, so that $1-\alpha\simeq 0$\note{(link this to CGL?)}.

Equation~\ref{equ:outflow_energy} is independent of the $\b{p}$ dependence of the distribution function $f_s$. 
However, some insights can be gained by considering a distribution that is isotropic in the comobile frame, 
for which we have the result $\bar{\b{p}} = h_0(T) \Gamma \bar{\b{v}}$, with $h_0$ the comobile 
enthalpy (as defined and pictured for a Maxwell-J\"uttner distribution in Fig.~\ref{fig_kappa_32}), 
and $\Gamma = (1-\bar{\b{v}}^2/c^2)^{-1/2}$. 
If, in addition, we neglect the contribution of the guide field, and assume an inflow plasma 
with non-relativistic temperatures and non-relativistic velocities,
Eq.~\ref{equ:outflow_energy} becomes\footnote{We
note that the non-relativistic limit of Eq.~\ref{equ:outflow_energy_bis}, with $h_{0,\out,s}\sim 1$ and 
$\bar{v}_{\out,s} \ll c$, is 
\begin{equation}\label{equ:NR_outflow_speed}
  \bar{v}_{\out,s} = \sqrt{2\sigma_{\ins,s}^\mathrm{cold}(B_0)} = \sqrt{2}V_\mathrm{s,A}^\mathrm{in}(B_0), 
\end{equation}
where $V_\mathrm{s,A}^\mathrm{in}(B_0)$ is the non-relativistic inflow Alfv\'en speed with only the mass of species $s$.
It comprises only the component $B_0$, so that if there is a guide field, this is the projection of the total Alfv\'en speed 
onto the outflow direction $\hat{\b{z}}$.}
\begin{equation}\label{equ:outflow_energy_bis}
 h_{0,\out,s}\Gamma_{\out,s}  = 1 + \sigma_{\ins,s}^\mathrm{cold}(B_0).
\end{equation}
We clearly see that for a relativistic inflow plasma, where $B^2/\mu_0 > nmc^2$ and hence $\sigma_{\ins,s}^\mathrm{cold}(B_0)>1$,
magnetic reconnection is expected to produce outflows with either relativistic bulk velocities ($\Gamma_{\out,s}>1$),
or relativistic temperatures ($h_{0,\out,s}>1$), or both. 
We also see that since $\sigma_s^\mathrm{cold}\propto 1/m_s$, 
electrons will be more accelerated/heated than ions, and that relativistic electrons 
($\sigma_\lec^\mathrm{cold}>1$) can be expected even at low ion magnetizations 
($\sigma_\ion^\mathrm{cold} = \sigma_\lec^\mathrm{cold}\times m_\lec/m_\ion \ll 1$).

%%%%%%%%%%%%
\subsubsection{Results from simulations}
%%%%%%%%%%%%
\label{sec:outflows_results_simulations}

We first check that the energy estimate of Eqs.~\ref{equ:outflow_energy} and~\ref{equ:outflow_energy_bis}
holds in all simulations,
which is indeed true up to a factor $\sim 6$.
An only approximate correspondence is to be expected because this relation assumes a simple geometry, 
and no energy exchange between the species.
For example, in Fig.~\ref{fig_wcewpe=3_NT=6000_cutalongZ_summary}, for $\wce/\wpe=3$, we measure 
in the inflow $1 + \sigma_{\ins,s}^\mathrm{cold}(B_0) = 2.1$ for ions and 1.7 for electrons, 
while we have at the outflow maximal velocity $\bar{p}_{\out,s}/\bar{v}_{\out,s} = 1.7$ for ions and 13 for electrons.
These orders of magnitude hold for all cases. 

A more refined analysis of the energy content of the outflow, split into its thermal and bulk contributions, can be performed. 
To do so, we decompose the particle energy flux as (see Appendix~\ref{sec:app_measure_relat_1}):
\begin{equation}\label{equ:particles_energy_flux}
\begin{aligned}
 n_s \langle &\gamma m_s c^2 \b{v}\rangle_s = n_s m_sc^2 h_{0,\out,s} \Gamma_{\out,s} \bar{\b{v}}_{\out,s} \\
                                           &= n_s m_sc^2 \bar{\b{v}}_{\out,s} \left[1 + (\Gamma_{\out,s}-1) + \Gamma_{\out,s}(h_{0,\out,s}-1)\right].
\end{aligned}
\end{equation}
On the right-hand side, the first term is the rest-mass energy flux, and is the same as that from the inflow. 
The second is the kinetic energy of a cold bulk flow of velocity $\bar{\b{v}}_{\out,s}$.
The third is the energy transported by thermal motions in the plasma rest-frame and by pressure work, 
and we will denote it as the enthalpy flux. 
We note that these definitions match those of \citet{Zenitani2009b}, 
who performed a similar analysis with two-fluid simulations of relativistic reconnection in pair plasmas.

We measure the maximum outflow velocity $\bar{v}_{\out,s}$, deduce the Lorentz factor $\Gamma_{\out,s}$,
measure the maximum in momentum $\bar{p}_{\out,s}$, 
and compute the enthalpy $h_{0,\out,s} = \bar{p}_{\out,s} / (\Gamma_{\out,s}\bar{v}_{\out,s})$.
From these values, we estimate in Table~\ref{tab:energy_outflow_flux} 
the balance of particle energy between each of the terms of Eq.~\ref{equ:particles_energy_flux}.

In all cases, a large fraction of the particle energy flux is in thermal kinetic energy, not in bulk flow kinetic energy. 
For electrons, we see that the thermal part clearly dominates more as one increases the relativistic nature of the inflow
(e.g., 69\% in the thermal part for the less relativistic case, 99\% for the most relativistic).
This is also the case for ions: from 80\% to 92\% in the thermal part as the ion magnetization increases.
The $T_\mathrm{bg,i}=2\times10^8$\,K case is exceptional with 
91\% in the thermal energy, but this large fraction is likely explained by interactions with the hot electrons
$T_\mathrm{bg,e}=3\times10^9$\,K.
We explain why thermally dominated outflows are expected at large inflow magnetization with a refined analytical model 
in Sect.~\ref{sec:outflow_analytical_estimate_harder}.

%%%%%%%%%%%%%%%%%%%%%%%%%%%%%%%
%%%%%%%%%%%%%%%%%%%%%%%%%%%%%%%
\subsection{Islands structure}
%%%%%%%%%%%%%%%%%%%%%%%%%%%%%%%
%%%%%%%%%%%%%%%%%%%%%%%%%%%%%%%
\label{sec:islands_no_guide_field}

Turning to the magnetic islands, we emphasize that they are magnetically isolated, 
have an M-shaped density distribution, and are hot with anisotropic temperatures.

After being expelled along $\pm z$ in the outflow, the particles meet the magnetic islands that separate each pair of X-points. 
The islands are initially formed by the tearing of the current sheet. 
They then consist only of particles of the current sheet, plus those of the background plasma that were inside the current sheet location.
Small at the beginning, they grow by collecting particles from the outflows at their periphery and by merging with other islands.
A remarkable property is that particles from the background plasma cannot enter inside the islands: 
they are scattered by the strong magnetic field structure surrounding the island, and circle around it.
Consequently, the particles at the island centers remain the same throughout the whole simulation, 
even after many island merging events. 
This matter is explored in more details in Melzani et al. (in prep.).
We stress here two main points.

\begin{figure}[tbp]
 \centering
 \includegraphics[width=\columnwidth]{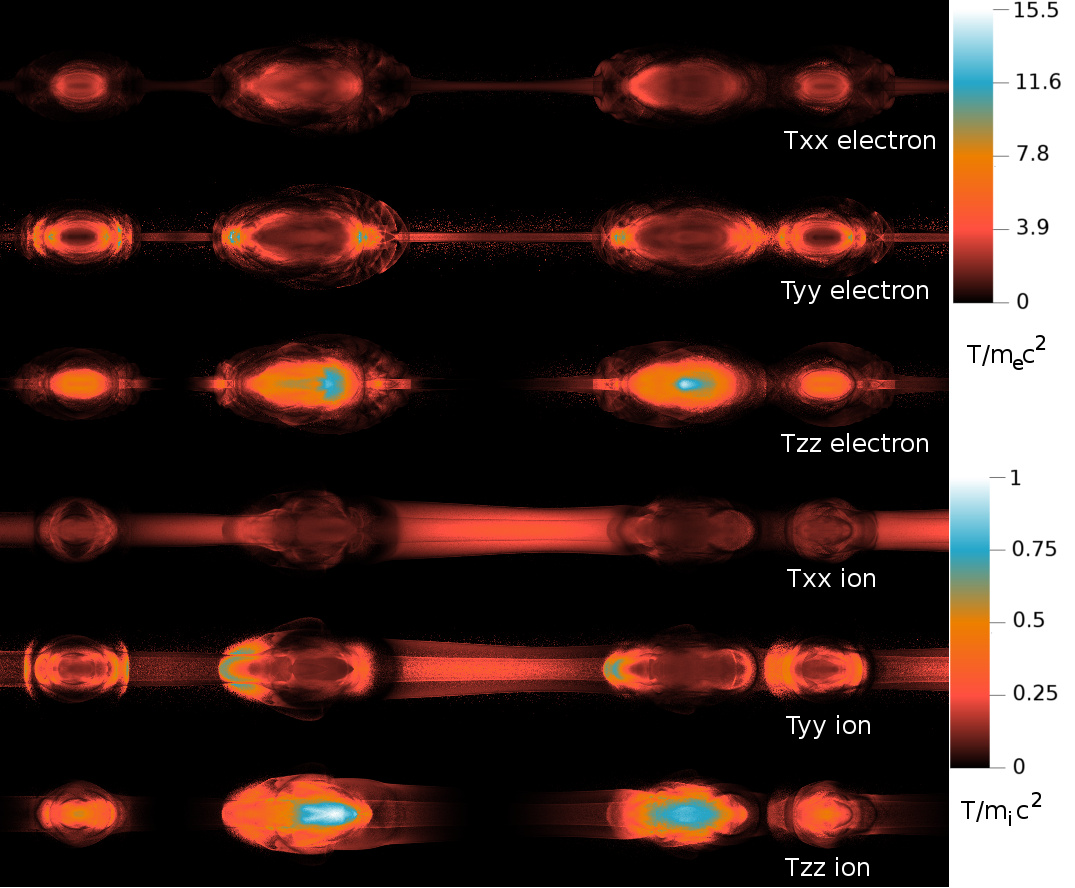}
 \caption{\label{fig:xcf_wcewpe=3_NT=6000_2D_pseudocolor_temperature_NT=6000}
          Temperatures for same simulation and time as in 
          Figs.~\ref{fig:xcf_wcewpe=3_NT=6000_2D_pseudocolor_illustration},
          \ref{fig_wcewpe=1_NT=21300_cutalongX_partNumber_velocity} upper-right,
          \ref{fig:wcewpe=3_cut_Z_NT6000_temperature_only}, and \ref{fig_wcewpe=3_NT=6000_cutalongZ_summary} right)
          Note the different units for ions and electrons. 
          Since $m_\ion c^2=25m_\lec c^2$, the ions are actually hotter than the electrons.
%           \modif{Put also the non-diagonal temperatures?}
          }
\end{figure}

First, the trapped particles are heated by the island contraction (which occurs when two islands merge),
so that the central temperatures are very high for both species, highly anisotropic 
(Figs.~\ref{fig_wcewpe=3_NT=6000_cutalongZ_summary} and~\ref{fig:xcf_wcewpe=3_NT=6000_2D_pseudocolor_temperature_NT=6000}),
with a dominance of $T_{zz}$.
The island centers are also where the currents are the strongest.

Second, as we said, 
most of the inflowing background particles populate a region around the center, 
and the central part of the island mainly consists of
particles originally from the current sheet. 
As a result, the central part is less dense than the surrounding part, and a cut along $z$ through an island center 
(Fig.~\ref{fig_wcewpe=3_NT=6000_cutalongZ_summary}) 
reveals for the particle density a M-shape, with a central dip and two shoulders.
This may explain observed density dips at the center of magnetic islands during magnetotail reconnection events
\citep{Khotyaintsev2010}, without invoking island merging or particle escape along the flux tube\note{As \citet{Markidis2013} does.}.

% As time goes on, the islands are populated by particles from the background plasma that pass in the diffusion region, in the outflow jets, 
% and then their course by circling around the islands.
% The magnetic and kinetic (bulk and thermal) energy of the outflows is then entirely stored in the form of magnetic and thermal energy,
% and the magnetic field is mainly compressed at the island entrance.

%%%%%%%%%%%%%%%%%%%%%%%%%%%%%%%
%%%%%%%%%%%%%%%%%%%%%%%%%%%%%%%
\subsection{Reconnection electric field and reconnection rate}
%%%%%%%%%%%%%%%%%%%%%%%%%%%%%%%
%%%%%%%%%%%%%%%%%%%%%%%%%%%%%%%
\label{sec:rec_electric_field}

\begin{figure}[tbp]
 \centering
 \def\svgwidth{\columnwidth}
 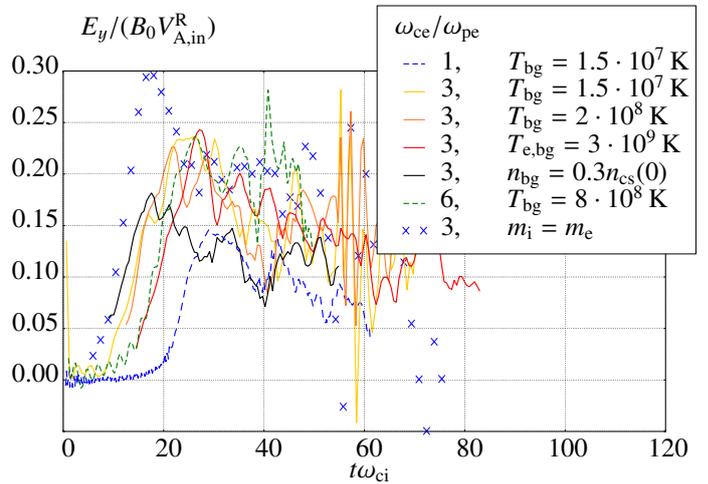
 \caption{\label{fig:reconnection_rates}Time evolution of the normalized reconnection electric field 
 $E_y/(B_0 V^\mathrm{R}_\mathrm{A,in})$, 
 measured at the X-point of various simulations.
 The velocity $V_\mathrm{A,in}^\mathrm{R}$ is given in Table~\ref{tab:param_magnetization}, 
 and $B_0=0.11$, 0.33 or 0.66 for $\wce/\wpe=1$, 3 or 6 respectively. 
 Time is normalized by the ion cyclotron pulsation, but
 note that the growth rate of the collisionless relativistic tearing mode is not proportional to 
 $\omega_\mathrm{ci}$ \citep{Petri2007}, 
 hence the different time lags and shapes. 
 In particular for $\wce/\wpe=6$, the time scale of the plot is divided by 3.
 For pairs, the timescale is $t\wce/25$.}
\end{figure}

The rate of variation of magnetic field flux across a X-point, 
$\dif\Phi_{B_z}/\dif t=(\dif/\dif t)\int_{x=0}^\text{X-point}B_z\dif x$, 
is equal in two-dimensional configurations to the $y$ component $E_y$ of the electric field at the X-point location.
In addition, $\dif\Phi_{B_z}/\dif t$ is in part determined by 
the outflow velocity, 
because the latter sets the rate at which magnetic field is extracted from 
around the X-point \citep[see e.g., in a resistive MHD context,][]{Borovsky2007,Cassak2007}. 
Since in non relativistic setups one expects $\bar{v}_\out \propto V_\mathrm{A}$,
the reconnection rate $E_y$ is usually normalized either to $B_0 V_\mathrm{A,0}^\mathrm{NR}$, 
with $V_\mathrm{A,0}^\mathrm{NR}$ the hybrid Alfv\'en speed of Eq.~\ref{equ:hybrid_NR_Alfven_speed},
or to $B_0 V_\mathrm{A,in}^\mathrm{NR}$, 
with $V_\mathrm{A,in}^\mathrm{NR}$ the Alfv\'en speed in the inflow of Eq.~\ref{equ:inflow_NR_Alfven_speed}.
These normalizations are chosen so that the normalized rate, $E^* = E_y/B_0 V_\mathrm{A}$, stays close to the same set of values.
For example it has been shown that it gives identical results when varying the mass ratio
(e.g., \citet{Hesse1999}, or \citet{Ricci2002,Ricci2003} for $m_\ion/m_\lec=25,180,1836$ 
with an implicit PIC code)\footnote{However, going down to $m_\ion/m_\lec=1$
leads to less systematic results.
For example, \citet{Fujimoto2009} reports $E^*=0.15$ for $m_\ion/m_\lec=100$ and 0.08 for pairs.
\citet{Liu2014} report close rates for $m_\ion/m_\lec=1$ and 25. 
Here we performed a simulation with $m_\ion/m_\lec=1$, and find a peak reconnection rate $E^* = 0.30$, larger than for $m_\ion/m_\lec=25$
(Fig.~\ref{fig:reconnection_rates}).}.

In the following we turn to our relativistic case and ask whether a normalization can be found that confines the 
range of values for $E^*$ in a narrow range, and relaxes to the above normalization in the non-relativistic case.
We argue here that the normalization by the hybrid Alfv\'en speed is not relevant, because it does not depend on 
the particle number density of the inflow, while the ratio $E_y/B_0$ clearly does. 
This is seen for the simulation with $n_\bg=0.3n_\cs(0)$, for which $E_y/B_0$ peaks at $0.13c$, compared to the 
otherwise identical simulation with $n_\bg=0.1n_\cs(0)$, where $E_y/B_0$ peaks at $0.20c$.
On another hand, the inflow Alfv\'en speed $V_\mathrm{A,in}^\mathrm{NR} \propto 1/\sqrt{n_\bg}$, and thus leads to 
closer normalized rates. We consequently exclude hybrid quantities for normalization.

In a relativistic configuration the non-relativistic Alfv\'en speed can increase to infinity. 
However, the ratio $E_y/B_0$ is also the $E\times B$ velocity of the incoming plasma, and cannot exceed the speed of light.
The normalizing Alfv\'en velocity should thus also saturate to some value,
which is why we choose to normalize the electric field by
\begin{equation}\label{equ:rec_rate_normalized}
 E^* = \frac{E_y}{B_0 V_\mathrm{A,in}^\mathrm{R}},
\end{equation}
with $V_\mathrm{A,in}^\mathrm{R}$ the relativistic Alfv\'en speed in the inflow (Eq.~\ref{equ:inflow_R_Alfven_speed}), that 
cannot exceed $c$.
The time evolution of $E^*$ is shown in Fig.~\ref{fig:reconnection_rates}.
Several comments can be made.

First, the rate $E^*$ is not sensitive to the background plasma temperatures, as can be seen for the simulations 
$\wce/\wpe=3$, $n_\bg=0.1n_\cs(0)$, no guide field and $T_\bg = 1.5\times10^7$, $2\times10^8$ and $3\times10^9$\,K.
This contrasts with the interpretation of \citet{Hesse2007} who attribute a lower rate to a larger inflow 
temperature. In addition to the temperatures, the magnetization of their simulation also changes, and may also affect the rates.
Coming back to our simulations, we note that we use very low background plasma $\beta$ 
($<10^{-2}$, Table~\ref{tab:param_magnetization}),
and that a weak plasma $\beta$ dependence is expected for higher values 
\citep[e.g.,][have rates $E^*$ multiplied by $\sim2$ when $\beta$ passes from 0.01 to 1]{TenBarge2013}.

Second, the reconnection rate for the simulation with a higher background particle density ($n_\bg=0.3n_\cs(0)$, $E^*=0.18$) 
remains lower than its counterpart with $n_\bg=0.1n_\cs(0)$ ($E^*=0.23$).
This is in line with the pair plasma simulations of \citet{Bessho2012} who found a similar rate for 
$n_\bg=0.1n_\cs(0)$ ($E^*=0.19$), and a higher rate for $n_\bg=0.01n_\cs(0)$ ($E^*=0.36$).
The reconnection rate thus increases with decreasing background plasma density, which is also coherent with 
the $\beta$ dependence mentioned above.

Finally, the normalization leads to very similar values of $E^*$ 
for the relativistic cases ($\wce/\wpe=3$ or 6), with $E^*=0.17$-0.24,
but to a significantly smaller rate for the less relativistic case ($\wce/\wpe=1$), with $E^*=0.14$.
More generally, the values for the relativistic cases are larger than those reported in the literature
for undriven, symmetric reconnection with zero guide field in \textit{non-relativistic} ion-electron plasmas.
We can quote for the peak values of $E^*$ (once normalized in the same way as here):
\citet{Birn2001,Pritchett2001}: $0.09$, 
\citet{Fujimoto2006,Fujimoto2009}: $0.15$,
\citet{Daughton2006}: $0.08$, 
\citet{Klimas2010}: $0.07$--$0.09$,
and the theoretical work of \citet{Hesse2009,Hesse2009b} predicting a maximal rate of $0.28$.
Our results thus suggest larger rates for relativistic reconnection, 
a fact already seen in relativistic simulations of pair plasmas with, e.g.,
\citet{Zenitani2007} ($E^*=0.2$), \citet{Cerutti2012b} ($E^*=0.17$),
or \citet{Bessho2012} ($E^*=0.19$ and 0.36).

In conclusion, the relativistic Alfv\'en speed of the inflow provides 
the best normalization for the reconnection electric field,
in that it is robust from non-relativistic to relativistic flows. 
Corresponding rates are in a close range (0.14--0.25),
which is higher than the rates found in non-relativistic simulations with the same normalization (0.07--0.15).
The rate does not depend on the inflow temperature at low $\beta$,
but is nevertheless not universal: it decreases with increasing background particle number density.
Generalization to the presence of a guide field is discussed in Sect.~\ref{sec:rec_electric_field_guide_field}.

%%%%%%%%%%%%%%%%%%%%%%%%%%%%%%%
%%%%%%%%%%%%%%%%%%%%%%%%%%%%%%%
\subsection{Hall field and dispersive waves}
%%%%%%%%%%%%%%%%%%%%%%%%%%%%%%%
%%%%%%%%%%%%%%%%%%%%%%%%%%%%%%%
\label{sec:Hall}
We can see in Fig.~\ref{fig_wcewpe=1_NT=21300_cutalongX_partNumber_velocity} that inside the ion diffusion region, 
but outside of the electron diffusion region, ions have a small fluid velocity, while electrons still $E\times B$ drift toward 
their diffusion region. This results in a net current roughly given by 
$q_\lec n_\lec \bar{\b{v}}_\lec = q_\lec n_\lec \b{E}\wedge\b{B}/B^2$, 
which is the Hall current. This current continues along the magnetic separatrices in the outflow direction, 
and is at the origin of a quadripolar magnetic field directed along $\pm\hat{\b{y}}$.
This Hall magnetic field, with the quadripolar structure, is present in our simulations. 
It has a weak intensity (between 1\% and 10\% of $B_0$).
The charge separation between electrons and ions (Fig.~\ref{fig_wcewpe=1_NT=21300_cutalongX_partNumber_velocity}) 
also leads to the creation of a Hall electric field directed along 
$+\hat{\b{x}}$ in the $x<0$ region and $-\hat{\b{x}}$ in the $x<0$ region.
Both the magnetic and electric Hall fields are absent in a simulation with pairs.

The difference in the dynamical response of ions and electrons
also allows the existence of waves with a quadratic dispersion relation,
$\omega\propto k^2$,
below ion scales \citep[either whistler waves or kinetic Alfv\'en waves, see][]{Rogers2001}.
Observations of the same reconnection rate for any simulation model allowing these waves 
\citep[PIC, electron-MHD, Hall-MHD, two-fluid with and without electron inertia, 
hybrid simulations, see ][]{Birn2001,Shay2007,Rogers2001},
as well as theoretical considerations, have led to the thesis that 
these waves are essential to allow for fast reconnection rates.
However, this view is questioned by a number of simulations 
that do not support quadratic dispersive waves, but still support fast rates
\citep[hybrid simulations with no Hall term, pair plasmas, or strong guide field regime, see][]{Karimabadi2004,Bessho2005,Daughton2006,Daughton2007,Liu2014}.
It is thus interesting to see whether our simulation data can provide any further insight into the matter.

A prediction of the dispersive wave physics is that the reconnection rate is controlled solely by the ion physics, and not by 
the electrons. According to \citet{Daughton2006}, it should be independent of the electron diffusion region length.
Their analysis we could, however, not reproduce
because the electron diffusion zone length is, in our case, limited by the standing islands. 
It cannot stretch to large values, and we are thus unable to conclude in favor or in disfavor of the dispersive wave paradigm.

However, we underline that the simulation with $m_\ion/m_\lec=1$ that we performed features an identical 
(and even slightly larger, Fig.~\ref{fig:reconnection_rates}) reconnection rate than simulations with $m_\ion/m_\lec=25$.
It thus points toward a negligible influence of the dispersive waves, or to another mechanism allowing fast rates in pair plasmas.

%%%%%%%%%%%%%%%%%%%%%%%%%%%%%%%
%%%%%%%%%%%%%%%%%%%%%%%%%%%%%%%
\subsection{Simulation-based scaling analysis}
%%%%%%%%%%%%%%%%%%%%%%%%%%%%%%%
%%%%%%%%%%%%%%%%%%%%%%%%%%%%%%%
\label{sec:outflow_analytical_estimate_harder}

The energy content of the outflows
and the balance between thermal and bulk inertia in Ohm's law 
were explored through the simulations in Sects.~\ref{Sec:Ohms_law} and~\ref{sec:outflows}.
The aim of the present section is to investigate these points 
with a simple analytical model in order to 
gain physical insight concerning these phenomena, 
and to extrapolate our simulation results to a larger parameter space.

We extend the analytical results of 
Sect.~\ref{sec:outflow_analytical_estimate},
where particle number and energy conservation allowed an estimation of the quantity $h_{0,\out,s}\Gamma_{\out,s}$
(Eqs.~\ref{equ:outflow_energy} or~\ref{equ:outflow_energy_bis}), 
by now also using the equation of conservation of momentum (Eq.~\ref{equ:fluid_2}).
% We rewrite the momentum equation into the form of Ohm's law, and use the analysis of Sect.~\ref{Sec:Ohms_law}.
% We exploit the constancy of the reconnection electric field $\b{E}_y$ in the inflow, the central region, and 
% in the outflows. 

%%%%%%%%%%%%%%%%%%%%%%%%%%%%%%%
\subsubsection{Thermal versus bulk electron inertia}

We first investigate the relative weight of thermal and bulk electron inertia.
At the center of the electron diffusion region, we learn from Sect.~\ref{Sec:Ohms_law} that the reconnection electric field 
is sustained by electron thermal and bulk inertia, with only the terms 
$\partial_x(n_\mathrm{e}\langle \delta{p}_{x}\delta{v}_{y}\rangle)$ and 
$\partial_z(n_\mathrm{e}\bar{p}_{z}\bar{v}_{y})$ contributing to either one of them, respectively.
\begin{itemize}
 \item Concerning thermal inertia, 
  the temperature tensor is defined via Eq.~\ref{equ:def_temperature}, so that  
  $\langle \delta{p}_{x}\delta{v}_{y}\rangle = c^2\Theta_{xy,\lec}/\Gamma_\lec$.
  We see in Fig.~\ref{fig:wcewpe=3_cut_Z_NT6000_temperature_only} that $\Theta_{xy,\lec}$ 
  is linear in the electron diffusion region. It vanishes at the center because there the distribution function 
  $f_\lec$ is symmetric with respect to $v_x$. It is maximal at the diffusion region edge with a value $\Theta^\mathrm{edge}_{xy,\lec}$.
  Consequently, we approximate the thermal inertia contribution 
  by $(c^2\Theta^\mathrm{edge}_{xy,\lec}/\Gamma^\mathrm{in}_\lec)/\delta_\lec$, where $\delta_\lec$ is the width of the electron diffusion region.
 \item For the bulk inertia term, we use the fact that $\bar{p}_{z}$ rises linearly from the center to its maximal 
  value denoted by $\bar{p}^\out_{z}$ over a distance $D_\lec/2$, and that $\bar{v}_{y}$ has a vanishing derivative at the center 
  (Fig.~\ref{fig_wcewpe=3_NT=6000_cutalongZ_summary}).
  Consequently, it can be estimated as $\bar{v}^\mathrm{center}_{y}\bar{p}^\out_{z}/D_\lec$.
\end{itemize}
All in all, from Ohm's law (Eq.~\ref{equ:Ohm_fluid_1}), the electric field at the center of the diffusion region is:
\begin{equation}
\label{equ:Erec_thermal_bulk}
\begin{aligned}
 E^\mathrm{center}_y &= \frac{m_\lec}{q_\lec n_\lec} \left( \frac{\partial}{\partial \b{x}}\cdot (n_\lec\langle \delta\b{p}_\lec\delta\b{v}_\lec\rangle)
      + \frac{\partial}{\partial \b{x}}\cdot (n_\lec\bar{\b{p}}_\lec\bar{\b{v}}_\lec) \right)_y \\
     &\sim \frac{m_\lec}{q_\lec} \left( 
     \frac{c^2\Theta^\mathrm{edge}_{xy,\lec}}{\delta_\lec\Gamma^\mathrm{in}_\lec} 
     + \frac{\bar{v}^\mathrm{center}_{y}\bar{p}^\out_{z}}{D_\lec} \right).
\end{aligned}
\end{equation}
The next step is to use the constancy of $E_y$, that is well respected in the simulations: 
$E^\mathrm{center}_y = E^\ins_y = \bar{v}_{\ins} B_0$. 
If we introduce the inertial length in the inflow, $d_\lec^\ins = c\sqrt{\epsilon_0 m_\lec/(n_\lec^\ins e^2)}$,
and the inflow magnetization $\sigma^\mathrm{cold}_{\ins,\lec} = B_0^2/(\mu_0m_\lec n_\lec^\ins c^2)$,
we ultimately obtain:
\begin{equation}
\label{equ:Erec_thermal_bulk_bis}
  \frac{\delta_\lec}{d_\lec^\ins} \left(\sigma^\mathrm{cold}_{\ins,\lec}\right)^{1/2} \frac{\bar{v}_{\ins}}{c}
     = \frac{\Theta^\mathrm{edge}_{xy,\lec}}{\Gamma^\mathrm{in}_\lec} 
     + \frac{\delta_\lec}{D_\lec}\frac{\bar{v}^\mathrm{center}_{y}\bar{p}^\out_{z}}{c^2}.
\end{equation}

We now proceed to derive approximate scaling relations 
for cases where either thermal or bulk inertia dominate the reconnection electric field.
\begin{itemize}
 \item First, if thermal inertia dominates over bulk inertia, then Eq.~\ref{equ:Erec_thermal_bulk_bis}
 gives 
 \begin{equation}\label{equ:scaling_theta_edge_thermally_dom}
  \frac{\Theta^\mathrm{edge}_{xy,\lec}}{\Gamma^\mathrm{in}_\lec} = \frac{\delta_\lec}{d_\lec^\ins}\left(\sigma^\mathrm{cold}_{\ins,\lec}\right)^{1/2} \frac{\bar{v}_{\ins,\lec}}{c} \propto \left(\sigma^\mathrm{cold}_{\ins,\lec}\right)^{1/2}.
 \end{equation}
 There are thus several factors contributing to $\Theta^\mathrm{edge}_{xy,\lec}$.
 The diffusion zone width $\delta_\lec$ is dynamically set during the reconnection process. It can be of the order 
 of the particles gyroradius at the center of the current sheet, 
 or of the plasma inertial length at the center of the current sheet. 
 In all our simulations we find that the latter assumption holds throughout time to within a factor 2
 (Sect.~\ref{sec:diff_width_2}), and in any case, $\delta_\lec / d_\lec^\ins$ is expected to be of order unity.
 
 The inflow speed is set by the reconnection electric field, $\bar{v}_{\ins} = E_y/B_0 = E^* V_\mathrm{A,in}^\mathrm{R}$
 with $E^*$ the normalized reconnection rate (which lies in the range 0.1-0.25, Sect.~\ref{sec:rec_electric_field}) 
 and $V_\mathrm{A,in}^\mathrm{R}$ the relativistic Alfv\'en speed in the inflow. For relativistic setups we thus have $\bar{v}_{\ins}\sim E^*c$.
 
 The inflow magnetization can be arbitrarily large. It is thus the main actor to produce relativistic temperatures, 
 and thermal inertia scales as $\Theta^\mathrm{edge}_{xy,\lec} \propto \left(\sigma^\mathrm{cold}_{\ins,\lec}\right)^{1/2}$.
 \item Second, the term corresponding to bulk inertia in Eq.~\ref{equ:Erec_thermal_bulk_bis}
       can be estimated with the help of Eq.~\ref{equ:outflow_energy} (with $\bar{p}_{\out,s}=\bar{p}^\out_{z}$,
       $\bar{v}_{\out,s}=\bar{v}^\out_{z}$, and neglecting the guide field):
       \begin{equation}
         \frac{\delta_\lec}{D_\lec}\frac{\bar{v}^\mathrm{center}_{y}\bar{p}^\out_{z}}{c^2} = \frac{\delta_\lec}{D_\lec} \frac{\bar{v}^\mathrm{center}_{y}\bar{v}^\out_{z}}{c^2} \left(\frac{\bar{p}_{\ins}}{\bar{v}_{\ins}} + \sigma^\mathrm{cold}_{\ins,\lec}\right).
       \end{equation}
       The ratio $\delta_\lec/D_\lec$ is of order $1/10$ in our simulations.
       If we neglect the term ${\bar{p}_{\ins}}/{\bar{v}_{\ins}}$, 
       which is of order unity for non-relativistic inflow temperatures, 
       we see that bulk inertia scales with $\sigma^\mathrm{cold}_{\ins,\lec}$.
\end{itemize}
In conclusion, 
thermal inertia scales at most as $\left(\sigma^\mathrm{cold}_{\ins,\lec}\right)^{1/2}$,
and bulk inertia as $\sigma^\mathrm{cold}_{\ins,\lec}$.
Consequently, regarding the non-ideal terms in Ohm's law in the electron diffusion region, 
we expect bulk inertia to outweight thermal inertia at large inflow electron magnetization.

%%%%%%%%%%%%%%%%%%%%%%%%%%%%%%%
\subsubsection{Energy content of the outflows}

We now turn to the energy content of the outflows, in order to see whether we can explain their thermally dominated character 
for relativistic runs.
The temperature in the outflows is dominated by $\Theta_{xx,\lec}$ or $\Theta_{yy,\lec}$, 
which we denote by $\Theta^\mathrm{out}_{\lec}$.
We first have to link $\Theta^\mathrm{out}_{\lec}$ to $\Theta^\mathrm{edge}_{xy,\lec}$.
The outflow temperature at the center of the diffusion region is roughly constant along $z$ 
throughout the area of linear increase of $\bar{v}_z$
(Fig.~\ref{fig_wcewpe=3_NT=6000_cutalongZ_summary}),
because particles on their way from the X-point to the exhaust 
mainly turn into the reconnected 
magnetic field and thus do not really gain thermal agitation, but convert it from one component of $\Theta$ to another.
We can thus assume $\Theta^\mathrm{out}_{\lec} = \Theta^\mathrm{center}_\lec$.
We now would like to assume $\Theta^\mathrm{edge}_{xy,\lec} \sim \Theta^\mathrm{center}_{xx,\lec}$.
This indeed holds for electrons in the case of Fig.~\ref{fig:wcewpe=3_cut_Z_NT6000_temperature_only}.
However, this does not hold in all simulations, 
and $\Theta^\mathrm{edge}_{xy,\lec}$ is between $1/10$ to 10 times $\Theta^\mathrm{center}_{xx,\lec}$.
This is due to the different origin of these components: 
$\Theta^\mathrm{center}_{xx,\lec}$ reflects particles in Speiser orbits going up and down along $x$ with 
a zero bulk $x$-velocity, while $\Theta^\mathrm{edge}_{xy,\lec}$ reflects the asymmetry of the distribution function 
with respect to $v_x$ due to the newly entering particles at the edge of the diffusion zone.

With the previous remark in mind, we still make the hypothesis 
$\Theta^\mathrm{edge}_{xy,\lec}\sim\Theta^\mathrm{center}_{xx,\lec}$.
Next, if we assume that thermal inertia contributes significantly in Ohm's law, 
we obtain with the scaling of Eq.~\ref{equ:scaling_theta_edge_thermally_dom}:
\begin{equation}\label{equ:theta_out_is_relativistic_simple}
 \Theta^\mathrm{out}_{\lec} \propto \left( \sigma^\mathrm{cold}_{\ins,\lec} \right)^{1/2}.
\end{equation}
For relativistic temperatures we have $h_{0,\out,\lec} \simeq 4\Theta^\mathrm{out}_\lec$ (Fig.~\ref{fig_kappa_32}),
so that with Eq.~\ref{equ:theta_out_is_relativistic_simple} we see that 
a relativistic inflow magnetization implies $h_{0,\out,\lec} \propto \left( \sigma^\mathrm{cold}_{\ins,\lec} \right)^{1/2}$.
On another hand, energy conservation gives, in its simplest form (Eq.~\ref{equ:outflow_energy_bis} 
with $\sigma_\lec^\mathrm{cold}(B_0)\gg1$):
\begin{equation}\label{equ:outflow_energy_ter}
 h_{0,\out,\lec}\Gamma_{\out,\lec}  \sim \sigma_{\ins,\lec}^\mathrm{cold}.
\end{equation}
Thus:
\begin{equation}\label{equ:outflow_energy_quatro}
  \Gamma_{\out,\lec} \propto \left( \sigma^\mathrm{cold}_{\ins,\lec} \right)^{1/2}.
\end{equation}
We finally turn to the ratio of energy fluxes in the outflow.
We see with Eq.~\ref{equ:particles_energy_flux} that the flux associated with 
kinetic bulk energy is $\Gamma_{\out,\lec}-1$.
With Eq.~\ref{equ:outflow_energy_quatro} (and for $\Gamma_{\out,\lec}\gg1$), this flux is 
$\Gamma_{\out,\lec} \propto \left( \sigma^\mathrm{cold}_{\ins,\lec} \right)^{1/2}$.
The flux associated with thermal kinetic energy and pressure work is 
$h_{0,\out,\lec}\Gamma_{\out,\lec}-1$, and with Eq.~\ref{equ:outflow_energy_ter} we have 
$h_{0,\out,\lec}\Gamma_{\out,\lec}-1 \sim \sigma_{\ins,\lec}^\mathrm{cold}$. 
The ratio of thermal to bulk energy fluxes is thus $\propto \left( \sigma^\mathrm{cold}_{\ins,\lec} \right)^{1/2}$,
meaning that relativistic inflow magnetization inevitably implies reconnection exhausts dominated by 
thermal energy -- which is what we see in our simulations (Table~\ref{tab:energy_outflow_flux}), 
even if the scalings derived here do not hold exactly because of the many assumptions involved.

In conclusion, we have shown that \textit{under the hypothesis of non-ideal effects sustained by thermal inertia}, 
relativistic inflow magnetizations produce thermally dominated outflows.
The physical reason is that the reconnection electric field $E_y$ 
is large in the inflow region,
so that thermal inertia must be high in order to sustain $E_y$ in the central region, 
which implies high temperatures.

However, we also demonstrated that thermal inertia is not expected to dominate for very relativistic inflows.
When this is the case, there is no constraints from Ohm's law on the temperature,
and we cannot conclude on the ratio of thermal to bulk energy fluxes.
Since this ratio is $(h_{0,\out,\lec}\Gamma_{\out,\lec}-1)/(\Gamma_{\out,\lec}-1) \sim h_{0,\out,\lec}$, 
a relativistic outflow temperature of the order of $m_\lec c^2$ suffices to give thermally dominated outflows.

For our simulations, thermal inertia contributes equally or less than bulk inertia (Sect.~\ref{Sec:Ohms_law}), 
but still significantly, so that the outflows are thermally dominated.

%%%%%%%%%%%%%%%%%%%%%%%%%%%%%%%%%%%%%%%%%%%%%%%%%%%%%%%%%%%%%%%%%%%%%%%%%%%%%%%%%%%%%%%%%%%%%%%%%%%%%%%%%%%%%%%%%%%%%%%%%%%%%%%%%%%%%%%%%%%%%%%%
%%%%%%%%%%%%%%%%%%%%%%%%%%%%%%%%%%%%%%%%%%%%%%%%%%%%%%%%%%%%%%%%%%%%%%%%%%%%%%%%%%%%%%%%%%%%%%%%%%%%%%%%%%%%%%%%%%%%%%%%%%%%%%%%%%%%%%%%%%%%%%%%
\section{Effects of a guide field}
%%%%%%%%%%%%%%%%%%%%%%%%%%%%%%%%%%%%%%%%%%%%%%%%%%%%%%%%%%%%%%%%%%%%%%%%%%%%%%%%%%%%%%%%%%%%%%%%%%%%%%%%%%%%%%%%%%%%%%%%%%%%%%%%%%%%%%%%%%%%%%%%
%%%%%%%%%%%%%%%%%%%%%%%%%%%%%%%%%%%%%%%%%%%%%%%%%%%%%%%%%%%%%%%%%%%%%%%%%%%%%%%%%%%%%%%%%%%%%%%%%%%%%%%%%%%%%%%%%%%%%%%%%%%%%%%%%%%%%%%%%%%%%%%%
\label{sec:results_guide_field}

Except in special configurations, the generic reconnection geometry involves asymptotic fields that are not antiparallel.
An angle different from 180$^\mathrm{o}$ can be described by the addition of a uniform guide magnetic field 
$\b{B}_\mathrm{G} = B_\mathrm{G}\hat{\b{y}}$ to the antiparallel configuration. 
Such configurations have been largely studied in the non-relativistic case \citep[e.g.,][]{Pritchett2004,Drake2005,Goldman2011,Le2013,Pahlen2013},
and feature significant differences with the antiparallel case.
Here we only focus on the reconnection rates and on the island structure, and postpone a study of other points to a future publication.
We present results from two simulations, with $B_\mathrm{G}=0.5B_0$ and $B_\mathrm{G}=B_0$.
\note{The flow structure is strongly disturbed by the Lorentz force associated with the guide field, and by the fact that 
particles are tied to the field lines everywhere.}

%%%%%%%%%%%%%%%%%%%%%%%%%%%%%%%
%%%%%%%%%%%%%%%%%%%%%%%%%%%%%%%
\subsection{Overall structure}
%%%%%%%%%%%%%%%%%%%%%%%%%%%%%%%
%%%%%%%%%%%%%%%%%%%%%%%%%%%%%%%
\label{sec:guide_field_overall_structure}

% \begin{figure}[tb]
%  \centering
%  \def\svgwidth{0.95\columnwidth}
%  \import{./images/}{drawing_reconnection_guide_field_all.pdf_tex}
%  \caption{\label{fig:drawing_reconnection_guide_field}Schematic representation of magnetic reconnection with the presence of a guide field.}
% \end{figure}

\begin{figure}[tb]
 \centering
 \includegraphics[width=\columnwidth]{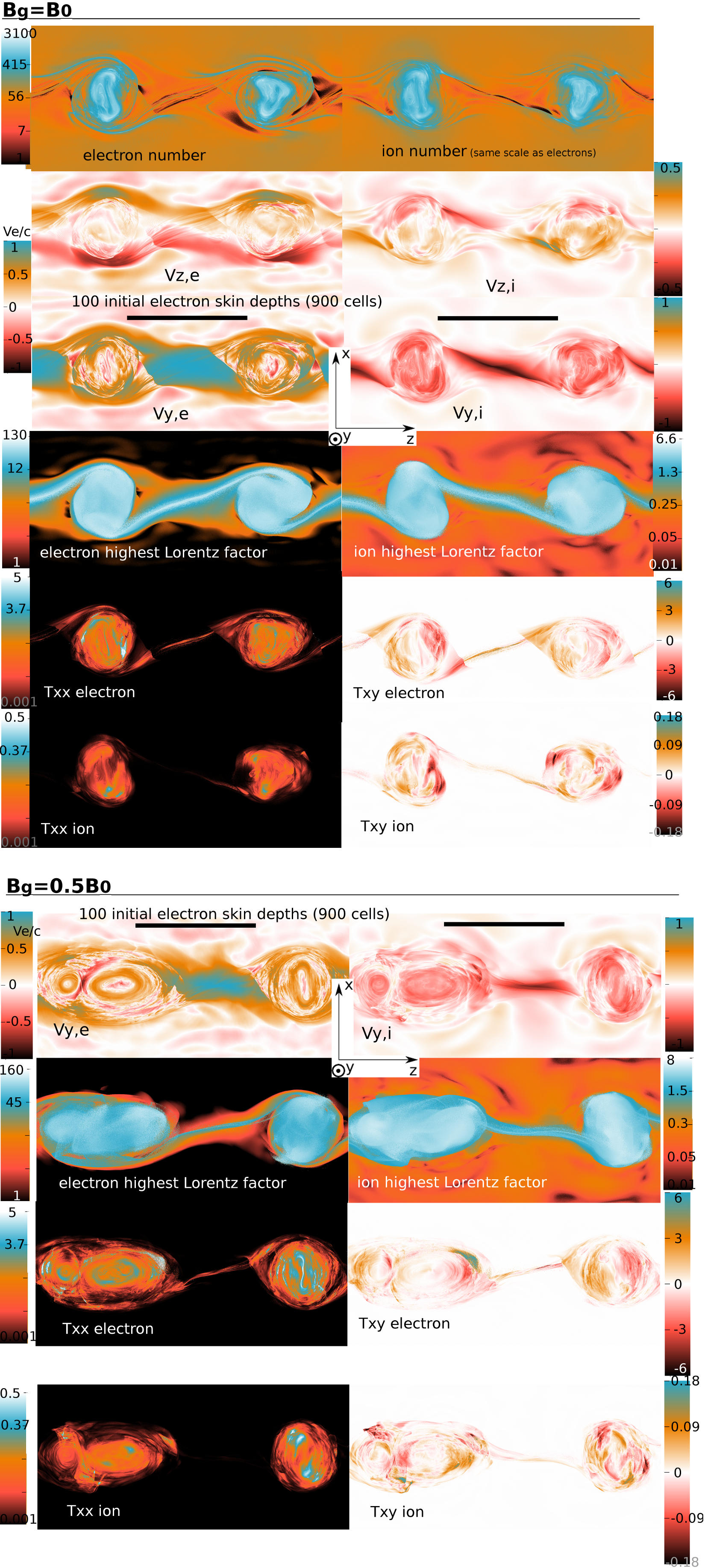}% width is 1650px.
 \caption{\label{fig:guide_field_overall_velocities}Top: Simulation with $\b{B}_\mathrm{G}=B_0\hat{\b{y}}$, 
 time $t=37\wci^{-1}=935\wce^{-1}$. %$NT=12500$, 
%  centered around $z=3200$.
 Bottom: Simulation with $\b{B}_\mathrm{G}=0.5B_0\hat{\b{y}}$, time $t=37\wci^{-1}=935\wce^{-1}$, same scale as for $\b{B}_\mathrm{G}=B_0\hat{\b{y}}$. %$NT=12500$,
 }
\end{figure}
% BIG image is : visit_wcewpe3_Bg1_NT12500_all_cut.jpg

% \begin{figure}[tb]
%  \centering
%  \includegraphics[width=\columnwidth]{images/visit_wcewpe3_Bg05_ONLY_NT12500_all_cut.jpg}% width is 1650px.
%  \caption{\label{fig:guide_field_Bg05}Simulation with $\b{B}_\mathrm{G}=0.5B_0\hat{\b{y}}$, time $t=37\wci^{-1}=935\wce^{-1}$ %$NT=12500$, 
%  with the same scale as in Fig.~\ref{fig:guide_field_overall_velocities}.
%  }
% \end{figure}

We first stress that because of the presence of the guide field, 
in both cases $B_\mathrm{G}=0.5B_0$ and $B_0$, the relation $B>E$ holds everywhere through time and space, 
and hence also the relation $B>E_\perp$ (with $E_\perp$ the component perpendicular to $\b{B}$).
Consequently, particle acceleration is not possible in directions perpendicular to $\b{B}$, and is only
allowed along the field lines at places where $\b{E}\cdot\b{B} \neq 0$.
Such parallel electric fields are allowed by the non-idealness of the plasma response, 
$\b{E}+\bar{\b{v}}_s\wedge\b{B}\neq 0$, and are indeed found at and around the X-points. \note{(Fig.~\ref{fig:guide_field_overall})}
% They allow particle acceleration there and strong currents.
% We thus have initially the creation of an electric field $\b{E}_\mathrm{rec}\propto-\hat{\b{y}}$ and of magnetic X- and O- points.

Just as in the zero guide field case, the plasma accelerated by $E_y$ is slowly deviated by
the reconnected $B_x$ component, which produces outflows directed along $\pm z$.
However, particles from these outflows feel the Lorentz force from the guide field, and their trajectories are tilted 
against the $z$ axis, as is evident in Fig.~\ref{fig:guide_field_overall_velocities}.
Reversing the guide field from $+\hat{\b{y}}$ to $-\hat{\b{y}}$ tilts them in the opposite direction.
We underline that while the tilt angle is indeed smaller for a smaller guide field, it also depends on the background plasma pressure, 
as shown by \citet{TenBarge2013}.

In the present case, $\b{E}_y \propto -\hat{\b{y}}$. 
Electrons are thus accelerated along the field lines in the $+\hat{\b{y}}$ direction.
Their motion along the field lines result in a projected ($x$-$z$ plane)
motion directed toward positive $z$ where $\b{B}\cdot\hat{\b{z}}>0$ (i.e., in the $x>0$ area),
or toward negative $z$ where $\b{B}\cdot\hat{\b{z}}<0$ (i.e., in the $x<0$ area).
It results in large and fast electron flows above and below the current sheet,
and to a rotation around the islands in a sense opposed to cyclotron gyration.
Ions are accelerated toward $-\hat{\b{y}}$ and counterstream with respect to the electrons.

\note{Plasma is dragged in from $\pm x$ and the electric field $E_y$ propagates upstream.
In the ideal region far from the current sheet, the ideal plasma response maintains an electric field perpendicular to $\b{B}$ (Fig.~\ref{fig:guide_field_overall}),
particles $E\times B$ drift along $\pm x$ toward the current sheet.
% This is summarized in Fig.~\ref{fig:drawing_reconnection_guide_field}.
In the present case, $\b{E}_y \propto -\hat{\b{y}}$. 
Electrons are thus accelerated along the field lines in the $+\hat{\b{y}}$ direction.
Since just outside the acceleration region $\b{B}$ is tilted with respect to $\hat{\b{y}}$, 
the motion of the electrons along the field lines result in a projected ($x$-$z$ plane)
motion directed toward positive $z$ where $\b{B}\cdot\hat{\b{z}}>0$ (i.e., in the $x>0$ area),
or toward negative $z$ where $\b{B}\cdot\hat{\b{z}}<0$ (i.e., in the $x<0$ area).
It results in large and fast electron flows above and below the current sheet.
Also, the projected electron motion is a rotation around the islands in a sense opposed to cyclotron gyration
in $\b{B}$.  
This is not a contradiction, because the electron Larmor radii are far smaller than the island scales.

Ions do the same, but they are accelerated by $\b{E}_y$ in the opposite direction ($-\hat{\b{y}}$), 
and thus counterstream against the electron 
along the same field lines. The motion projected in the $x$-$z$ plane is consequently opposed to that of electrons.
Their Larmor radii are also smaller than the island scales.

Reversing the guide field from $+\hat{\b{y}}$ to $-\hat{\b{y}}$ does not change the 
sign of $E_y$, and thus not the direction of acceleration of the particles along the field lines.
However, the field lines above and below the midplane are now tilted with an angle opposed to that in the previous case, 
so that the particle motions in the $x$-$z$ plane is exactly opposed with respect to those of the $\b{B}_\mathrm{G}\propto+\hat{\b{y}}$ case.
This is indeed what happens in a simulation where we set $\b{B}_\mathrm{G}\propto-\hat{\b{y}}$.
\citet{Pahlen2013} also argue in this direction.

A consequence of the dynamics as described above is that only a few electrons reach the central part of the current sheet
when compared to the zero guide field case. }

\begin{figure}[tb]
 \centering
 \def\svgwidth{0.9\columnwidth}
 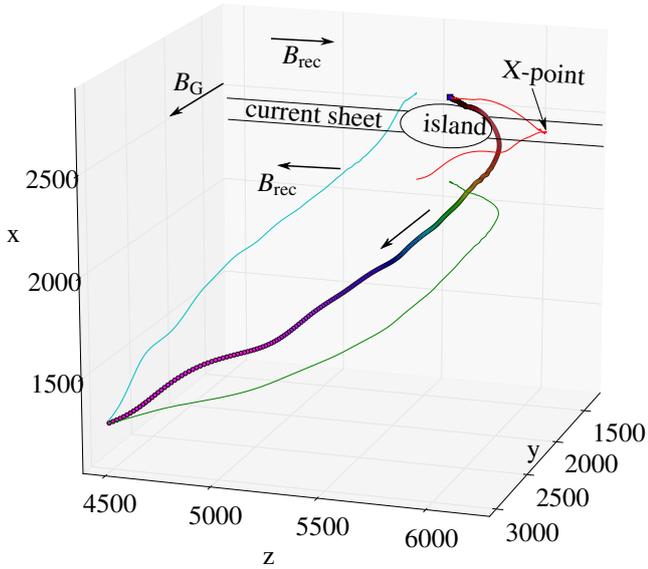
 \caption{\label{fig:traj_3D_guide_filed}Typical trajectory for a particle, here an electron, 
 for simulation with $\b{B}_\mathrm{G}=B_0\hat{\b{y}}$.
 Axis scales are given in cell numbers, with 9 cells representing one initial electron inertial length. Dot colors are the particle
 Lorentz factor, from 1 to 60 here. Solid lines are projections onto the $x$-$y$, $y$-$z$, and $z$-$x$ planes.}
\end{figure}

Particles reaching the central part are accelerated along $y$ by $E_y$ and,  
because they always feel a magnetic field $B>E$, they are guided by the magnetic field and spend more 
time in the acceleration region for strong $B_\mathrm{G}$ (Fig.~\ref{fig:traj_3D_guide_filed}). 
Consequently, $\bar{v}_y$ is larger than with no guide field under similar conditions, 
and reaches large values on a larger area (compare $\bar{v}_y$ in Figs.~\ref{fig:xcf_wcewpe=3_NT=6000_2D_pseudocolor_illustration}
and~\ref{fig:guide_field_overall_velocities}).

\note{Another consequence of the presence of a guide field concerns the relative width of the ion and electron diffusion regions.
Ions have a Larmor radius $m_\ion/m_\lec$ times larger than that of electrons, so that they feel less the guide magnetic field.
They behave more like in the case with $B_\mathrm{G}=0$, and more ions reach the center of the current sheet.
Ions and electrons thus follow a different dynamics, and as a result the electron ``diffusion zone'' is wider than that for the ions: 
ions reach smaller distances from the center of the current sheet. This is in strong contrast to the $B_\mathrm{G}=0$ case.
We stress that this is for a mass ratio of 25. With larger masses, ions would behave even more like in the $B_\mathrm{G}=0$ case,
with a diffusion zone scaling as $\sqrt{m_\ion}$ that could very well be larger than that for electrons. }

% \begin{figure}[tb]
%  \centering
%  %\def\svgwidth{\columnwidth}
%  %\import{./images/}{summary_rec_electric_field.pdf}
%  \includegraphics[width=\columnwidth]{images/visit_NT=12400_e_parallel_lec_Vxz_2.jpg}
%  \caption{\label{fig:guide_field_overall}Parallel electric field ($\b{E}\cdot\b{B}/B$) and projection of electron velocities
%  on the $x-z$ plane, for simulation with $\b{B}_\mathrm{G}=B_0\hat{\b{y}}$, at $t=37\wci^{-1}=935\wce^{-1}$ %($t=248$, $NT=12400$).
%  For clarity, electron velocities with $v>0.5c$ (which are found inside the islands) are not shown.
%  We can see around the X-point that significant electron velocities coincide with the $\b{E}\cdot\b{B}/B\neq0$ blue area.}
% \end{figure}

% %%%%%%%%%%%%%%%%%%%%%%%%%%%%%%%
% %%%%%%%%%%%%%%%%%%%%%%%%%%%%%%%
% \subsection{X-point structure}
% %%%%%%%%%%%%%%%%%%%%%%%%%%%%%%%
% %%%%%%%%%%%%%%%%%%%%%%%%%%%%%%%
% \label{sec:outflows_guide_field}

% \begin{figure}[tb]
%  \centering
%  \def\svgwidth{\columnwidth}
% %  \import{./images/}{}
%  \caption{\label{fig:cut_along_z_Bguide}Cut along $z$ for the simulation with $B_\mathrm{G}=B_0$.}
% \end{figure}

%%%%%%%%%%%%%%%%%%%%%%%%%%%%%%%
%%%%%%%%%%%%%%%%%%%%%%%%%%%%%%%
\subsection{Islands structure}
%%%%%%%%%%%%%%%%%%%%%%%%%%%%%%%
%%%%%%%%%%%%%%%%%%%%%%%%%%%%%%%
\label{sec_islands_guide_field}

The structure of the islands bears resemblance with the no guide field case: 
their centers is composed of particles initially in the current sheet, 
with background particles only circling at the periphery.
They are the hottest and strongest current-carrying part of the simulations.
There are, however, important differences.

First, the inclination of the outflows makes the island asymmetric, with electrons rotating around in a direction 
opposite to that of ions (when looking at the motion projection in the $x$-$z$ plane).
Second, as islands form and contract, the guide magnetic field is compressed and increases in strength. 
In the simulation with $B_\mathrm{G}=0.5B_0$, it passes from 0.5$B_0$ initially to 1.8-2.1$B_0$ in the islands,
while in the simulation  with $B_\mathrm{G}=B_0$, it passes from $B_0$ initially to 2.1-2.4$B_0$ 
in the islands.
Third, because of the strong magnetic field along $y$, temperatures are isotropized along $x$ and $z$
(in the no guide field case we had $\Theta_{zz}\sim 2\Theta_{xx}$). 
Here $\Theta_{zz} \simeq \Theta_{yy} \simeq \Theta_{xx}$
equal to up to 4 for electrons and 0.2 for ions.
This is a value close to the average $(\Theta_{zz}+\Theta_{yy}+\Theta_{xx})/3$ of the zero guide field case.
Off-diagonal terms are an order of magnitude smaller. 

%%%%%%%%%%%%%%%%%%%%%%%%%%%%%%%
%%%%%%%%%%%%%%%%%%%%%%%%%%%%%%%
\subsection{Reconnection electric field and reconnection rate}
%%%%%%%%%%%%%%%%%%%%%%%%%%%%%%%
%%%%%%%%%%%%%%%%%%%%%%%%%%%%%%%
\label{sec:rec_electric_field_guide_field}

\begin{figure}[tbp]
 \centering
 \def\svgwidth{\columnwidth}
 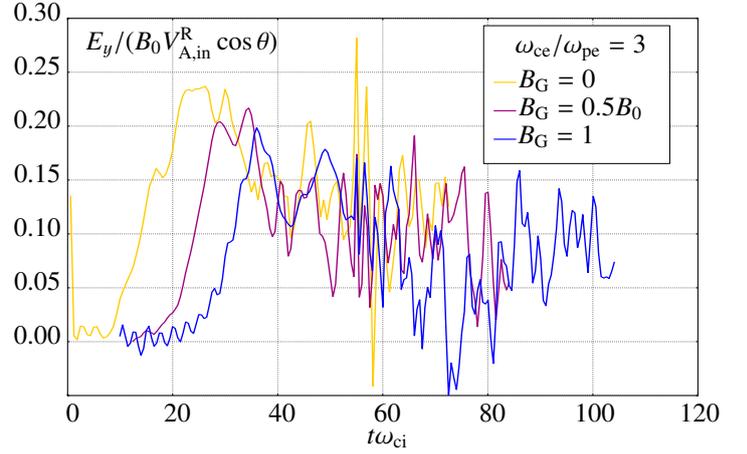
 \caption{\label{fig:reconnection_rates_guide_field}
 Time evolution of the normalized reconnection electric field 
 $E_y/(B_0 V^\mathrm{R}_\mathrm{A,in}\cos\theta)$, 
 measured at the X-point of various simulations.
 The velocity $V_\mathrm{A,in}^\mathrm{R}\cos\theta$ is given in Table~\ref{tab:param_magnetization}.
 Time is normalized by the ion cyclotron pulsation.
 Here $B_0=0.33$, $V_\mathrm{A,in}^\mathrm{R}=0.88c$, $\cos\theta=(1+B_\mathrm{G}^2/B_0^2)^{-1/2} = 1,\,0.89,\,0.71$.}
\end{figure}

In the normalization of Sect.~\ref{sec:rec_electric_field}, $E^* = E_y/(B_0V_\mathrm{A,in}^\mathrm{R})$, the Alfv\'en velocity includes 
the total magnetic field $\b{B}=B_0(\hat{\b{z}} + \alpha\hat{\b{y}})$. 
In the presence of a guide field ($\alpha\neq0$), Alfv\'en waves propagating along the magnetic field do so in a 
direction that makes an angle $\theta = \arctan\alpha$ with the outflow direction $\hat{\b{z}}$.
As we explained in Sect.~\ref{sec:rec_electric_field}, the reconnection electric field is in part set by the velocity at which the field lines 
are extracted from the X-point, i.e., by the outflow velocity. This outflow velocity is mostly set by the projection of the Alfv\'en 
speed onto the outflow direction (Eq.~\ref{equ:NR_outflow_speed} in the non-relativistic limit, Eq.~\ref{equ:outflow_energy}
in the general case). Consequently, it seems justified to normalize the electric field with the projection of the Alfv\'en speed 
onto the $\hat{\b{z}}$ direction, i.e., to use
$E^* = E_y/(B_0V_\mathrm{A,in}^\mathrm{R}\cos\theta)$.

Figure~\ref{fig:reconnection_rates_guide_field} shows the time evolution of $E^*$ for three simulations with $B_\mathrm{G}/B_0=0$, 0.5, 1, 
with otherwise identical parameters. 
The peak reconnection rate decreases when the guide field increases: 0.24, 0.22 and 0.20.
This is also the kind of dependence found in PIC simulations of ion-electron non-relativistic reconnection 
\citep{Ricci2003,Huba2005,TenBarge2013,Liu2014}, relativistic pair reconnection \citep{Hesse2007},
or two-fluid simulations of relativistic reconnection \citep{Zenitani2009b}\footnote{We 
note, however, that in asymmetric reconnection the rate increases with the guide field strength, see \citet{Aunai2013}, 
\citet{Hesse2013}.}.
Consequently, relativistic reconnection in ion-electron plasmas does not bring new effects
in this respect.
However, we underline that the normalization used here allows 
to minimize the scatter in $E^*$ for the various simulations.

%%%%%%%%%%%%%%%%%%%%%%%%%%%%%%%%%%%%%%%%%%%%%%%%%%%%%%%%%%%%%%%%%%%%%%%%%%%%%%%%%%%%%%%%%%%%%%%%%%%%%%%%%%%%%%%%%%%%%%%%%%%%%%%%%%%%%%%%%%%%%%%%
%%%%%%%%%%%%%%%%%%%%%%%%%%%%%%%%%%%%%%%%%%%%%%%%%%%%%%%%%%%%%%%%%%%%%%%%%%%%%%%%%%%%%%%%%%%%%%%%%%%%%%%%%%%%%%%%%%%%%%%%%%%%%%%%%%%%%%%%%%%%%%%%
\section{Summary and discussion}\label{sec:ccl}
%%%%%%%%%%%%%%%%%%%%%%%%%%%%%%%%%%%%%%%%%%%%%%%%%%%%%%%%%%%%%%%%%%%%%%%%%%%%%%%%%%%%%%%%%%%%%%%%%%%%%%%%%%%%%%%%%%%%%%%%%%%%%%%%%%%%%%%%%%%%%%%%
%%%%%%%%%%%%%%%%%%%%%%%%%%%%%%%%%%%%%%%%%%%%%%%%%%%%%%%%%%%%%%%%%%%%%%%%%%%%%%%%%%%%%%%%%%%%%%%%%%%%%%%%%%%%%%%%%%%%%%%%%%%%%%%%%%%%%%%%%%%%%%%%

%%%%%%%%%%%%%%%%%%%%%%%%%%%%%%%
%%%%%%%%%%%%%%%%%%%%%%%%%%%%%%%
\subsection{Summary}
%%%%%%%%%%%%%%%%%%%%%%%%%%%%%%%
%%%%%%%%%%%%%%%%%%%%%%%%%%%%%%%

We investigate magnetic reconnection in low $\beta$ ion-electron plasmas (mass ratio of 25) with 2D PIC simulations,
under relativistic conditions, i.e., the magnetic energy of the inflowing plasma exceeds its rest-mass energy.  
The simulations start from a Harris kinetic equilibrium with no localized perturbation.
% , that we explicit in Sect.~\ref{sec:pb_setup}, Appendix~\ref{app:relat_Harris_again},
% and in \citet{Melzani2013}. It implies relativistic temperatures and drift velocities in the current sheet,
% and particles are loaded with the method of \citet{Melzani2013}. 
For diagnostics and analytical models, we use momentum and energy fluid equations based on lab-frame quantities
(Appendix~\ref{sec:app_measure_relat_1}).
% that we derive in Appendix~\ref{sec:app_measure_relat_1} from the relativistic Vlasov equation (Eqs.~\ref{equ:conservation_part_number_lab}, \ref{equ:fluid_2}, \ref{equ:energy_relat_lab_all_species}). 
They have the advantage of being valid whatever the particle distribution function, 
while the usual relativistic fluid equations using comobile quantities are restricted to isotropic 
comobile distribution functions.

% Sect.~\ref{sec:inflow} 
For antiparallel reconnection, the structure of the diffusion region
has a width (in the inflow direction) $\delta_s$ given by 
the respective inertial length $d_s$ of the species $s$, measured at the center of the diffusion region.
A large inflow temperature increases this width.
  \note{For all cases with low inflow plasma $\beta$ ($\leq2.5\times10^{-3}$) we find:
  $0.5\leq\delta_\ion/d_\ion\leq1$ and $1\leq\delta_\lec/d_\lec\leq1.5$.
  For higher inflow temperatures, and thus larger $\beta$, the width is increased by the large thermal 
  velocity of the incoming particles.
  This is similar to non-relativistic studies.}
At the entrance of the diffusion regions for simulations at low 
background $\beta\leq 2.5\times10^{-3}$ we find sharp transitions in the fluid quantities that were not reported elsewhere.
We argue that they are not shocks. Instead, they occur when the inflowing particles have a thermal velocity 
far smaller than their bulk $E\times B$ velocity,
so that they enter the diffusion region with the same velocity and bounce back at the same location.
We stress that these sharp transitions are not a feature of relativistic reconnection,
as they depend only on the cold nature of the inflow. 
However, the phenomenon should be more common in relativistic reconnection 
because then the inflow bulk velocity $v_{E\times B}\sim E/B$ is large.

We explicit the balance of terms in the relativistic Ohm's law for antiparallel reconnection. % in Sect.~\ref{Sec:Ohms_law},
The ion diffusion region is dominated by bulk inertia (as defined in Eq.~\ref{equ:Ohm_fluid_1}).
In the electron diffusion region, bulk inertia contributes 
equally or more than thermal inertia. 
This latter result challenges the thermal-inertia-dominated paradigm that holds for 
non-relativistic or mildly relativistic antiparallel reconnection.
We show analytically %in Sect.~\ref{sec:outflow_analytical_estimate_harder} 
that a significant contribution of bulk inertia is to be expected whenever the inflow magnetization 
$\sigma_\lec^\mathrm{cold}$ (cold meaning that the temperature is not taken into account, see Eq.~\ref{equ:sigma_s_cold})
of the \textit{electrons} is large, because then bulk inertia 
$\partial_z\bar{p}_z\bar{v}_y \sim \bar{p}_z c/D \propto \sigma_\lec^\mathrm{cold} / D$
can exceed thermal inertia $\partial_x\delta{p}_x\delta{v}_y \propto (\sigma_\lec^\mathrm{cold})^{1/2}/\delta$. 
This is a new result that should hold for any large electron magnetization.

For the reconnection outflows we show analytically from mass and energy conservation that reconnection 
is expected to produce relativistic outflow temperatures and/or relativistic outflow bulk velocities. 
From simulations we find that outflow thermal energy dominates over bulk kinetic energy 
(from 70\% to 99\%, for simulations with increasing background magnetization).
A more refined analytical analysis shows that 
this is expected if the reconnection electric field is sustained by thermal inertia.
If bulk inertia dominates over thermal inertia, as expected at very large inflow magnetization, 
then our simple analytical model does not allow to conclude on the cold or hot nature of the outflows.
Also, our simulations do not probe high enough electron magnetizations to reach this regime:
at $\sigma_\lec^\mathrm{cold} = 90$, thermal inertia contributes as 50\% of the reconnection electric field,
and this fraction goes down to 25-40\% at $\sigma_\lec^\mathrm{cold} = 360$, 
which is significant enough for the hypothesis of $E_\mathrm{rec}$ sustained by thermal inertia to hold.

% In Sect.~\ref{sec:islands_no_guide_field} 
For the islands we show that, with or without a guide field, their centers consist mainly 
of particles initially in the current sheet that were gathered inside the island during the tearing instability, 
that do not mix with the background plasma even after many island merging events.
Particles of the background plasma cannot reach the inner parts because of the strong magnetic field surrounding the islands, 
and thus circle around the central part. As a result, the central part is less dense than its immediate surrounding.
% (Fig.~\ref{fig_wcewpe=3_NT=6000_cutalongZ_summary}).
This may explain observed density dips at the center of magnetic islands during magnetotail reconnection events
\citep{Khotyaintsev2010}, without invoking island merging or particle escape along the flux tube\note{As \citet{Markidis2013} does.}.
Islands are also the hottest parts of the flow, with fully anisotropic temperatures
in the antiparallel case, and distributions close to gyrotropic with a guide field.% (Fig.~\ref{fig_wcewpe=3_NT=6000_cutalongZ_summary}).

% Section~\ref{sec:rec_electric_field} studies the reconnection electric field. 
We argue that the reconnection rates are to be normalized by the asymptotic 
magnetic field and relativistic Alfv\'en speed in the inflow,
projected onto the outflow direction if there is a guide field: 
$E^*=E_y/(B_0 V_\mathrm{A,in}^\mathrm{R}\cos\theta)$.
This leads to rates in a narrow range: $E^*$ peaks between 0.14--0.25. % (Fig.~\ref{fig:reconnection_rates}).
However, we stress that there is no universal value for $E^*$ as defined here or elsewhere. 
First, because other studies show that it depends on the inflow plasma $\beta$ (increasing with decreasing $\beta$). 
Here we find no dependence on the background plasma temperature, 
but smaller rates for larger particle number densities\note{($E^*=0.18$ instead of $E^*=0.23$ for a three times less dense background)}.
Second, we find larger rates for the relativistic setups (0.18--0.25) than for the mildly relativistic case (0.15). 
These rates are also larger than those reported in the literature for ion-electron non-relativistic reconnection 
\citep[0.07--0.15 for][]{Birn2001,Pritchett2001,Fujimoto2006,Fujimoto2009,Daughton2006,Klimas2010}.
This points toward relativistic reconnection being slightly faster than non-relativistic reconnection. 
This trend is reinforced by simulations in relativistic pair plasmas \citep[$E^*=0.3$ in our case or, e.g., 0.17--0.36 for][]{Zenitani2007,Bessho2012,Cerutti2012b}.
We note that this is against the interpretation of \citet{Hesse2007} of a smaller rate for more relativistic setups.
Third, we confirm that a guide field leads to a smaller normalized rate.

We explore the consequences of adding a guide magnetic field.
The flow structure is strongly disturbed for two reasons: the Lorentz force associated with the guide field, and  
the relation $E<B$ everywhere.
The acceleration region is now defined by the condition $\b{E}\cdot\b{B}\neq 0$.
Inflowing ion and electron Larmor radii are smaller than the island scales or magnetic gradient scales,
and remain so even after the acceleration phase by $E_\mathrm{rec}$ 
because this phase conserves the perpendicular-to-$\b{B}$ momentum.
Particles thus remain tied to the field lines everywhere,
including in the acceleration region
where they spend more time before being deviated in the outflows.

%%%%%%%%%%%%%%%%%%%%%%%%%%%%%%%
%%%%%%%%%%%%%%%%%%%%%%%%%%%%%%%
\subsection{Discussion and astrophysical outlook}
%%%%%%%%%%%%%%%%%%%%%%%%%%%%%%%
%%%%%%%%%%%%%%%%%%%%%%%%%%%%%%%
\label{sec:astro_outlook}
This study may serve as micro-physics input for analyses on larger scales concerning magnetic energy conversion 
in relativistic ion-electron plasmas, as should be encountered in the coronae of AGN and microquasar accretion flows, 
in the lobe of radio galaxies, or inside relativistic jets from AGNs or GRBs.
We now discuss such applications, and give estimates for key parameters in these objects:
magnetic field $B$, 
electron number density $n_\lec$, 
magnetizations $\sigma_s^\mathrm{cold}$ (where cold means that only the rest mass energy is taken into account,
Eq.~\ref{equ:sigma_s_cold}), with $s=\ion,\,\lec$ for ions or electrons, 
and Alfv\'en speeds $V_\mathrm{A}^\mathrm{R}$.
The properties of magnetic reconnection as studied here depend only on the inflow magnetization
and temperatures, regardless of the real values of $B$ and $n_\lec$. 
This is true at least as long as 
effects such as pair creation and annihilation, radiative braking, or Compton drag on the electrons, can be neglected
% \citep[see][for a discussion on these effects]{Melzani2014b}.
(see Melzani et al., in prep, for a discussion on these effects).

% %%%%%%%%%%%%%%%%%%%%%%%%%%%%%%%
% \subsubsection{Main differences with non-relativistic reconnection}
% In non-relativistic setups, the outflow energy fluxes are mainly in the bulk kinetic energy, 
% while here they are dominated by thermal energy. 
% We find in all cases reconnection rates larger than those found in non-relativistic studies 
% ($E^*\sim 0.07$--0.15 for a range of references, $0.14$--$0.25$ here).
% Finally, relativistic magnetic reconnection produces more easily high-energy particles, with possibly power-law tails.

%%%%%%%%%%%%%%%%%%%%%%%%%%%%%%%
\subsubsection{Toward a new regime: non-dissipative reconnection?}
Our finding of a reconnection electric field sustained equally or more by bulk inertia than by thermal inertia for 
large inflow electron magnetization ($\sigma_\lec^\mathrm{cold} \geq 90$),
and the extrapolation of Sect.~\ref{sec:outflow_analytical_estimate_harder}, 
indicate that bulk inertia might end up dominating at even larger inflow electron magnetizations.
This was also envisioned by \citet{Hesse2007}, and 
reconnection in such a regime would bear significant differences with the standard picture. 
However, we nuance the assertion of \citet{Hesse2007} that reconnection would then be a reversible process:
as underlined in Sect.~\ref{sec:outflow_analytical_estimate_harder}, the reconnection outflows may be 
thermally dominated even when bulk inertia dominates Ohm's law.
A definite answer to these questions requires very high magnetizations, 
that we can hardly afford with a PIC code, and that may require relativistic gyrokinetic codes.

Highly magnetized environments, such as magnetar magnetospheres 
\citep[with magnetizations exceeding $\sigma_\lec^\mathrm{cold}=10^{13}$,][]{Lyutikov2013},
pulsar winds near the termination shock \citep[$\sigma_\lec^\mathrm{cold}=10^{13}$,][]{Bucciantini2011,Sironi2011b},
other objects with $\sigma_\lec^\mathrm{cold}\ggg 1$ discussed in what follows,
are likely to support this reconnection regime.

%%%%%%%%%%%%%%%%%%%%%%%%%%%%%%%
\subsubsection{Large scale transient outflow production, the example of microquasars}
We have shown that the reconnection outflows are thermally dominated, with a bulk Lorentz factor not necessarily increasing with 
the inflow magnetization and featuring low values (1.63 at most, Table~\ref{tab:energy_outflow_flux}).
However, applications to large scale outflows triggered by reconnection events require some care.
The outflows studied in the present manuscript originate from the electron diffusion region, and feature 
ion/electron decoupling.
On larger distances, if not bounded by the islands and by our periodic setup, electrons and ions are expected to 
couple and to follow the ideal MHD dynamic.
The scale on which they can propagate is fixed by larger scales than simulated here.

On another hand, it is expected and observed \citep{Khotyaintsev2005} that 
magnetic energy conversion takes place also along the magnetic 
separatrices far away from the dissipation region, 
on length scales of hundreds of ion inertial lengths. 
This conversion occurs through instabilities that produce thermal and non-thermal electrons 
\citep{Drake2005,Egedal2009,Egedal2012}, 
and through the complex structure of collisionless non-linear waves
(slow shock, compound wave, rotational wave)
by which the magnetized inflowing plasma transits to the hot and unmagnetized outflow 
on MHD scales \citep{Liu2012,Higashimori2012}.
It is this large scale outflow that should be identified to the transient reconnection-driven jets 
in microquasar models \citep{deGouveia2005,deGouveia2010,Kowal2011,McKinney2012,Dexter2013}.
In the magnetosphere close to the black hole, \citet{deGouveia2005} estimates on the basis of an analytical model,
$n_\lec \sim 5\times10^{15}\,\mathrm{cm^{-3}}$, $B \sim 7\times10^7$\,G, which gives 
electron and ion magnetizations $\sigma_\lec^\mathrm{cold}\sim 10^5$ and $\sigma_\ion^\mathrm{cold} \sim 60$, 
and an Alfv\'en speed $V_\mathrm{A}^\mathrm{R} \sim c$.
This is clearly in the relativistic case.
The energy content of the large scale outflows in this case has not been studied, 
but we can expect from the 
collisionless slow shocks, or rotational discontinuities 
at the separatrices, to produce a thermally dominated outflow. 
It may not be so for other jet production mechanisms, and could 
help discriminating in favor or against reconnection scenarios.

Another unknown is what becomes of the ambient plasma that is expelled by the first reconnected field lines,
ahead of the dipolarization front. In our study, it would correspond to half of a magnetic island, 
ejected out of the simulation box. The ambient plasma would be the plasma from the current sheet trapped in 
the island. As we demonstrate, this plasma does not mix with the reconnected plasma and is simply compressed 
and heated \citep[see][for a 3D case where instabilities imply magnetic to kinetic energy conversion]{Vapirev2013}. 
In an open configuration, it would be at the head of the large scale outflow.

%%%%%%%%%%%%%%%%%%%%%%%%%%%%%%%
\subsubsection{Plasma heating in AGN and microquasar coronae and in galaxy radio lobes}
Photon emission in the hard state of microquasars and AGNs is believed to come from inverse-Compton scattering of seed 
photons by the electrons of a corona. To achieve this, these electrons must reach temperatures of the order of 
$10^9$\,K, i.e., $\Theta_\lec = T_\lec/m_\lec c^2 \sim 0.2$.
A non-thermal population of electrons is also required by the observation of MeV photons \citep{Poutanen2014}.
A proposed mechanism for plasma heating is by magnetic reconnection
\citep{Matteo1998,Merloni2001,Reis2013}.
The plasma Alfv\'en speed estimated by these authors lies in the range $0.03c$--$0.3c$. 
Associated electron magnetizations are $\sigma_\lec^\mathrm{cold} \sim 1.7$-180, which is in the range of the present study.
A crucial question is the energy distribution between ions and electrons:
if most of the magnetic energy goes to ions, and because of the low collisionality of these dilute environments, 
a large temperature difference can be sustained \citep{Matteo1997}.
Our study shows that ions are slightly more heated than electrons: 
this can be seen with the temperatures of Figs.~\ref{fig:wcewpe=3_cut_Z_NT6000_temperature_only}, \ref{fig_wcewpe=3_NT=6000_cutalongZ_summary},
\ref{fig:xcf_wcewpe=3_NT=6000_2D_pseudocolor_temperature_NT=6000}, and \ref{fig:guide_field_overall_velocities}.
More generally, the kinetic energy of the particles trapped in the magnetic islands 
is distributed as 55\% for ions and 45\% for electrons, 
and the kinetic energy of the particles from the background plasma that are accelerated 
when reaching the current sheet is also distributed as 60\% for ions and 40\% for electrons 
% \citep[for details see][]{Melzani2014b}. 
(for details see Melzani et al., in prep.).
The energy distribution by acceleration processes far downstream of the diffusion region requires another study.

Similar questions arise concerning the heating of the lobes of radio galaxies \citep{Kronberg2004}.
There, $n\sim 3\times10^{-6}\,\mathrm{cm^{-3}}$ for the number densities,
and $B \sim 5\,\mathrm{\mu G}$ for the equipartition magnetic field with values that can be locally ten times higher,
which gives magnetizations $\sigma_\lec^\mathrm{cold} \sim 0.8$-80
and Alfv\'en speeds $\sim 0.02c$-$0.2c$.
Our conclusion for the energy repartition between ions and electrons also holds.

% %%%%%%%%%%%%%%%%%%%%%%%%%%%%%%%
% \subsubsection{Magnetic energy dissipation in GRBs}
% A possible model for the production and properties of gamma-ray burst outflows is that magnetic energy in the jet is converted 
% into kinetic energy via reconnection \citep{Drenkhahn2002,McKinney2012b}. 
% \citet{McKinney2012b} estimates that the dissipation occurs suddendly at a typical radius of $10^{14}$\,cm, where reconnection 
% becomes collisionless and fast. 

%%%%%%%%%%%%%%%%%%%%%%%%%%%%%%%
\subsubsection{Flares and ``mini''-jets in extragalactic jets and in GRBs}
Flare-like activity in the GeV-TeV range is observed from extragalactic jets, and may possibly be explained by 
local reconnection events inside the jet, that produce smaller jets (the reconnection exhausts) 
which in turn radiate the expected photons \citep{Giannios2009}. 
This $\gamma$-ray emission region may be located close to the black-hole \citep[$<0.05$\,pc,][]{Giroletti2004},
where $B\sim 0.02$-$0.2$\,G and the plasma magnetization is high.
For example, \citet{Giannios2009} take $\sigma^\mathrm{cold}_\ion = 100$, which leads to $\sigma^\mathrm{cold}_\lec = 2\times10^5$
and $V_\mathrm{A}\sim c$. 
This is in the regime where bulk inertia should dominate in Ohm's law.
Also, \citet{Giannios2009} estimate from energy considerations, that the blobs emitted from the reconnection exhausts 
should be $\sim10^{14}$\,cm, i.e., based on its estimated particle density, $10^{10}$ ion inertial 
lengths.\note{$d_i=1000$cm with $n=80cm^{-3}$} 
Here again, the physics far from the dissipation region should play an important role in producing such large scale structures.

% Similar models are invoked for gamma-ray bursts \citep{Lyutikov2006c,Lazar2009}, and reconnection then occurs in a pair plasma.
% From a MHD jet model, \citet{McKinney2012b} estimate 
% $B=10^8$\,G, 
% $n = 2\times10^6\,\mathrm{cm^{-3}}$, and magnetizations of the order of 10 only.

%%%%%%%%%%%%%%%%%%%%%%%%%%%%%%%
\subsubsection{Radio emission from extragalactic jets}
Another case for relativistic magnetic reconnection is inside 
jets from AGNs, on scales of 10-100\,kpc. Radio spectra may be explained by radiation linked to reconnection events
\citep{Romanova1992}.\note{Romanova uses 
$B\sim 300\mu G$, $n\sim 10^{-3}cm^{-3}$ -- $\sigma_e = 10$, but with no ref. to observations.}
Observations of AGN jets indicate 
$B\sim 10$-$30\mathrm{\mu G}$, $n\sim 0.8$-$5\times 10^{-8}\mathrm{cm^{-3}}$, and
electron magnetizations in the range $\sigma_\lec^\mathrm{cold} \sim 500$-2500
\citep{Schwartz2006}, which implies ion magnetizations $\sigma_\ion^\mathrm{cold} \sim 0.3$-1.3 and Alfv\'en speeds 
$\sim0.5$-$0.8c$.
Again, our results apply in these cases, and in particular 
the electron magnetizations are in the very relativistic range where bulk inertia should dominate in Ohm's law.

%%%%%%%%%%%%%%%%%%%%%%%%%%%%%%%
\subsubsection{High-energy particle production}
The proposed normalization of the reconnection rate, $E^* = E_\mathrm{rec} / (B_0 V_\mathrm{A,in}^\mathrm{R}\cos\theta)$ with 
$V_\mathrm{A,in}^\mathrm{R}$ the relativistic inflow Alfv\'en speed, leads to $E^*$ in a close range (0.14-0.25)
and, because it relies only on inflow quantities, allows for an easy prediction of the reconnection electric field.
In particular, the ratio $E_\mathrm{rec}/B_0$ is a key quantity to estimate 
the time scale of energy dissipation or the maximal energy gain for particles.
It is interesting to notice that for very relativistic plasmas, $V_\mathrm{A,in}^\mathrm{R}$ saturates at $c$,
so that $E_\mathrm{rec}/B_0$ saturates at $\sim 0.2c$.
It may imply that the hardness of the high-energy tails saturates.
We explore these matters in a forthcoming paper (Melzani et al., in prep.).
Briefly, we find for a given species a power-law tail 
whenever its background magnetization is relativistic (above a few),
with an index depending mainly on the inflow magnetization.

%%%%%%%%%%%%%%%%%%%%%%%%%%%%%%%
\subsubsection{Other complications}
% Initial conditions, forcing, asymmetries, guide field, turbulence, and 3D
We finally point out that the present study is oversimplified in many respects.
Magnetic reconnection in magnetized coronae and jets likely often implies asymmetric plasmas from each side of 
the current sheet, guide fields \citep{Aunai2013,Hesse2013,Eastwood2013}, 
and also normal fields (along $\hat{\b{x}}$ here) reminiscent from the 
ambient magnetic field. The last point has been studied in the context of the Earth magnetotail 
\citep{Pritchett2005b,Pritchett2010,Sitnov2011}.
Also, the initial conditions chosen in the simulations are arbitrary and do not necessarily reflect the 
real environments. 
Explored alternatives to the Harris sheet include X-point collapse \citep[e.g.,][]{Pahlen2013} 
or force-free equilibrium \citep[e.g.,][]{Liu2014}, and show little differences with the Harris case.
However, three dimensional initial configurations should also be considered, because in a real environment most of the energy dissipation 
may occur at 3D nulls, involving for example spine-fan reconnection, or 
at quasi-separatrix layers \citep{Birn2007,Pontin2011}. Few kinetic simulations of such setups exist \citep{Baumann2012,Olshevsky2013}.
A related point is the external forcing, i.e., the large scale plasma flow that can increase the 
magnetic field gradients and trigger reconnection. Studies 
\citep{Pei2001,Pritchett2005,Ohtani2009,Klimas2010} show that the reconnection rate $E^*$ is 
then fixed by the boundary conditions, and is thus larger than the spontaneous rate. 
The timescale of the forcing also proves to be of importance \citep{Pei2001}.
These considerations, as well as some of the points made earlier on, 
highlight the multiscale nature of reconnection in the context of concrete astrophysical objects 
-- and demonstrate the need for corresponding multiscale simulation studies, a field still in its infancy
\citep[e.g., ][]{Horiuchi2010,Innocenti2013}

Another central question is the validity of the 2D findings in three dimensions. 
Magnetic islands then become extended filaments, modulated or broken by instabilities in the third dimension or 
by a lack of coherence of the tearing instability \citep{Daughton2011,Kagan2012,Markidis2013}.
It may imply more mixing of the current sheet particles with those of the background plasma.
Concerning the validity of our claims on Ohm's law,
3D PIC simulations in non-relativistic plasmas have shown that anomalous resistivity due to microinstabilities 
remains a negligible dissipation mechanism in the diffusion region \citep{Liu2013b,Karimabadi2013}, 
where the reconnection electric field is still sustained by thermal electron inertia. 
The scaling analysis of Sect.~\ref{sec:outflow_analytical_estimate_harder} should thus remain valid,
as well as the conclusion that bulk inertia dominates at high inflow magnetization.

%%%%%%%%%%%%%%%%%%%%%%%%%%%%%%%%%%%%%%%%%%%%%%%%%%%%%%%%%%%%%%%%%%%%%%%%
%%%%%%%%%%%%%%%%%%%%%%%%%%%%%%%%%%%%%%%%%%%%%%%%%%%%%%%%%%%%%%%%%%%%%%%%
\begin{acknowledgements}
%   M.~Melzani would like to thank H.~Baty, B.~Cerutti, and M.~E.~Innocenti for interesting discussions.
   The simulations were performed using HPC ressources from GENCI (Grand \'Equipement National de Calcul Intensif)
   at CINES, CCRT and IDRIS, under the allocation x2013046960.
   Tests were conducted at p\^ole scientifique de mod\'elisation num\'erique,
   PSMN, at the \'Ecole Normale Sup\'erieure de Lyon, whose staff we thank for their steady technical support.
   This work has been financially supported by the Programme National Hautes Energies (PNHE).
%
% This work was granted access to the HPC ressources of CCRT and CINES (TGCC, IDRIS)
% under the allocation ... made by GENCI (Grand \'Equipement National de Calcul Intensif).
%    or
% This work was performed using HPC ressources from GENCI-CCRT/IDRIS/CINES/TGCC (grant ...).
%
\end{acknowledgements}
%
%
%
%
%
%
%%%%%%%%%%%%%%%%%%%%%%%%%%%%%%%%%%%%%%%%%%%%%%%%%%%%%%%%%%%%%%%%%%%%%%%%%%%%%%%%%%%%%%%%%%%%%%%%%%%%%%%%%%%%%%%%%%%%%%%%%%%%%%%%%%%%%%%%%%%%%%%%
%%%%%%%%%%%%%%%%%%%%%%%%%%%%%  Bibliography  %%%%%%%%%%%%%%%%%%%%%%%%%%%
%%%%%%%%%%%%%%%%%%%%%%%%%%%%%%%%%%%%%%%%%%%%%%%%%%%%%%%%%%%%%%%%%%%%%%%%%%%%%%%%%%%%%%%%%%%%%%%%%%%%%%%%%%%%%%%%%%%%%%%%%%%%%%%%%%%%%%%%%%%%%%%%
%
\bibliographystyle{apj} 
\bibliography{biblio_particles.bib}
\addcontentsline{toc}{section}{References}
\appendix

%%%%%%%%%%%%%%%%%%%%%%%%%%%%%%%%%%%%%%%%%%%%%%%%%%%%%%%%%%%%%%%%%%%%%%%%%%%%%%%%%%%%%%%%%%%%%%%%%%%%%%%%%%%%%%%%%%%%%%%%%%%%%%%%%%%%%%%%%%%%%%%%
%%%%%%%%%%%%%%%%%%%%%%%%%%%%%%%%%%%%%%%%%%%%%%%%%%%%%%%%%%%%%%%%%%%%%%%%%%%%%%%%%%%%%%%%%%%%%%%%%%%%%%%%%%%%%%%%%%%%%%%%%%%%%%%%%%%%%%%%%%%%%%%%
\section{Relativistic Harris equilibrium}
%%%%%%%%%%%%%%%%%%%%%%%%%%%%%%%%%%%%%%%%%%%%%%%%%%%%%%%%%%%%%%%%%%%%%%%%%%%%%%%%%%%%%%%%%%%%%%%%%%%%%%%%%%%%%%%%%%%%%%%%%%%%%%%%%%%%%%%%%%%%%%%%
%%%%%%%%%%%%%%%%%%%%%%%%%%%%%%%%%%%%%%%%%%%%%%%%%%%%%%%%%%%%%%%%%%%%%%%%%%%%%%%%%%%%%%%%%%%%%%%%%%%%%%%%%%%%%%%%%%%%%%%%%%%%%%%%%%%%%%%%%%%%%%%%
\label{app:relat_Harris_again}

We derived the equilibrium relations for relativistic temperatures and current drift speeds, as well as for 
arbitrary ion to electron mass ratio and temperature ratio, in \citet{Melzani2013}.

Each species follow a Maxwell-J\"uttner distribution :
\begin{equation}\label{equ:Jutt_Harris}
 f_s(\b{x},\widetilde{\b{p}}) = \frac{\mu_s\,n_\mathrm{cs}(x)}{4\pi \Gamma_sK_2(\mu_s)} \exp\left\{ -\mu_s\Gamma_s\left(\sqrt{1+\tilde{p}^2}  - U_s \tilde{p}_y/c\right) \right\},
\end{equation}
with $s=\ion$ for ions or $\lec$ for electrons, $\mu_s=1/\Theta_s=m_sc^2/T_s$, $\widetilde{\b{p}}=\b{p}/c=\gamma\b{v}/c$, 
and $K_2$ the modified Bessel function of the second kind.
We note that $f_s$ is indeed normalized with respect to $\widetilde{\b{p}}$ to $n_\mathrm{cs}(x)$ \citep{Melzani2013},
so that the particle density in the simulation frame is $n_\mathrm{cs}(x)$, and that in the comobile plasma frame 
of each species is $n_\mathrm{0,cs}(x) = n_\mathrm{cs}(x)/\Gamma_s$.
Loading the distribution~\ref{equ:Jutt_Harris} when both $\Theta_s$ and $U_s$ are relativistic is non-trivial, and 
we use the method detailed in \citet{Melzani2013}.

For the special case where ions and electrons have the same temperatures, fulfilling Vlasov and Maxwell's equations leads to:
\begin{subequations}
\label{equ:haris_equal_temperatures}
\begin{align}
 n_\mathrm{cs}(x)      &= \frac{n_\mathrm{cs}(0)}{\mathrm{cosh}^2 (x/L)}, \label{equ:haris_equal_temperatures_1} \\
 \Theta_\ion           &= (m_\lec/m_\ion)\Theta_\lec, \label{equ:haris_equal_temperatures_2}\\
 \Theta_\lec           &= \frac{1}{4} \left( \frac{\omega_\mathrm{ce}}{\omega_{0,\pe}} \right)^2, \label{equ:haris_equal_temperatures_3} \\
 \frac{\Gamma_sU_s}{c} &= -2 \Theta_s \frac{d_{0,\lec}}{L} \frac{\omega_{0,\pe}}{\omega_{\mathrm{c}s}}\mathrm{sgn}(q_s) \label{equ:haris_equal_temperatures_4},
\end{align}
\end{subequations}
with $\mathrm{sgn}(q_s)$ the sign of the charge $q_s$, $\omega_{\mathrm{c}s} = eB_0/m_s$ the cyclotron pulsation 
defined in the asymptotic magnetic field $B_0$ ($e>0$ here), $\omega_{0,\pe}=\sqrt{n_{0,\cs}(0)e^2/(\epsilon_0m_\lec)}$ 
the electron plasma pulsation defined 
by the comobile number density $n_{0,\cs}(0)=n_\mathrm{cs}(0)/\Gamma_s$ at the center of the current sheet, 
and $d_{0,\lec} = c/\omega_{0,\pe}$ the associated inertial length.

Inserting Eq.~\ref{equ:haris_equal_temperatures_3} into Eq.~\ref{equ:haris_equal_temperatures_4}, the latter becomes
\begin{equation}
\label{equ:gamma_U_Harris}
\frac{\Gamma_sU_s}{c} = -\frac{1}{2} \frac{d_{0,\lec}}{L} \frac{\wce}{\omega_{0,\pe}}\mathrm{sgn}(q_s),
\end{equation}
so that we see that
Eqs.~\ref{equ:haris_equal_temperatures_1}-\ref{equ:haris_equal_temperatures_3} and Eq.~\ref{equ:gamma_U_Harris} are well suited to express 
the equilibrium relations in term of the comobile quantities $\wce/\omega_{0,\pe}$ and $L/d_{0,\lec}$ only.
Some manipulations are needed to express everything in terms of the lab-frame quantities $\wce/\wpe$ and $L/d_\lec$,
where $\wpe$ is the counterpart of $\omega_{0,\pe}$ in the simulation frame ($\wpe = \sqrt{\Gamma_\lec} \omega_{0,\pe}$),
and $d_\lec=c/\wpe$. 
To do so, we note in Eq.~\ref{equ:haris_equal_temperatures_4} that $\omega_{0,\pe}d_{0,\lec} = c = \wpe d_\lec$, 
and we express Eq.~\ref{equ:haris_equal_temperatures_3} as $\Theta_\lec = \Gamma_e (\wce/\wpe)^2/4$.
With this, we obtain 
\begin{subequations}
\begin{align}
 \frac{U_s}{c} &= \frac{1}{2} \frac{\wce}{\wpe} \frac{d_\lec}{L},   \label{equ:harris_equil_lab_frame_1} \\
 \Theta_\lec &= \frac{\Gamma_e}{4} \left(\frac{\wce}{\wpe}\right)^2 \label{equ:harris_equil_lab_frame_2} .
\end{align}
\end{subequations}
We see from Eq.~\ref{equ:harris_equil_lab_frame_1} that we have the condition $\wce/\wpe < 2L/d_\lec$.
If this is not the case, the equilibrium cannot be achieved. Why it is so can be seen by rewriting 
Eq.~\ref{equ:haris_equal_temperatures_4} as $\Theta_\lec = (L/2d_\lec)(\wce/\wpe)\Gamma_eU_e/c$:
satisfying Vlasov equation is possible only if  
$\Theta_\lec < (L/2d_\lec)(\wce/\wpe)\Gamma_e$, but this is not possible if the pressure balance condition 
(which is Eq.~\ref{equ:harris_equil_lab_frame_2}) requires a temperature exceeding this limit to balance the magnetic field pressure.

\begin{figure}[tbp]
 \centering
 \def\svgwidth{0.6\columnwidth}
 \begin{tiny}
 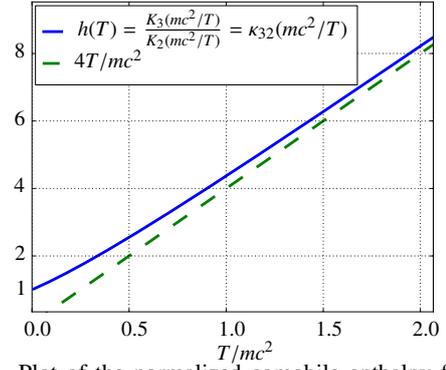
 \end{tiny}
 \caption{\label{fig_kappa_32}Plot of the normalized comobile enthalpy for species $s$: 
          $h_{0s} = (n_{0s}\langle\gamma m_s c^2\rangle_s+P_s)/(n_{0s}m_sc^2)$,
          for a Maxwell-J\"uttner distribution of temperature $T_s$.
          For the background plasma temperatures considered in this study, $T = (1.5\times10^7,\,2\times10^8,\,3\times10^9)\,\mathrm{K}$, 
          we have for electrons
          $\Theta_\lec = (2.5\times10^{-3},\,3.4\times10^{-2},\,0.51)$ and corresponding $h_{0\lec} = (1.006,\,1.086,\,2.57)$, 
          while for ions $h_{0\ion}\sim1$ always holds.}
\end{figure}

As a final note, we express the thermal Larmor radius of the particles,
defined as $\langle(\gamma v_\perp)^2\rangle^{1/2}/\wce$ where $\langle\cdot\rangle$ is an average over the 
distribution function, at current sheet center:
\begin{equation}\label{equ:rce_over_de}
 \frac{\langle r_\mathrm{ce}\rangle}{d_\lec} = \frac{\wpe}{\wce} \sqrt{\Theta_\lec} \sqrt{\kappa_{32}(\mu_\lec)} = \sqrt{\Gamma_\lec \kappa_{32}(\mu_\lec)/2},
\end{equation}
where the first part of the equality is general \citep{Melzani2013}, and the second is obtained for the Harris equilibrium using 
Eq.~\ref{equ:harris_equil_lab_frame_2} for the ratio $\wce/\wpe$. 
The function $\kappa_{32}$ is the plasma comobile enthalpy and is plotted in Fig.~\ref{fig_kappa_32}.
The thermal Larmor radius is consequently temperature dependent via $\kappa_{32}(1/\Theta_\lec)$.

%%%%%%%%%%%%%%%%%%%%%%%%%%%%%%%%%%%%%%%%%%%%%%%%%%%%%%%%%%%%%%%%%%%%%%%%%%%%%%%%%%%%%%%%%%%%%%%%%%%%%%%%%%%%%%%%%%%%%%%%%%%%%%%%%%%%%%%%%%%%%%%%
%%%%%%%%%%%%%%%%%%%%%%%%%%%%%%%%%%%%%%%%%%%%%%%%%%%%%%%%%%%%%%%%%%%%%%%%%%%%%%%%%%%%%%%%%%%%%%%%%%%%%%%%%%%%%%%%%%%%%%%%%%%%%%%%%%%%%%%%%%%%%%%%
\section{From Vlasov to fluid equations}\label{sec:app_measure_relat_1}
%%%%%%%%%%%%%%%%%%%%%%%%%%%%%%%%%%%%%%%%%%%%%%%%%%%%%%%%%%%%%%%%%%%%%%%%%%%%%%%%%%%%%%%%%%%%%%%%%%%%%%%%%%%%%%%%%%%%%%%%%%%%%%%%%%%%%%%%%%%%%%%%
%%%%%%%%%%%%%%%%%%%%%%%%%%%%%%%%%%%%%%%%%%%%%%%%%%%%%%%%%%%%%%%%%%%%%%%%%%%%%%%%%%%%%%%%%%%%%%%%%%%%%%%%%%%%%%%%%%%%%%%%%%%%%%%%%%%%%%%%%%%%%%%%

%%%%%%%%%%%%%%%%%%%%%%%%%%%%%%%
%%%%%%%%%%%%%%%%%%%%%%%%%%%%%%%
\subsection{Fluid equations}
%%%%%%%%%%%%%%%%%%%%%%%%%%%%%%%
%%%%%%%%%%%%%%%%%%%%%%%%%%%%%%%

Fluid equations employed in numerical codes are usually expressed in term of comobile quantities, 
such as, for species $s$, the comobile particle number density $n_{0s}$, the comobile enthalpy $h_{0s}$, the comobile pressure $P_{0s}$, and in term of 
the fluid velocity $\bar{\b{v}}_s$ and its associated Lorentz factor $\Gamma_s$. 
The conservation of particle number and of momentum for each species, and of total energy, then read \citep[see e.g.,][]{Mihalas1984,Barkov2013}:
\begin{subequations}\label{equ:fluid_3}
 \begin{align}
 &\frac{\partial}{\partial t} (\Gamma_sn_{0s}) + \frac{\partial}{\partial \b{x}}\cdot (\Gamma_sn_{0s} \bar{\b{v}}_s) = 0, \label{equ:fluid_3_a} \\
 &\begin{aligned}
   \frac{\partial}{\partial t} (\Gamma_s^2 n_{0s} h_{0s} & {\bar{\b{v}}}_s) + \frac{\partial}{\partial \b{x}}\cdot (n_{0s}h_{0s}\Gamma_s\bar{\b{v}}_s\Gamma_s\bar{\b{v}}_s) \\
         &= -\frac{1}{m_s}\frac{\partial P_{0s}}{\partial \b{x}} + \frac{q_s}{m_s}\Gamma_s n_{0s}(\b{E}+{\bar{\b{v}}}_s\wedge\b{B}),
 \end{aligned} \label{equ:fluid_3_b} \\
 &\begin{aligned}
   \frac{\partial}{\partial t} &\left\{\sum_s(\Gamma_s^2n_{0s}h_{0s}m_sc^2 - P_{0s}) + \frac{E^2}{2\mu_0c^2} + \frac{B^2}{2\mu_0}\right\} \\
     &+ \frac{\partial}{\partial \b{x}}\cdot \left\{\sum_s(\Gamma_s^2n_{0s}h_{0s}m_sc^2\bar{\b{v}}_s) + \frac{\b{E}\wedge\b{B}}{\mu_0} \right\} = 0.
 \end{aligned} \label{equ:fluid_3_c}
 \end{align}
\end{subequations}
These equations are, however, not well suited for the analysis of particle simulations. 
First, because accessible quantities are those defined in the simulation (or lab) frame, while those in the comobile frame 
of the plasma must be obtained by a boost at the local mean speed $\bar{\b{v}}_s$. 
Second, because they assume a comobile particle distribution that is isotropic in momentum space in order
to use a scalar pressure $P_{0s}$ instead of the full pressure tensor, and, as we show below, in order 
to write relations such as $\bar{\b{p}}_s = \langle\gamma\b{v}\rangle_s = h_{0s}\Gamma_s\bar{\b{v}}_s$ for the mean momentum.
This is not the case in the out-of-equilibrium plasmas that we study.

Instead, we derive the fluid equations directly from the collisionless Vlasov equation. The latter reads
\begin{equation}\label{equ:Vlasov}
  \frac{\partial f_s(\b{x},\b{p},t)}{\partial t} + \b{v}\cdot\frac{\partial f_s}{\partial {\b{x}}} + \frac{q_s}{m_s} \left(\b{E}+\b{v}\wedge\b{B}\right)\cdot\frac{\partial f_s}{\partial \b{p}} = 0,
\end{equation}
where $\b{p} = \gamma\b{v}$, and $f_s$ is the distribution function in the simulation or lab frame.
We will denote by $f_{0s}$ its counterpart in the comobile frame.

The first moment (with 1) of Eq.~\ref{equ:Vlasov} gives the equation of conservation of the number of particles:
\begin{equation}
\label{equ:conservation_part_number_lab}
 \frac{\partial}{\partial t} n_{\mathrm{lab},s} + \frac{\partial}{\partial \b{x}}\cdot (n_{\mathrm{lab},s} \bar{\b{v}}_s) = 0.
\end{equation}
Note that 
$n_{\mathrm{lab},s} = \Gamma_s n_{0s}$, 
so that we indeed recover Eq.~\ref{equ:fluid_3_a}.

The second moment (with $\b{p}$) gives the equation of conservation of momentum:
\begin{equation}\label{equ:fluid_2}
\begin{aligned}
 \frac{\partial}{\partial t} & (n_{\mathrm{lab},s}\bar{\b{p}}_s) + \frac{\partial}{\partial \b{x}}\cdot (n_{\mathrm{lab},s}\bar{\b{p}}_s\bar{\b{v}}_s) \\
  &= -\frac{\partial}{\partial \b{x}}\cdot (n_{\mathrm{lab},s}\langle \delta\b{p}_s\delta\b{v}_s\rangle_s) + \frac{q_s}{m_s}n_{\mathrm{lab},s}(\b{E}+\bar{\b{v}}_s\wedge\b{B}).
\end{aligned}
 \end{equation}
We used the definition $\delta\b{p} = \b{p}-\bar{\b{p}}_s$, where $\b{p}=\gamma \b{v}$ is the momentum, and similarly for $\delta\b{v}$.
Also, $\langle\cdot\rangle_s$ denotes an average in $\b{p}$ over the distribution function $f_s$.
In order to recover Eq.~\ref{equ:fluid_3_b}, 
we use the relation $\bar{\b{p}}_s = h_{0s}\Gamma_s\bar{\b{v}}_s$ for the mean momentum, 
which is valid only if the comobile distribution $f_{0s}$ is isotropic in $\b{p}$
\citep{Melzani2013}.
Also, we note that the stress tensor is defined as 
${\Pi}_{ij} = n_{\mathrm{lab},s} m_s \langle p_iv_j\rangle_s = n_{\mathrm{lab},s} m_s (\bar{p}_i\bar{v}_j + \langle \delta p_i \delta v_j\rangle_s)$,
and is also equal (again if $f_{0s}$ is isotropic) to 
$\Pi_{ij} = P_{0s}\delta_{ij} + \Gamma^2_{\bar{\b{v}}_s} h_{0s}n_{0s}m_s\bar{v}_{s,i}\bar{v}_{s,j}$.
Inserting these expressions into Eq.~\ref{equ:fluid_2} does lead to Eq.~\ref{equ:fluid_3_b}.
We also note that Eq.~\ref{equ:fluid_2} can be put into a conservative form by using the conservation of momentum for the 
electromagnetic field, which reads
\begin{equation}\label{equ:field_momentum}
 \frac{\partial}{\partial t}\epsilon_0(\b{E}\wedge\b{B}) - \frac{\partial}{\partial\b{x}}\cdot\b{T} = -[\rho\b{E} + \b{j}\wedge\b{B}], 
\end{equation}
where $T_{ij} = \epsilon_0(E_iE_j - E^2\delta_{ij}/2) + \mu_0^{-1}(B_iB_j - B^2\delta_{ij}/2)$ is Maxwell stress tensor,
$\rho = \sum_s q_sn_{\mathrm{lab},s}$ is the charge density, and $\b{j} = \sum_s q_sn_{\mathrm{lab},s} \bar{\b{v}}_s$ is the current 
density. One thus has to sum Eq.~\ref{equ:fluid_2} over all species and then use Eq.~\ref{equ:field_momentum}, to obtain:
\begin{equation}\label{equ:fluid_2_cons}
\begin{aligned}
 \frac{\partial}{\partial t} & \left(\epsilon_0\b{E}\wedge\b{B} + \sum_sm_sn_{\mathrm{lab},s}\bar{\b{p}}_s\right) \\
 &+ \frac{\partial}{\partial \b{x}}\cdot \left(-\b{T} + \sum_sm_sn_{\mathrm{lab},s}\left[\bar{\b{p}}_s\bar{\b{v}}_s +\langle \delta\b{p}_s\delta\b{v}_s\rangle_s\right]\right) = 0.
\end{aligned}
\end{equation}
\note{Expressed with comobile quantities as in Eq.~\ref{equ:fluid_3_b}, this same equation reads:
\begin{equation}\label{equ:fluid_2_cons_comobile}
\begin{aligned}
 \frac{\partial}{\partial t} & \left(\epsilon_0\b{E}\wedge\b{B} + \sum_sm_s\Gamma_s^2 n_{0s} h_{0s} {\bar{\b{v}}}_s \right) \\
 &+ \frac{\partial}{\partial \b{x}}\cdot \left(-\b{T} + \sum_sm_s n_{0s}h_{0s}\Gamma_s\bar{\b{v}}_s\Gamma_s\bar{\b{v}}_s + \sum_sP_{0s}\b{I} \right) = 0.
\end{aligned}
\end{equation}}

Finally, multiplying Vlasov equation~\ref{equ:Vlasov} by $\gamma m_s c^2$ and integrating over $\b{p}$ 
gives the equation of conservation of energy:
\begin{equation}
\label{equ:energy_relat_lab}
\begin{aligned}
  \frac{\partial}{\partial t} (n_{\mathrm{lab},s}\langle\gamma m_sc^2\rangle_s) + \frac{\partial}{\partial \b{x}}\cdot (n_{\mathrm{lab},s} \langle & \b{v}\gamma m_sc^2\rangle_s) \\
   & = q_s n_{\mathrm{lab},s}\langle \b{E}\cdot\b{v}\rangle_s.
\end{aligned}
\end{equation}
The right hand side accounts for the coupling between the species and the electromagnetic fields, and thus possibly with other species 
via collective interactions. 
The non-relativistic limit of this equation is easily obtained by making the difference between 
Eqs.~\ref{equ:energy_relat_lab} and~\ref{equ:conservation_part_number_lab}.
Also, Eq.~\ref{equ:energy_relat_lab} can be put into a useful conservative form by expressing 
its right hand side through the energy equation for the fields, which is: 
\begin{equation}
\begin{aligned}
\frac{\partial}{\partial t} & \left(\frac{E^2}{2\mu_0c^2}+\frac{B^2}{2\mu_0}\right)+\frac{\partial}{\partial \b{x}}\cdot\frac{\b{E}\wedge\b{B}}{\mu_0} = -\b{E}\cdot\b{j} \\
 & = -\b{E} \cdot \sum_s \int\dif^3\b{p}f_s(\b{x},\b{p})q_s\b{v} \\
 & = - \sum_s n_\mathrm{lab,s} q_s \langle \b{v}\cdot\b{E}\rangle_s.
\end{aligned}
\end{equation}
We thus have to sum Eq.~\ref{equ:energy_relat_lab} for all species,
to obtain:
\begin{equation}
\label{equ:energy_relat_lab_all_species}
\begin{aligned}
   \frac{\partial}{\partial t} &\left\{\sum_s(n_\mathrm{lab,s}\langle\gamma m_sc^2\rangle_s) + \frac{E^2}{2\mu_0c^2} + \frac{B^2}{2\mu_0}\right\} \\
     &+ \frac{\partial}{\partial \b{x}}\cdot \left\{\sum_s(n_\mathrm{lab,s} \langle \b{v}\gamma m_sc^2\rangle_s) + \frac{\b{E}\wedge\b{B}}{\mu_0} \right\} = 0.
 \end{aligned}
\end{equation}
To recover Eq.~\ref{equ:fluid_3_c}, we use the relation 
$n_{\mathrm{lab},s}\langle\gamma m_sc^2\rangle_s = \Gamma_s^2n_{0s}h_{0s}m_sc^2 - P_{0s}$
\citep[][Table\,5]{Melzani2013}, and the relations previously used 
for $\bar{\b{p}}_s$ and $n_{\mathrm{lab},s}$.

%%%%%%%%%%%%%%%%%%%%%%%%%%%%%%%
%%%%%%%%%%%%%%%%%%%%%%%%%%%%%%%
\subsection{Measure of the temperature tensor}
%%%%%%%%%%%%%%%%%%%%%%%%%%%%%%%
%%%%%%%%%%%%%%%%%%%%%%%%%%%%%%%
The kinetic temperature is a second order moment of the particle distribution function and is, as such,
not always suited to characterize the velocity distribution of a population strongly out of equilibrium.
We nevertheless use it as an indication of the thermal energy content of the population, and of the momentum flux transfers, 
the latter being especially relevant for Ohm's law.
We define the temperature tensor as the ratio of the pressure tensor $n_{\mathrm{lab},s}m_s\langle \delta\b{p}_s\delta\b{v}_s\rangle_s$
(appearing in the equation of conservation of momentum, Eq.~\ref{equ:fluid_2}),
to the comobile particle number density $n_{0s}$.
Since $n_{\mathrm{lab},s} = \Gamma_s n_{0,s}$, the temperature tensor is 
\begin{equation}\label{equ:def_temperature}
 \frac{T_{ij,s}}{m_sc^2} = \Gamma_s\frac{\langle \delta p_{i}\delta v_{j} \rangle_s}{c^2}.
\end{equation}

For the special case of a Maxwell-J\"uttner distribution function of temperature $\Theta_s = 1/\mu_s = T_s/(m_sc^2)$, 
of drift velocity $\beta_{0,s}$ and associated Lorentz factor $\Gamma_{0,s}$, given in the simulation frame by 
$f_s(\b{x},\tilde{\b{p}}) = n_{\mathrm{lab},s}(\b{x})g_s(\tilde{\b{p}})$, $\tilde{\b{p}}=\gamma\b{v}/c$, and
\begin{equation}\label{equ:Jutt}
 g_s(\tilde{\b{p}}) = \frac{\mu_s}{4\pi K_2(\mu_s)\Gamma_{0,s}} \exp\left\{ -\mu\Gamma_{0,s}\left(\sqrt{1+\tilde{p}^2}  - \bar{\b{v}}_s/c\cdot \tilde{\b{p}}\right) \right\},
\end{equation}
we do have \citep{Melzani2013} $\langle \delta p_i\delta v_j\rangle_s/c^2 = \Theta_s\delta_{ij}/\Gamma_s$
where $\delta_{ij}$ is Kronecker's delta.
We also note that this distribution is normalized to unity: $\iiint\!\dif^3\tilde{\b{p}}\,g_s(\tilde{\b{p}}) = 1$.
Based upon these considerations, we measure the temperatures with expression~\ref{equ:def_temperature}.

% \bibliographystyle{apj} 
% % 
% \bibliography{biblio_particles.bib}

\end{document}

%% file: fig_particle_trajectory_3D_n=3274873044_arranged_2.pdf_tex
%% Creator: Inkscape inkscape 0.48.4, www.inkscape.org
%% Accompanies image file 'fig_particle_trajectory_3D_n=3274873044_arranged_2.pdf' (pdf, eps, ps)
\begingroup%
  \makeatletter%
  \providecommand\color[2][]{%
    \errmessage{(Inkscape) Color is used for the text in Inkscape, but the package 'color.sty' is not loaded}%
    \renewcommand\color[2][]{}%
  }%
  \providecommand\transparent[1]{%
    \errmessage{(Inkscape) Transparency is used (non-zero) for the text in Inkscape, but the package 'transparent.sty' is not loaded}%
    \renewcommand\transparent[1]{}%
  }%
  \providecommand\rotatebox[2]{#2}%
  \ifx\svgwidth\undefined%
    \setlength{\unitlength}{942.96699219bp}%
    \ifx\svgscale\undefined%
      \relax%
    \else%
      \setlength{\unitlength}{\unitlength * \real{\svgscale}}%
    \fi%
  \else%
    \setlength{\unitlength}{\svgwidth}%
  \fi%
  \global\let\svgwidth\undefined%
  \global\let\svgscale\undefined%
  \makeatother%
  \begin{picture}(1,0.94937045)%
    \put(0,0){\includegraphics[width=\unitlength]{fig_particle_trajectory_3D_n=3274873044_arranged_2.pdf}}%
    \put(0.89244952,0.10990252){\makebox(0,0)[lb]{\smash{y}}}%
    \put(0.94309656,0.27546246){\makebox(0,0)[lb]{\smash{400}}}%
    \put(0.90112083,0.23556969){\makebox(0,0)[lb]{\smash{600}}}%
    \put(0.85799491,0.19472852){\makebox(0,0)[lb]{\smash{800}}}%
    \put(0.81006206,0.15290473){\makebox(0,0)[lb]{\smash{1000}}}%
    \put(0.76484041,0.11006242){\makebox(0,0)[lb]{\smash{1200}}}%
    \put(0.71850398,0.06616393){\makebox(0,0)[lb]{\smash{1400}}}%
    \put(0.36823578,0.06055707){\makebox(0,0)[lb]{\smash{z}}}%
    \put(0.15142802,0.18278897){\makebox(0,0)[lb]{\smash{4800}}}%
    \put(0.22822065,0.15768053){\makebox(0,0)[lb]{\smash{5000}}}%
    \put(0.30595127,0.13220769){\makebox(0,0)[lb]{\smash{5200}}}%
    \put(0.38481822,0.10636245){\makebox(0,0)[lb]{\smash{5400}}}%
    \put(0.46484665,0.0801366){\makebox(0,0)[lb]{\smash{5600}}}%
    \put(0.54606237,0.05352167){\makebox(0,0)[lb]{\smash{5800}}}%
    \put(0.62845061,0.02650892){\makebox(0,0)[lb]{\smash{6000}}}%
    \put(0.06581135,0.28875309){\makebox(0,0)[lb]{\smash{1600}}}%
    \put(0.06207228,0.36267443){\makebox(0,0)[lb]{\smash{1800}}}%
    \put(0.05803902,0.43782525){\makebox(0,0)[lb]{\smash{2000}}}%
    \put(0.05417397,0.51423653){\makebox(0,0)[lb]{\smash{2200}}}%
    \put(0.05024354,0.59194024){\makebox(0,0)[lb]{\smash{2400}}}%
    \put(0.04624609,0.67096949){\makebox(0,0)[lb]{\smash{2600}}}%
    \put(-0.00015271,0.46809928){\makebox(0,0)[lb]{\smash{x}}}%
    \put(0.825353,0.66592117){\color[rgb]{0,0,0}\rotatebox{-16.03527096}{\makebox(0,0)[lb]{\smash{island}}}}%
    \put(0.28794863,0.63718593){\color[rgb]{0,0,0}\rotatebox{36.04481208}{\makebox(0,0)[lb]{\smash{Current sheet}}}}%
    \put(0.79352451,0.7779423){\color[rgb]{0,0,0}\makebox(0,0)[lb]{\smash{$B_\mathrm{rec}$}}}%
    \put(0.80115624,0.47367269){\color[rgb]{0,0,0}\makebox(0,0)[lb]{\smash{$B_\mathrm{rec}$}}}%
  \end{picture}%
\endgroup%

%% file: fig_wcewpe=3_NT=0000000006000_cutX_various_slides_pdftex_ALL.pdf_tex
%% Creator: Inkscape inkscape 0.48.4, www.inkscape.org
%% Accompanies image file 'fig_wcewpe=3_NT=0000000006000_cutX_various_slides_pdftex_ALL.pdf' (pdf, eps, ps)
\begingroup%
  \makeatletter%
  \providecommand\color[2][]{%
    \errmessage{(Inkscape) Color is used for the text in Inkscape, but the package 'color.sty' is not loaded}%
    \renewcommand\color[2][]{}%
  }%
  \providecommand\transparent[1]{%
    \errmessage{(Inkscape) Transparency is used (non-zero) for the text in Inkscape, but the package 'transparent.sty' is not loaded}%
    \renewcommand\transparent[1]{}%
  }%
  \providecommand\rotatebox[2]{#2}%
  \ifx\svgwidth\undefined%
    \setlength{\unitlength}{1838.95957031bp}%
    \ifx\svgscale\undefined%
      \relax%
    \else%
      \setlength{\unitlength}{\unitlength * \real{\svgscale}}%
    \fi%
  \else%
    \setlength{\unitlength}{\svgwidth}%
  \fi%
  \global\let\svgwidth\undefined%
  \global\let\svgscale\undefined%
  \makeatother%
  \begin{picture}(1,0.59637291)%
    \put(0,0){\includegraphics[width=\unitlength]{fig_wcewpe=3_NT=0000000006000_cutX_various_slides_pdftex_ALL.pdf}}%
    \put(0.52540692,0.324999){\makebox(0,0)[lb]{\smash{1800}}}%
    \put(0.6164366,0.324999){\makebox(0,0)[lb]{\smash{1900}}}%
    \put(0.70734739,0.324999){\makebox(0,0)[lb]{\smash{2000}}}%
    \put(0.79837714,0.324999){\makebox(0,0)[lb]{\smash{2100}}}%
    \put(0.88940689,0.324999){\makebox(0,0)[lb]{\smash{2200}}}%
    \put(0.98043663,0.324999){\makebox(0,0)[lb]{\smash{2300}}}%
    \put(0.7537417,0.32505511){\makebox(0,0)[lb]{\smash{cells}}}%
    \put(0.88178525,0.41531118){\makebox(0,0)[lb]{\smash{$B_z$}}}%
    \put(0.88178525,0.39967904){\makebox(0,0)[lb]{\smash{$10E_y$}}}%
    \put(0.88178525,0.38404686){\makebox(0,0)[lb]{\smash{$v_x$ ion}}}%
    \put(0.88178525,0.36928474){\makebox(0,0)[lb]{\smash{$[\textbf{E}\wedge\textbf{B}/B^2]_x$}}}%
    \put(0.00840333,0.46669501){\makebox(0,0)[lb]{\smash{200}}}%
    \put(0.00840333,0.49783908){\makebox(0,0)[lb]{\smash{100}}}%
    \put(0.01636011,0.52898319){\makebox(0,0)[lb]{\smash{0}}}%
    \put(0.00843263,0.56012722){\makebox(0,0)[lb]{\smash{100}}}%
    \put(0.00819775,0.59127132){\makebox(0,0)[lb]{\smash{200}}}%
    \put(0.11603571,0.32367452){\makebox(0,0)[lb]{\smash{1900}}}%
    \put(0.20594776,0.32367452){\makebox(0,0)[lb]{\smash{2000}}}%
    \put(0.29597721,0.32367452){\makebox(0,0)[lb]{\smash{2100}}}%
    \put(0.38600672,0.32367452){\makebox(0,0)[lb]{\smash{2200}}}%
    \put(0.47603617,0.32367452){\makebox(0,0)[lb]{\smash{2300}}}%
    \put(0.25260438,0.32352521){\makebox(0,0)[lb]{\smash{cells}}}%
    \put(0.01044182,0.34836227){\makebox(0,0)[lb]{\smash{0.6}}}%
    \put(0.01039148,0.36419703){\makebox(0,0)[lb]{\smash{0.4}}}%
    \put(0.0106767,0.38003179){\makebox(0,0)[lb]{\smash{0.2}}}%
    \put(0.01021105,0.39586636){\makebox(0,0)[lb]{\smash{0.0}}}%
    \put(0.0104292,0.41170115){\makebox(0,0)[lb]{\smash{0.2}}}%
    \put(0.01014398,0.42753587){\makebox(0,0)[lb]{\smash{0.4}}}%
    \put(0.01019432,0.44337063){\makebox(0,0)[lb]{\smash{0.6}}}%
    \put(0.11422287,0.04210876){\makebox(0,0)[lb]{\smash{1900}}}%
    \put(0.20577967,0.04210876){\makebox(0,0)[lb]{\smash{2000}}}%
    \put(0.29745545,0.04210876){\makebox(0,0)[lb]{\smash{2100}}}%
    \put(0.3891312,0.04210876){\makebox(0,0)[lb]{\smash{2200}}}%
    \put(0.48080696,0.04210876){\makebox(0,0)[lb]{\smash{2300}}}%
    \put(0.25259488,0.04276637){\makebox(0,0)[lb]{\smash{cells}}}%
    \put(0.0047759,0.19000757){\makebox(0,0)[lb]{\smash{200}}}%
    \put(0.0047759,0.2171814){\makebox(0,0)[lb]{\smash{100}}}%
    \put(0.01283498,0.24435524){\makebox(0,0)[lb]{\smash{0}}}%
    \put(0.00480564,0.27152906){\makebox(0,0)[lb]{\smash{100}}}%
    \put(0.00456774,0.2987029){\makebox(0,0)[lb]{\smash{200}}}%
    \put(0.00684059,0.05664609){\makebox(0,0)[lb]{\smash{0.6}}}%
    \put(0.00678961,0.07397867){\makebox(0,0)[lb]{\smash{0.4}}}%
    \put(0.00707849,0.09131124){\makebox(0,0)[lb]{\smash{0.2}}}%
    \put(0.00660693,0.10864387){\makebox(0,0)[lb]{\smash{0.0}}}%
    \put(0.00682784,0.12597645){\makebox(0,0)[lb]{\smash{0.2}}}%
    \put(0.00653896,0.14330902){\makebox(0,0)[lb]{\smash{0.4}}}%
    \put(0.00658994,0.16064158){\makebox(0,0)[lb]{\smash{0.6}}}%
    \put(0.00662392,0.17797415){\makebox(0,0)[lb]{\smash{0.8}}}%
    \put(0.61693339,0.04019109){\makebox(0,0)[lb]{\smash{1900}}}%
    \put(0.70856859,0.04019109){\makebox(0,0)[lb]{\smash{2000}}}%
    \put(0.80032281,0.04019109){\makebox(0,0)[lb]{\smash{2100}}}%
    \put(0.89207696,0.04019109){\makebox(0,0)[lb]{\smash{2200}}}%
    \put(0.98383111,0.04019109){\makebox(0,0)[lb]{\smash{2300}}}%
    \put(0.75561907,0.04249237){\makebox(0,0)[lb]{\smash{cells}}}%
    \put(0.50918003,0.48771597){\makebox(0,0)[lb]{\smash{400}}}%
    \put(0.50918003,0.5084787){\makebox(0,0)[lb]{\smash{200}}}%
    \put(0.51723905,0.52924139){\makebox(0,0)[lb]{\smash{0}}}%
    \put(0.50897187,0.55000433){\makebox(0,0)[lb]{\smash{200}}}%
    \put(0.5088189,0.57076708){\makebox(0,0)[lb]{\smash{400}}}%
    \put(0.50895484,0.59152981){\makebox(0,0)[lb]{\smash{600}}}%
    \put(0.5112617,0.33921303){\makebox(0,0)[lb]{\smash{1.0}}}%
    \put(0.51139764,0.368617){\makebox(0,0)[lb]{\smash{0.5}}}%
    \put(0.51101107,0.39976108){\makebox(0,0)[lb]{\smash{0.0}}}%
    \put(0.51114701,0.4309052){\makebox(0,0)[lb]{\smash{0.5}}}%
    \put(0.51129139,0.46030917){\makebox(0,0)[lb]{\smash{1.0}}}%
    \put(0.50867012,0.20380441){\makebox(0,0)[lb]{\smash{100}}}%
    \put(0.51672931,0.2440578){\makebox(0,0)[lb]{\smash{0}}}%
    \put(0.50870004,0.28431147){\makebox(0,0)[lb]{\smash{100}}}%
    \put(0.51073482,0.05563988){\makebox(0,0)[lb]{\smash{0.6}}}%
    \put(0.51068387,0.07353826){\color[rgb]{1,1,1}\makebox(0,0)[lb]{\smash{0.4}}}%
    \put(0.51097273,0.09143663){\color[rgb]{1,1,1}\makebox(0,0)[lb]{\smash{0.2}}}%
    \put(0.51050127,0.10933489){\color[rgb]{1,1,1}\makebox(0,0)[lb]{\smash{0.0}}}%
    \put(0.51072218,0.12723325){\color[rgb]{1,1,1}\makebox(0,0)[lb]{\smash{0.2}}}%
    \put(0.5104333,0.14513166){\color[rgb]{1,1,1}\makebox(0,0)[lb]{\smash{0.4}}}%
    \put(0.51048428,0.16303002){\color[rgb]{1,1,1}\makebox(0,0)[lb]{\smash{0.6}}}%
    \put(0.51051826,0.18092842){\makebox(0,0)[lb]{\smash{0.8}}}%
    \put(0.53360281,0.01843594){\makebox(0,0)[lb]{\smash{30}}}%
    \put(0.60946092,0.01843594){\makebox(0,0)[lb]{\smash{20}}}%
    \put(0.68531903,0.01843594){\makebox(0,0)[lb]{\smash{10}}}%
    \put(0.76039545,0.01843591){\makebox(0,0)[lb]{\smash{0}}}%
    \put(0.83431632,0.01843591){\makebox(0,0)[lb]{\smash{10}}}%
    \put(0.91005548,0.01843591){\makebox(0,0)[lb]{\smash{20}}}%
    \put(0.69249219,0.00224771){\makebox(0,0)[lb]{\smash{electron inertial lengths $c/\omega_{\mathrm{p}e}(t=0,x=0)$}}}%
    \put(0.86390653,0.51857431){\makebox(0,0)[lb]{\smash{electron density}}}%
    \put(0.86390653,0.50468228){\makebox(0,0)[lb]{\smash{ion density}}}%
    \put(0.86390653,0.49079028){\makebox(0,0)[lb]{\smash{current $j_y$}}}%
    \put(0.86297209,0.47772769){\makebox(0,0)[lb]{\smash{$3000[\textbf{E}+\textbf{v}_\mathrm{e}\wedge\textbf{B}]_y$}}}%
    \put(0.88178525,0.44657548){\makebox(0,0)[lb]{\smash{$v_y$ electron}}}%
    \put(0.88178525,0.43094333){\makebox(0,0)[lb]{\smash{$v_x$ electron}}}%
    \put(0.03014835,0.01772715){\makebox(0,0)[lb]{\smash{30}}}%
    \put(0.10600646,0.01772715){\makebox(0,0)[lb]{\smash{20}}}%
    \put(0.18186457,0.01772715){\makebox(0,0)[lb]{\smash{10}}}%
    \put(0.25694099,0.01772712){\makebox(0,0)[lb]{\smash{0}}}%
    \put(0.33086186,0.01772712){\makebox(0,0)[lb]{\smash{10}}}%
    \put(0.40660102,0.01772712){\makebox(0,0)[lb]{\smash{20}}}%
    \put(0.18903772,0.00153894){\makebox(0,0)[lb]{\smash{electron inertial lengths $c/\omega_{\mathrm{p}e}(t=0,x=0)$}}}%
    \put(0.38120447,0.40751729){\makebox(0,0)[lb]{\smash{$B_z$}}}%
    \put(0.38120447,0.39188515){\makebox(0,0)[lb]{\smash{$100E_y$}}}%
    \put(0.38120447,0.37625297){\makebox(0,0)[lb]{\smash{$v_x$ ion}}}%
    \put(0.38120447,0.36149085){\makebox(0,0)[lb]{\smash{$[\textbf{E}\wedge\textbf{B}/B^2]_x$}}}%
    \put(0.38120447,0.4387816){\makebox(0,0)[lb]{\smash{$v_y$ electron}}}%
    \put(0.38120447,0.42314945){\makebox(0,0)[lb]{\smash{$v_x$ electron}}}%
    \put(0.36149082,0.51878822){\makebox(0,0)[lb]{\smash{electron density}}}%
    \put(0.36149082,0.50489619){\makebox(0,0)[lb]{\smash{ion density}}}%
    \put(0.36149082,0.49100419){\makebox(0,0)[lb]{\smash{current $j_y$}}}%
    \put(0.36055638,0.4779416){\makebox(0,0)[lb]{\smash{$6000[\textbf{E}+\textbf{v}_\mathrm{e}\wedge\textbf{B}]_y$}}}%
    \put(0.3685294,0.23255295){\makebox(0,0)[lb]{\smash{electron density}}}%
    \put(0.3685294,0.21866092){\makebox(0,0)[lb]{\smash{ion density}}}%
    \put(0.3685294,0.20476893){\makebox(0,0)[lb]{\smash{current $j_y$}}}%
    \put(0.36759497,0.19170634){\makebox(0,0)[lb]{\smash{$3000[\textbf{E}+\textbf{v}_\mathrm{e}\wedge\textbf{B}]_y$}}}%
    \put(0.87237385,0.22962021){\makebox(0,0)[lb]{\smash{electron density}}}%
    \put(0.87237385,0.21572818){\makebox(0,0)[lb]{\smash{ion density}}}%
    \put(0.87237385,0.20183619){\makebox(0,0)[lb]{\smash{current $j_y$}}}%
    \put(0.87143941,0.1887736){\makebox(0,0)[lb]{\smash{$6000[\textbf{E}+\textbf{v}_\mathrm{e}\wedge\textbf{B}]_y$}}}%
    \put(0.05275024,0.4761943){\color[rgb]{0,0,0}\makebox(0,0)[lb]{\smash{$\omega_\mathrm{ce}/\omega_\mathrm{pe}=1$}}}%
    \put(0.05275024,0.45566514){\color[rgb]{0,0,0}\makebox(0,0)[lb]{\smash{$T_\mathrm{bg,i~or~e}=1.5\times10^7\,\mathrm{K}$}}}%
    \put(0.38355067,0.12538787){\makebox(0,0)[lb]{\smash{$B_z$}}}%
    \put(0.38355067,0.10975573){\makebox(0,0)[lb]{\smash{$10E_y$}}}%
    \put(0.38355067,0.09412355){\makebox(0,0)[lb]{\smash{$v_x$ ion}}}%
    \put(0.38355067,0.07936143){\makebox(0,0)[lb]{\smash{$[\textbf{E}\wedge\textbf{B}/B^2]_x$}}}%
    \put(0.38355067,0.15665217){\makebox(0,0)[lb]{\smash{$v_y$ electron}}}%
    \put(0.38355067,0.14102003){\makebox(0,0)[lb]{\smash{$v_x$ electron}}}%
    \put(0.04832941,0.19576595){\color[rgb]{0,0,0}\makebox(0,0)[lb]{\smash{$\omega_\mathrm{ce}/\omega_\mathrm{pe}=3$}}}%
    \put(0.04832941,0.17523681){\color[rgb]{0,0,0}\makebox(0,0)[lb]{\smash{$T_\mathrm{bg,i~or~e}=2\times10^8\,\mathrm{K}$}}}%
    \put(0.55307315,0.20630429){\color[rgb]{0,0,0}\makebox(0,0)[lb]{\smash{$\omega_\mathrm{ce}/\omega_\mathrm{pe}=3$}}}%
    \put(0.55307315,0.18751527){\color[rgb]{0,0,0}\makebox(0,0)[lb]{\smash{$T_\mathrm{bg,i}=2\times10^8\,\mathrm{K}$}}}%
    \put(0.55166755,0.16874573){\color[rgb]{0,0,0}\makebox(0,0)[lb]{\smash{$T_\mathrm{bg,e}=3\times10^9\,\mathrm{K}$}}}%
    \put(0.55453709,0.47845263){\color[rgb]{0,0,0}\makebox(0,0)[lb]{\smash{$\omega_\mathrm{ce}/\omega_\mathrm{pe}=3$}}}%
    \put(0.55453709,0.45792346){\color[rgb]{0,0,0}\makebox(0,0)[lb]{\smash{$T_\mathrm{bg,i~or~e}=1.5\times10^7\,\mathrm{K}$}}}%
    \put(0.8897413,0.12773406){\makebox(0,0)[lb]{\smash{$B_z$}}}%
    \put(0.8897413,0.11210192){\makebox(0,0)[lb]{\smash{$10E_y$}}}%
    \put(0.8897413,0.09646974){\makebox(0,0)[lb]{\smash{$v_x$ ion}}}%
    \put(0.8897413,0.08170762){\makebox(0,0)[lb]{\smash{$[\textbf{E}\wedge\textbf{B}/B^2]_x$}}}%
    \put(0.8897413,0.15899837){\makebox(0,0)[lb]{\smash{$v_y$ electron}}}%
    \put(0.8897413,0.14336622){\makebox(0,0)[lb]{\smash{$v_x$ electron}}}%
    \put(0.52502218,0.04025286){\makebox(0,0)[lb]{\smash{1800}}}%
    \put(0.02117777,0.04230578){\makebox(0,0)[lb]{\smash{1800}}}%
    \put(0.82813779,0.58182594){\color[rgb]{0,0,0}\makebox(0,0)[lb]{\smash{ion diffusion region}}}%
    \put(0.78286338,0.56057097){\color[rgb]{0,0,0}\makebox(0,0)[lb]{\smash{electron diffusion region}}}%
  \end{picture}%
\endgroup%

%% file: scale_diff_region_nice_withothers.pdf_tex
%% Creator: Inkscape inkscape 0.48.4, www.inkscape.org
%% Accompanies image file 'scale_diff_region_nice_withothers.pdf' (pdf, eps, ps)
\begingroup%
  \makeatletter%
  \providecommand\color[2][]{%
    \errmessage{(Inkscape) Color is used for the text in Inkscape, but the package 'color.sty' is not loaded}%
    \renewcommand\color[2][]{}%
  }%
  \providecommand\transparent[1]{%
    \errmessage{(Inkscape) Transparency is used (non-zero) for the text in Inkscape, but the package 'transparent.sty' is not loaded}%
    \renewcommand\transparent[1]{}%
  }%
  \providecommand\rotatebox[2]{#2}%
  \ifx\svgwidth\undefined%
    \setlength{\unitlength}{483.42109375bp}%
    \ifx\svgscale\undefined%
      \relax%
    \else%
      \setlength{\unitlength}{\unitlength * \real{\svgscale}}%
    \fi%
  \else%
    \setlength{\unitlength}{\svgwidth}%
  \fi%
  \global\let\svgwidth\undefined%
  \global\let\svgscale\undefined%
  \makeatother%
  \begin{picture}(1,1.47166426)%
    \put(0,0){\includegraphics[width=\unitlength]{scale_diff_region_nice_withothers.pdf}}%
    \put(0.05408878,0.78857704){\makebox(0,0)[lb]{\smash{0}}}%
    \put(0.23095063,0.78857704){\makebox(0,0)[lb]{\smash{20}}}%
    \put(0.41534344,0.78857704){\makebox(0,0)[lb]{\smash{40}}}%
    \put(0.60028571,0.78857704){\makebox(0,0)[lb]{\smash{60}}}%
    \put(0.7849371,0.78857704){\makebox(0,0)[lb]{\smash{80}}}%
    \put(0.94568651,0.78857704){\makebox(0,0)[lb]{\smash{100}}}%
    \put(0.49089419,0.7663945){\makebox(0,0)[lb]{\smash{$t/T_\mathrm{pe}$}}}%
    \put(0.02963093,1.13661352){\makebox(0,0)[lb]{\smash{0}}}%
    \put(0.01408419,1.2055927){\makebox(0,0)[lb]{\smash{50}}}%
    \put(-0.00091309,1.27457192){\makebox(0,0)[lb]{\smash{100}}}%
    \put(-0.00091309,1.34355113){\makebox(0,0)[lb]{\smash{150}}}%
    \put(-0.0018181,1.41253029){\makebox(0,0)[lb]{\smash{200}}}%
    \put(0.01489223,0.84896663){\makebox(0,0)[lb]{\smash{10}}}%
    \put(0.01398722,0.92117922){\makebox(0,0)[lb]{\smash{20}}}%
    \put(0.01408419,0.99339204){\makebox(0,0)[lb]{\smash{30}}}%
    \put(0.01340543,1.06560466){\makebox(0,0)[lb]{\smash{40}}}%
    \put(0.34663307,1.09424438){\makebox(0,0)[lb]{\smash{width $\delta_\mathrm{e}$}}}%
    \put(0.34663307,1.05132762){\makebox(0,0)[lb]{\smash{inertial length $d_\mathrm{e}$}}}%
    \put(0.17021747,0.04304056){\makebox(0,0)[lb]{\smash{10}}}%
    \put(0.30137864,0.04304056){\makebox(0,0)[lb]{\smash{20}}}%
    \put(0.43304081,0.04304056){\makebox(0,0)[lb]{\smash{30}}}%
    \put(0.56431505,0.04304056){\makebox(0,0)[lb]{\smash{40}}}%
    \put(0.69626821,0.04304056){\makebox(0,0)[lb]{\smash{50}}}%
    \put(0.827801,0.04304056){\makebox(0,0)[lb]{\smash{60}}}%
    \put(0.95956016,0.04304056){\makebox(0,0)[lb]{\smash{70}}}%
    \put(0.02529552,0.42404163){\makebox(0,0)[lb]{\smash{40}}}%
    \put(0.02581266,0.46489511){\makebox(0,0)[lb]{\smash{60}}}%
    \put(0.02574801,0.5057487){\makebox(0,0)[lb]{\smash{80}}}%
    \put(0.01097699,0.54660234){\makebox(0,0)[lb]{\smash{100}}}%
    \put(0.01097699,0.58745598){\makebox(0,0)[lb]{\smash{120}}}%
    \put(0.01097699,0.62830964){\makebox(0,0)[lb]{\smash{140}}}%
    \put(0.01097699,0.66916332){\makebox(0,0)[lb]{\smash{160}}}%
    \put(0.01097699,0.71001696){\makebox(0,0)[lb]{\smash{180}}}%
    \put(0.02394589,0.10644459){\makebox(0,0)[lb]{\smash{10}}}%
    \put(0.02304088,0.14742708){\makebox(0,0)[lb]{\smash{20}}}%
    \put(0.02313785,0.1884095){\makebox(0,0)[lb]{\smash{30}}}%
    \put(0.0224591,0.22939189){\makebox(0,0)[lb]{\smash{40}}}%
    \put(0.02313785,0.27037438){\makebox(0,0)[lb]{\smash{50}}}%
    \put(0.02297624,0.31135683){\makebox(0,0)[lb]{\smash{60}}}%
    \put(0.02326714,0.35233926){\makebox(0,0)[lb]{\smash{70}}}%
    \put(0.81260859,1.18320335){\color[rgb]{0,0,0}\makebox(0,0)[lb]{\smash{\textbf{a}}}}%
    \put(0.80692862,0.85180815){\color[rgb]{0,0,0}\makebox(0,0)[lb]{\smash{\textbf{a}}}}%
    \put(0.85426087,0.45327858){\color[rgb]{0,0,0}\makebox(0,0)[lb]{\smash{\textbf{b}}}}%
    \put(0.85426434,0.10153119){\color[rgb]{0,0,0}\makebox(0,0)[lb]{\smash{\textbf{b}}}}%
    \put(0.68786345,0.32421286){\makebox(0,0)[lb]{\smash{width $\delta_\mathrm{e}$}}}%
    \put(0.68786345,0.28129609){\makebox(0,0)[lb]{\smash{inertial length $d_\mathrm{e}$}}}%
    \put(0.19750245,0.67156766){\makebox(0,0)[lb]{\smash{width $\delta_\mathrm{i}$}}}%
    \put(0.19750245,0.6286509){\makebox(0,0)[lb]{\smash{inertial length $d_\mathrm{i}$}}}%
    \put(0.19117523,1.41663869){\makebox(0,0)[lb]{\smash{width $\delta_\mathrm{i}$}}}%
    \put(0.19117523,1.37372192){\makebox(0,0)[lb]{\smash{inertial length $d_\mathrm{i}$}}}%
    \put(0.513311,0.0058543){\makebox(0,0)[lb]{\smash{$t/T_\mathrm{pe}$}}}%
  \end{picture}%
\endgroup%

%% file: visit_ballistic_motions_with_profiles.pdf_tex
%% Creator: Inkscape inkscape 0.48.4, www.inkscape.org
%% Accompanies image file 'visit_ballistic_motions_with_profiles.pdf' (pdf, eps, ps)
\begingroup%
  \makeatletter%
  \providecommand\color[2][]{%
    \errmessage{(Inkscape) Color is used for the text in Inkscape, but the package 'color.sty' is not loaded}%
    \renewcommand\color[2][]{}%
  }%
  \providecommand\transparent[1]{%
    \errmessage{(Inkscape) Transparency is used (non-zero) for the text in Inkscape, but the package 'transparent.sty' is not loaded}%
    \renewcommand\transparent[1]{}%
  }%
  \providecommand\rotatebox[2]{#2}%
  \ifx\svgwidth\undefined%
    \setlength{\unitlength}{1510.15625bp}%
    \ifx\svgscale\undefined%
      \relax%
    \else%
      \setlength{\unitlength}{\unitlength * \real{\svgscale}}%
    \fi%
  \else%
    \setlength{\unitlength}{\svgwidth}%
  \fi%
  \global\let\svgwidth\undefined%
  \global\let\svgscale\undefined%
  \makeatother%
  \begin{picture}(1,0.6076358)%
    \put(0,0){\includegraphics[width=\unitlength]{visit_ballistic_motions_with_profiles.pdf}}%
    \put(0.8146894,0.56538168){\makebox(0,0)[lb]{\smash{current $j_y$}}}%
    \put(0.81369604,0.51750305){\makebox(0,0)[lb]{\smash{$\bar{v}_x$ electron}}}%
    \put(0.81422578,0.47429143){\makebox(0,0)[lb]{\smash{$\bar{v}_x$ ion}}}%
    \put(0.62930672,0.03596043){\color[rgb]{0,0,0}\makebox(0,0)[lb]{\smash{cut}}}%
    \put(0.90451537,0.00865087){\color[rgb]{0,0,0}\makebox(0,0)[lb]{\smash{$x$}}}%
  \end{picture}%
\endgroup%

%% file: svg_wcewpe=3_NT=6001_Ohms_law_smooth3D=7.pdf_tex
%% Creator: Inkscape inkscape 0.48.4, www.inkscape.org
%% Accompanies image file 'svg_wcewpe=3_NT=6001_Ohms_law_smooth3D=7.pdf' (pdf, eps, ps)
\begingroup%
  \makeatletter%
  \providecommand\color[2][]{%
    \errmessage{(Inkscape) Color is used for the text in Inkscape, but the package 'color.sty' is not loaded}%
    \renewcommand\color[2][]{}%
  }%
  \providecommand\transparent[1]{%
    \errmessage{(Inkscape) Transparency is used (non-zero) for the text in Inkscape, but the package 'transparent.sty' is not loaded}%
    \renewcommand\transparent[1]{}%
  }%
  \providecommand\rotatebox[2]{#2}%
  \ifx\svgwidth\undefined%
    \setlength{\unitlength}{1041.12109375bp}%
    \ifx\svgscale\undefined%
      \relax%
    \else%
      \setlength{\unitlength}{\unitlength * \real{\svgscale}}%
    \fi%
  \else%
    \setlength{\unitlength}{\svgwidth}%
  \fi%
  \global\let\svgwidth\undefined%
  \global\let\svgscale\undefined%
  \makeatother%
  \begin{picture}(1,0.7912462)%
    \put(0,0){\includegraphics[width=\unitlength]{svg_wcewpe=3_NT=6001_Ohms_law_smooth3D=7.pdf}}%
    \put(0.20011468,0.10784626){\makebox(0,0)[lb]{\smash{1950}}}%
    \put(0.35682936,0.10784626){\makebox(0,0)[lb]{\smash{2000}}}%
    \put(0.51375416,0.10784626){\makebox(0,0)[lb]{\smash{2050}}}%
    \put(0.67067893,0.10784626){\makebox(0,0)[lb]{\smash{2100}}}%
    \put(0.82760375,0.10784626){\makebox(0,0)[lb]{\smash{2150}}}%
    \put(0.92981684,0.10962944){\makebox(0,0)[lb]{\smash{cells}}}%
    \put(-0.00075977,0.4002651){\makebox(0,0)[lb]{\smash{0.}}}%
    \put(-0.00075977,0.59342119){\makebox(0,0)[lb]{\smash{0.1}}}%
    \put(0.64338965,0.73853231){\makebox(0,0)[lb]{\smash{$E_y$}}}%
    \put(0.64338965,0.6934035){\makebox(0,0)[lb]{\smash{$-[\bar{\b{v}}_\mathrm{e}\wedge\b{B}]_y$}}}%
    \put(0.64338965,0.64879093){\makebox(0,0)[lb]{\smash{$-[n_\mathrm{e}^{-1}\nabla\cdot(n_\mathrm{e}\bar{\b{v}}_\mathrm{e}\bar{\b{p}}_\mathrm{e})]_y$}}}%
    \put(0.64338965,0.60546916){\makebox(0,0)[lb]{\smash{$-[n_\mathrm{e}^{-1}\nabla\cdot(n_\mathrm{e}\delta\b{v}_\mathrm{e}\delta\b{p}_\mathrm{e})]_y$}}}%
    \put(0.64338971,0.56305724){\makebox(0,0)[lb]{\smash{all (should be }}}%
    \put(0.12843768,0.04445742){\makebox(0,0)[lb]{\smash{-15}}}%
    \put(0.25908827,0.04430734){\makebox(0,0)[lb]{\smash{-10}}}%
    \put(0.39364842,0.04445742){\makebox(0,0)[lb]{\smash{-5}}}%
    \put(0.52472491,0.04430733){\makebox(0,0)[lb]{\smash{0}}}%
    \put(0.65567564,0.04445741){\makebox(0,0)[lb]{\smash{5}}}%
    \put(0.78284439,0.04430733){\makebox(0,0)[lb]{\smash{10}}}%
    \put(0.9137351,0.04445741){\makebox(0,0)[lb]{\smash{15}}}%
    \put(0.3120047,0.0027183){\makebox(0,0)[lb]{\smash{electron inertial lengths $c/\omega_{\mathrm{pe}}(x=0,t=0)$}}}%
    \put(0.64328577,0.53297287){\makebox(0,0)[lb]{\smash{equal to $E_y$)}}}%
    \put(0.48684788,0.55595794){\color[rgb]{0,0,0}\makebox(0,0)[lb]{\smash{\tiny elec.}}}%
    \put(0.48725796,0.53073072){\color[rgb]{0,0,0}\makebox(0,0)[lb]{\smash{\tiny diff.}}}%
    \put(0.48725796,0.50678656){\color[rgb]{0,0,0}\makebox(0,0)[lb]{\smash{\tiny region}}}%
    \put(0.37384801,0.53101223){\color[rgb]{0,0,0}\makebox(0,0)[lb]{\smash{\tiny ion}}}%
    \put(0.37425809,0.50578502){\color[rgb]{0,0,0}\makebox(0,0)[lb]{\smash{\tiny diff.}}}%
    \put(0.37425809,0.48184085){\color[rgb]{0,0,0}\makebox(0,0)[lb]{\smash{\tiny region}}}%
    \put(0.14602708,0.355941){\color[rgb]{0,0,0}\makebox(0,0)[lb]{\smash{\tiny bulk inertia}}}%
    \put(0.11899652,0.43187714){\color[rgb]{0,0,0}\makebox(0,0)[lb]{\smash{\tiny thermal inertia}}}%
  \end{picture}%
\endgroup%

%% file: svg_wcewpe=3_NT=6000_cutalongX_temperature.pdf_tex
%% Creator: Inkscape inkscape 0.48.4, www.inkscape.org
%% Accompanies image file 'svg_wcewpe=3_NT=6000_cutalongX_temperature.pdf' (pdf, eps, ps)
\begingroup%
  \makeatletter%
  \providecommand\color[2][]{%
    \errmessage{(Inkscape) Color is used for the text in Inkscape, but the package 'color.sty' is not loaded}%
    \renewcommand\color[2][]{}%
  }%
  \providecommand\transparent[1]{%
    \errmessage{(Inkscape) Transparency is used (non-zero) for the text in Inkscape, but the package 'transparent.sty' is not loaded}%
    \renewcommand\transparent[1]{}%
  }%
  \providecommand\rotatebox[2]{#2}%
  \ifx\svgwidth\undefined%
    \setlength{\unitlength}{865.61777344bp}%
    \ifx\svgscale\undefined%
      \relax%
    \else%
      \setlength{\unitlength}{\unitlength * \real{\svgscale}}%
    \fi%
  \else%
    \setlength{\unitlength}{\svgwidth}%
  \fi%
  \global\let\svgwidth\undefined%
  \global\let\svgscale\undefined%
  \makeatother%
  \begin{picture}(1,0.83906245)%
    \put(0,0){\includegraphics[width=\unitlength]{svg_wcewpe=3_NT=6000_cutalongX_temperature.pdf}}%
    \put(0.09048301,0.54845928){\makebox(0,0)[lb]{\smash{0.5}}}%
    \put(0.08966172,0.60888131){\makebox(0,0)[lb]{\smash{0.0}}}%
    \put(0.08995054,0.66930297){\makebox(0,0)[lb]{\smash{0.5}}}%
    \put(0.0902574,0.72972463){\makebox(0,0)[lb]{\smash{1.0}}}%
    \put(0.09054621,0.79023654){\makebox(0,0)[lb]{\smash{1.5}}}%
    \put(0.12657036,0.09518413){\makebox(0,0)[lb]{\smash{1950}}}%
    \put(0.3312249,0.09518413){\makebox(0,0)[lb]{\smash{2000}}}%
    \put(0.53613216,0.09518413){\makebox(0,0)[lb]{\smash{2050}}}%
    \put(0.74103974,0.09518413){\makebox(0,0)[lb]{\smash{2100}}}%
    \put(0.93077677,0.09680734){\makebox(0,0)[lb]{\smash{cells}}}%
    \put(0.22241453,0.03973247){\makebox(0,0)[lb]{\smash{10}}}%
    \put(0.39772841,0.03991297){\makebox(0,0)[lb]{\smash{5}}}%
    \put(0.56226601,0.03973237){\makebox(0,0)[lb]{\smash{0}}}%
    \put(0.73323867,0.03991288){\makebox(0,0)[lb]{\smash{5}}}%
    \put(0.8996626,0.03973237){\makebox(0,0)[lb]{\smash{10}}}%
    \put(0.26642843,0.00288362){\makebox(0,0)[lb]{\smash{electron inertial lengths $c/\omega_\mathrm{pe}(x=0,t=0)$}}}%
    \put(0.09066352,0.17158907){\makebox(0,0)[lb]{\smash{0.2}}}%
    \put(0.09055521,0.22193522){\makebox(0,0)[lb]{\smash{0.1}}}%
    \put(0.08966172,0.27228134){\makebox(0,0)[lb]{\smash{0.0}}}%
    \put(0.09002273,0.32262751){\makebox(0,0)[lb]{\smash{0.1}}}%
    \put(0.09013103,0.37297364){\makebox(0,0)[lb]{\smash{0.2}}}%
    \put(0.08986027,0.42331977){\makebox(0,0)[lb]{\smash{0.3}}}%
    \put(0.02302106,0.17161685){\makebox(0,0)[lb]{\smash{5}}}%
    \put(0.01014615,0.221963){\makebox(0,0)[lb]{\smash{2.5}}}%
    \put(0.00925267,0.27230916){\makebox(0,0)[lb]{\smash{0.0}}}%
    \put(0.00961369,0.32265529){\makebox(0,0)[lb]{\smash{2.5}}}%
    \put(0.00972199,0.37300142){\makebox(0,0)[lb]{\smash{5}}}%
    \put(0.00945123,0.42334755){\makebox(0,0)[lb]{\smash{7.5}}}%
    \put(0.00910827,0.47369368){\makebox(0,0)[lb]{\smash{10}}}%
    \put(0.80892755,0.79340397){\makebox(0,0)[lb]{\smash{$\Theta_{xx}$}}}%
    \put(0.80892755,0.7608389){\makebox(0,0)[lb]{\smash{$\Theta_{yy}$}}}%
    \put(0.80892755,0.72478857){\makebox(0,0)[lb]{\smash{$\Theta_{zz}$}}}%
    \put(0.90275674,0.7935904){\makebox(0,0)[lb]{\smash{$\Theta_{xy}$}}}%
    \put(0.90275674,0.76102547){\makebox(0,0)[lb]{\smash{$\Theta_{xz}$}}}%
    \put(0.90275674,0.724975){\makebox(0,0)[lb]{\smash{$\Theta_{zy}$}}}%
    \put(0.80892755,0.6872482){\makebox(0,0)[lb]{\smash{$j_y$}}}%
    \put(0.17546907,0.75973806){\color[rgb]{0,0,0}\makebox(0,0)[lb]{\smash{electrons}}}%
    \put(0.17447378,0.39534296){\color[rgb]{0,0,0}\makebox(0,0)[lb]{\smash{ions}}}%
    \put(0.83016716,0.44331516){\makebox(0,0)[lb]{\smash{$\Theta_{xx}$}}}%
    \put(0.83016716,0.41075009){\makebox(0,0)[lb]{\smash{$\Theta_{yy}$}}}%
    \put(0.83016716,0.37469975){\makebox(0,0)[lb]{\smash{$\Theta_{zz}$}}}%
    \put(0.92399635,0.44350159){\makebox(0,0)[lb]{\smash{$\Theta_{xy}$}}}%
    \put(0.92399635,0.41093666){\makebox(0,0)[lb]{\smash{$\Theta_{xz}$}}}%
    \put(0.92399635,0.37488618){\makebox(0,0)[lb]{\smash{$\Theta_{zy}$}}}%
    \put(0.83016716,0.33715939){\makebox(0,0)[lb]{\smash{$j_y$}}}%
  \end{picture}%
\endgroup%

%% file: fig_wcewpe=3_NT=6000_cutalongZ_summary.pdf_tex
%% Creator: Inkscape inkscape 0.48.4, www.inkscape.org
%% Accompanies image file 'fig_wcewpe=3_NT=6000_cutalongZ_summary.pdf' (pdf, eps, ps)
\begingroup%
  \makeatletter%
  \providecommand\color[2][]{%
    \errmessage{(Inkscape) Color is used for the text in Inkscape, but the package 'color.sty' is not loaded}%
    \renewcommand\color[2][]{}%
  }%
  \providecommand\transparent[1]{%
    \errmessage{(Inkscape) Transparency is used (non-zero) for the text in Inkscape, but the package 'transparent.sty' is not loaded}%
    \renewcommand\transparent[1]{}%
  }%
  \providecommand\rotatebox[2]{#2}%
  \ifx\svgwidth\undefined%
    \setlength{\unitlength}{1807.38066406bp}%
    \ifx\svgscale\undefined%
      \relax%
    \else%
      \setlength{\unitlength}{\unitlength * \real{\svgscale}}%
    \fi%
  \else%
    \setlength{\unitlength}{\svgwidth}%
  \fi%
  \global\let\svgwidth\undefined%
  \global\let\svgscale\undefined%
  \makeatother%
  \begin{picture}(1,1.02941682)%
    \put(0,0){\includegraphics[width=\unitlength]{fig_wcewpe=3_NT=6000_cutalongZ_summary.pdf}}%
    \put(0.53328915,0.87835618){\makebox(0,0)[lb]{\smash{2400}}}%
    \put(0.60291963,0.87835618){\makebox(0,0)[lb]{\smash{2600}}}%
    \put(0.67255013,0.87835618){\makebox(0,0)[lb]{\smash{2800}}}%
    \put(0.74219109,0.87835618){\makebox(0,0)[lb]{\smash{3000}}}%
    \put(0.81182159,0.87835618){\makebox(0,0)[lb]{\smash{3200}}}%
    \put(0.88145206,0.87835618){\makebox(0,0)[lb]{\smash{3400}}}%
    \put(0.95108255,0.87835618){\makebox(0,0)[lb]{\smash{3600}}}%
    \put(0.51756457,0.89679474){\makebox(0,0)[lb]{\smash{2}}}%
    \put(0.51718309,0.91349131){\makebox(0,0)[lb]{\smash{0}}}%
    \put(0.51739617,0.93011261){\makebox(0,0)[lb]{\smash{2}}}%
    \put(0.51703874,0.94680604){\makebox(0,0)[lb]{\smash{4}}}%
    \put(0.51718996,0.96346854){\makebox(0,0)[lb]{\smash{6}}}%
    \put(0.51720371,0.9801276){\makebox(0,0)[lb]{\smash{8}}}%
    \put(0.51404872,0.99678672){\makebox(0,0)[lb]{\smash{10}}}%
    \put(0.51422743,1.01340802){\makebox(0,0)[lb]{\smash{12}}}%
    \put(0.53270461,0.70859413){\makebox(0,0)[lb]{\smash{2400}}}%
    \put(0.60238966,0.70859413){\makebox(0,0)[lb]{\smash{2600}}}%
    \put(0.6720747,0.70859413){\makebox(0,0)[lb]{\smash{2800}}}%
    \put(0.74177022,0.70859413){\makebox(0,0)[lb]{\smash{3000}}}%
    \put(0.81145525,0.70859413){\makebox(0,0)[lb]{\smash{3200}}}%
    \put(0.88114029,0.70859413){\makebox(0,0)[lb]{\smash{3400}}}%
    \put(0.95082533,0.70859413){\makebox(0,0)[lb]{\smash{3600}}}%
    \put(0.51252621,0.73354431){\makebox(0,0)[lb]{\smash{0.2}}}%
    \put(0.51214473,0.75438439){\makebox(0,0)[lb]{\smash{0.0}}}%
    \put(0.51232345,0.77522457){\makebox(0,0)[lb]{\smash{0.2}}}%
    \put(0.51208974,0.79606488){\makebox(0,0)[lb]{\smash{0.4}}}%
    \put(0.51213099,0.81690486){\makebox(0,0)[lb]{\smash{0.6}}}%
    \put(0.51215847,0.83774504){\makebox(0,0)[lb]{\smash{0.8}}}%
    \put(0.53086062,0.54919439){\makebox(0,0)[lb]{\smash{2400}}}%
    \put(0.6010961,0.54919439){\makebox(0,0)[lb]{\smash{2600}}}%
    \put(0.67133159,0.54919439){\makebox(0,0)[lb]{\smash{2800}}}%
    \put(0.74158023,0.54919439){\makebox(0,0)[lb]{\smash{3000}}}%
    \put(0.81181572,0.54919439){\makebox(0,0)[lb]{\smash{3200}}}%
    \put(0.88205124,0.54919439){\makebox(0,0)[lb]{\smash{3400}}}%
    \put(0.95228672,0.54919439){\makebox(0,0)[lb]{\smash{3600}}}%
    \put(0.51479985,0.6156675){\makebox(0,0)[lb]{\smash{0}}}%
    \put(0.50216071,0.64127424){\makebox(0,0)[lb]{\smash{2000}}}%
    \put(0.5020051,0.66688091){\makebox(0,0)[lb]{\smash{4000}}}%
    \put(0.54097785,0.04293122){\makebox(0,0)[lb]{\smash{2400}}}%
    \put(0.61106234,0.04293122){\makebox(0,0)[lb]{\smash{2600}}}%
    \put(0.68114684,0.04293122){\makebox(0,0)[lb]{\smash{2800}}}%
    \put(0.75124431,0.04293122){\makebox(0,0)[lb]{\smash{3000}}}%
    \put(0.82132881,0.04293122){\makebox(0,0)[lb]{\smash{3200}}}%
    \put(0.89141331,0.04293122){\makebox(0,0)[lb]{\smash{3400}}}%
    \put(0.96149781,0.04293122){\makebox(0,0)[lb]{\smash{3600}}}%
    \put(0.75772145,0.03236516){\makebox(0,0)[lb]{\smash{cells}}}%
    \put(0.50848734,0.09096928){\makebox(0,0)[lb]{\smash{0.5}}}%
    \put(0.50809399,0.12785453){\makebox(0,0)[lb]{\smash{0.0}}}%
    \put(0.50823231,0.16473979){\makebox(0,0)[lb]{\smash{0.5}}}%
    \put(0.56350017,0.01283247){\makebox(0,0)[lb]{\smash{120}}}%
    \put(0.6248241,0.01283247){\makebox(0,0)[lb]{\smash{100}}}%
    \put(0.68826177,0.01283247){\makebox(0,0)[lb]{\smash{80}}}%
    \put(0.7495857,0.01283247){\makebox(0,0)[lb]{\smash{60}}}%
    \put(0.81090964,0.01283247){\makebox(0,0)[lb]{\smash{40}}}%
    \put(0.87223357,0.01283247){\makebox(0,0)[lb]{\smash{20}}}%
    \put(0.93276215,0.01283239){\makebox(0,0)[lb]{\smash{0}}}%
    \put(0.53096359,0.38118039){\makebox(0,0)[lb]{\smash{2400}}}%
    \put(0.60146958,0.38118039){\makebox(0,0)[lb]{\smash{2600}}}%
    \put(0.67197556,0.38118039){\makebox(0,0)[lb]{\smash{2800}}}%
    \put(0.74249468,0.38118039){\makebox(0,0)[lb]{\smash{3000}}}%
    \put(0.81300067,0.38118039){\makebox(0,0)[lb]{\smash{3200}}}%
    \put(0.88350665,0.38118039){\makebox(0,0)[lb]{\smash{3400}}}%
    \put(0.95401264,0.38118039){\makebox(0,0)[lb]{\smash{3600}}}%
    \put(0.51206379,0.39184888){\makebox(0,0)[lb]{\smash{20}}}%
    \put(0.51206379,0.42786922){\makebox(0,0)[lb]{\smash{10}}}%
    \put(0.51603622,0.46388957){\makebox(0,0)[lb]{\smash{0}}}%
    \put(0.51209405,0.49990991){\makebox(0,0)[lb]{\smash{10}}}%
    \put(0.51185199,0.53593026){\makebox(0,0)[lb]{\smash{20}}}%
    \put(0.52971039,0.21307597){\makebox(0,0)[lb]{\smash{2400}}}%
    \put(0.6004067,0.21307597){\makebox(0,0)[lb]{\smash{2600}}}%
    \put(0.67110302,0.21307597){\makebox(0,0)[lb]{\smash{2800}}}%
    \put(0.74181253,0.21307597){\makebox(0,0)[lb]{\smash{3000}}}%
    \put(0.81250884,0.21307597){\makebox(0,0)[lb]{\smash{3200}}}%
    \put(0.88320515,0.21307597){\makebox(0,0)[lb]{\smash{3400}}}%
    \put(0.95390147,0.21307597){\makebox(0,0)[lb]{\smash{3600}}}%
    \put(0.50897953,0.24815333){\makebox(0,0)[lb]{\smash{0.5}}}%
    \put(0.50858616,0.28091732){\makebox(0,0)[lb]{\smash{0.0}}}%
    \put(0.50872448,0.31368123){\makebox(0,0)[lb]{\smash{0.5}}}%
    \put(0.50887148,0.34644517){\makebox(0,0)[lb]{\smash{1.0}}}%
    \put(0.03992572,0.711619){\makebox(0,0)[lb]{\smash{4200}}}%
    \put(0.11646713,0.711619){\makebox(0,0)[lb]{\smash{4400}}}%
    \put(0.19300838,0.711619){\makebox(0,0)[lb]{\smash{4600}}}%
    \put(0.26954978,0.711619){\makebox(0,0)[lb]{\smash{4800}}}%
    \put(0.34616365,0.711619){\makebox(0,0)[lb]{\smash{5000}}}%
    \put(0.4227051,0.711619){\makebox(0,0)[lb]{\smash{5200}}}%
    \put(0.00702061,0.75357053){\makebox(0,0)[lb]{\smash{0.00}}}%
    \put(0.00713097,0.78565975){\makebox(0,0)[lb]{\smash{0.05}}}%
    \put(0.00702061,0.81774898){\makebox(0,0)[lb]{\smash{0.10}}}%
    \put(0.00713097,0.8498382){\makebox(0,0)[lb]{\smash{0.15}}}%
    \put(0.03519525,0.55054387){\makebox(0,0)[lb]{\smash{4200}}}%
    \put(0.11188857,0.55054387){\makebox(0,0)[lb]{\smash{4400}}}%
    \put(0.18858173,0.55054387){\makebox(0,0)[lb]{\smash{4600}}}%
    \put(0.26527505,0.55054387){\makebox(0,0)[lb]{\smash{4800}}}%
    \put(0.34205914,0.55054387){\makebox(0,0)[lb]{\smash{5000}}}%
    \put(0.41875251,0.55054387){\makebox(0,0)[lb]{\smash{5200}}}%
    \put(0.01653371,0.58617863){\makebox(0,0)[lb]{\smash{0}}}%
    \put(0.00413663,0.61176886){\makebox(0,0)[lb]{\smash{1000}}}%
    \put(0.00389456,0.63735912){\makebox(0,0)[lb]{\smash{2000}}}%
    \put(0.0039205,0.66294934){\makebox(0,0)[lb]{\smash{3000}}}%
    \put(0.31933667,0.6704945){\makebox(0,0)[lb]{\smash{electron density}}}%
    \put(0.31933667,0.65901552){\makebox(0,0)[lb]{\smash{ion density}}}%
    \put(0.31933667,0.64753654){\makebox(0,0)[lb]{\smash{current $j_y$}}}%
    \put(0.03609999,0.38072802){\makebox(0,0)[lb]{\smash{4200}}}%
    \put(0.11251878,0.38072802){\makebox(0,0)[lb]{\smash{4400}}}%
    \put(0.18893738,0.38072802){\makebox(0,0)[lb]{\smash{4600}}}%
    \put(0.26535618,0.38072802){\makebox(0,0)[lb]{\smash{4800}}}%
    \put(0.34186573,0.38072802){\makebox(0,0)[lb]{\smash{5000}}}%
    \put(0.41828455,0.38072802){\makebox(0,0)[lb]{\smash{5200}}}%
    \put(0.00993793,0.41536575){\makebox(0,0)[lb]{\smash{1.0}}}%
    \put(0.00968307,0.45079881){\makebox(0,0)[lb]{\smash{0.0}}}%
    \put(0.00996839,0.48623187){\makebox(0,0)[lb]{\smash{1.0}}}%
    \put(0.00972633,0.52166495){\makebox(0,0)[lb]{\smash{2.0}}}%
    \put(0.20062269,0.5205469){\makebox(0,0)[lb]{\smash{$\bar{p}_z$ elec.}}}%
    \put(0.20062269,0.50641218){\makebox(0,0)[lb]{\smash{$5\bar{p}_z$ ions}}}%
    \put(0.20062269,0.49227747){\makebox(0,0)[lb]{\smash{$B_x/B_0$}}}%
    \put(0.20062269,0.47814278){\makebox(0,0)[lb]{\smash{elec. density}}}%
    \put(0.04554284,0.04324148){\makebox(0,0)[lb]{\smash{4200}}}%
    \put(0.12221525,0.04324148){\makebox(0,0)[lb]{\smash{4400}}}%
    \put(0.19888763,0.04324148){\makebox(0,0)[lb]{\smash{4600}}}%
    \put(0.27556004,0.04324148){\makebox(0,0)[lb]{\smash{4800}}}%
    \put(0.35232322,0.04324148){\makebox(0,0)[lb]{\smash{5000}}}%
    \put(0.42899562,0.04324148){\makebox(0,0)[lb]{\smash{5200}}}%
    \put(0.25858888,0.03267542){\makebox(0,0)[lb]{\smash{cells}}}%
    \put(0.01091779,0.06074725){\makebox(0,0)[lb]{\smash{0.4}}}%
    \put(0.01121172,0.0839369){\makebox(0,0)[lb]{\smash{0.2}}}%
    \put(0.0107321,0.10712655){\makebox(0,0)[lb]{\smash{0.0}}}%
    \put(0.01095687,0.1303162){\makebox(0,0)[lb]{\smash{0.2}}}%
    \put(0.01066294,0.15350585){\makebox(0,0)[lb]{\smash{0.4}}}%
    \put(0.01071481,0.1766955){\makebox(0,0)[lb]{\smash{0.6}}}%
    \put(0.01074939,0.19988516){\makebox(0,0)[lb]{\smash{0.8}}}%
    \put(0.42567487,0.18681764){\makebox(0,0)[lb]{\smash{$\bar{v}_y$ elec.}}}%
    \put(0.42567487,0.17091235){\makebox(0,0)[lb]{\smash{$\bar{v}_y$ ion}}}%
    \put(0.42567487,0.1541218){\makebox(0,0)[lb]{\smash{$10E_y$}}}%
    \put(0.04983081,0.01314265){\makebox(0,0)[lb]{\smash{80}}}%
    \put(0.11494377,0.01314265){\makebox(0,0)[lb]{\smash{100}}}%
    \put(0.1820321,0.01314265){\makebox(0,0)[lb]{\smash{120}}}%
    \put(0.24912045,0.01314265){\makebox(0,0)[lb]{\smash{140}}}%
    \put(0.31620881,0.01314265){\makebox(0,0)[lb]{\smash{160}}}%
    \put(0.38329716,0.01314265){\makebox(0,0)[lb]{\smash{180}}}%
    \put(0.45026448,0.01314265){\makebox(0,0)[lb]{\smash{200}}}%
    \put(0.14913696,0.00169132){\makebox(0,0)[lb]{\smash{electron inertial lengths $c/\omega_\mathrm{pe}(x=0,t=0)$}}}%
    \put(0.03730152,0.21159578){\makebox(0,0)[lb]{\smash{4200}}}%
    \put(0.11400483,0.21159578){\makebox(0,0)[lb]{\smash{4400}}}%
    \put(0.19070801,0.21159578){\makebox(0,0)[lb]{\smash{4600}}}%
    \put(0.26741132,0.21159578){\makebox(0,0)[lb]{\smash{4800}}}%
    \put(0.34420542,0.21159578){\makebox(0,0)[lb]{\smash{5000}}}%
    \put(0.42090878,0.21159578){\makebox(0,0)[lb]{\smash{5200}}}%
    \put(0.01020637,0.22623485){\makebox(0,0)[lb]{\smash{1.0}}}%
    \put(0.01034469,0.26373082){\makebox(0,0)[lb]{\smash{0.5}}}%
    \put(0.00995154,0.30122679){\makebox(0,0)[lb]{\smash{0.0}}}%
    \put(0.01008986,0.33872275){\makebox(0,0)[lb]{\smash{0.5}}}%
    \put(0.08467728,0.25666028){\makebox(0,0)[lb]{\smash{$\bar{v}_z$ elec.}}}%
    \put(0.08467728,0.24164028){\makebox(0,0)[lb]{\smash{$\bar{v}_z$ ion}}}%
    \put(0.03979756,0.88292079){\makebox(0,0)[lb]{\smash{4200}}}%
    \put(0.11641263,0.88292079){\makebox(0,0)[lb]{\smash{4400}}}%
    \put(0.19302751,0.88292079){\makebox(0,0)[lb]{\smash{4600}}}%
    \put(0.26964256,0.88292079){\makebox(0,0)[lb]{\smash{4800}}}%
    \put(0.34633007,0.88292079){\makebox(0,0)[lb]{\smash{5000}}}%
    \put(0.42294517,0.88292079){\makebox(0,0)[lb]{\smash{5200}}}%
    \put(0.01393494,0.90507949){\makebox(0,0)[lb]{\smash{0.0}}}%
    \put(0.01404534,0.93298447){\makebox(0,0)[lb]{\smash{0.5}}}%
    \put(0.01416259,0.96088953){\makebox(0,0)[lb]{\smash{1.0}}}%
    \put(0.01427298,0.9888291){\makebox(0,0)[lb]{\smash{1.5}}}%
    \put(0.01396946,1.01669964){\makebox(0,0)[lb]{\smash{2.0}}}%
    \put(0.26815621,0.99182248){\makebox(0,0)[lb]{\smash{$\Theta_{xx}$}}}%
    \put(0.26815621,0.9789828){\makebox(0,0)[lb]{\smash{$\Theta_{yy}$}}}%
    \put(0.26815621,0.96614327){\makebox(0,0)[lb]{\smash{$\Theta_{zz}$}}}%
    \put(0.31309429,0.99191176){\makebox(0,0)[lb]{\smash{$\Theta_{xy}$}}}%
    \put(0.31309429,0.97907216){\makebox(0,0)[lb]{\smash{$\Theta_{xz}$}}}%
    \put(0.31309429,0.96623255){\makebox(0,0)[lb]{\smash{$\Theta_{zy}$}}}%
    \put(0.26815621,0.95347535){\makebox(0,0)[lb]{\smash{$j_y$}}}%
    \put(0.25400716,1.00697749){\color[rgb]{0,0,0}\makebox(0,0)[lb]{\smash{electrons}}}%
    \put(0.80765298,0.44720873){\makebox(0,0)[lb]{\smash{$\bar{p}_z$ elec.}}}%
    \put(0.80765298,0.43307401){\makebox(0,0)[lb]{\smash{$5\bar{p}_z$ ions}}}%
    \put(0.80765298,0.41893929){\makebox(0,0)[lb]{\smash{$10B_x/B_0$}}}%
    \put(0.80765298,0.40480461){\makebox(0,0)[lb]{\smash{elec. density}}}%
    \put(0.94116483,0.18478484){\makebox(0,0)[lb]{\smash{$\bar{v}_y$ elec.}}}%
    \put(0.94116483,0.16887959){\makebox(0,0)[lb]{\smash{$\bar{v}_y$ ion}}}%
    \put(0.94116483,0.15208898){\makebox(0,0)[lb]{\smash{$10E_y$}}}%
    \put(0.58978997,0.35130748){\makebox(0,0)[lb]{\smash{$\bar{v}_z$ elec.}}}%
    \put(0.58978997,0.33628748){\makebox(0,0)[lb]{\smash{$\bar{v}_z$ ion}}}%
    \put(0.8751945,0.26323518){\makebox(0,0)[lb]{\smash{$(\b{E}\wedge\b{B})_z/B^2$}}}%
    \put(0.8751945,0.24998565){\makebox(0,0)[lb]{\smash{$10[E_y + (\bar{\b{v}}_\mathrm{e}\wedge\b{B})_y]$}}}%
    \put(0.8751945,0.23652006){\makebox(0,0)[lb]{\smash{elec. density}}}%
    \put(0.36925182,0.27142043){\makebox(0,0)[lb]{\smash{$(\b{E}\wedge\b{B})_z/B^2$}}}%
    \put(0.36925182,0.2581709){\makebox(0,0)[lb]{\smash{$10[E_y + (\bar{\b{v}}_\mathrm{e}\wedge\b{B})_y]$}}}%
    \put(0.36925182,0.24470531){\makebox(0,0)[lb]{\smash{elec. density}}}%
    \put(0.67842086,0.50588056){\color[rgb]{0,0,0}\makebox(0,0)[lb]{\smash{X-point}}}%
    \put(0.22219565,0.34322944){\color[rgb]{0,0,0}\makebox(0,0)[lb]{\smash{X-point}}}%
    \put(0.08707252,0.5986261){\color[rgb]{0,0,0}\makebox(0,0)[lb]{\smash{island}}}%
    \put(0.43685463,0.60216711){\color[rgb]{0,0,0}\makebox(0,0)[lb]{\smash{island}}}%
    \put(0.75099387,0.9968237){\makebox(0,0)[lb]{\smash{$\Theta_{xx}$}}}%
    \put(0.75099387,0.98398403){\makebox(0,0)[lb]{\smash{$\Theta_{yy}$}}}%
    \put(0.75099387,0.97114449){\makebox(0,0)[lb]{\smash{$\Theta_{zz}$}}}%
    \put(0.79593195,0.99691299){\makebox(0,0)[lb]{\smash{$\Theta_{xy}$}}}%
    \put(0.79593195,0.98407339){\makebox(0,0)[lb]{\smash{$\Theta_{xz}$}}}%
    \put(0.79593195,0.97123378){\makebox(0,0)[lb]{\smash{$\Theta_{zy}$}}}%
    \put(0.75099387,0.95847657){\makebox(0,0)[lb]{\smash{$j_y$}}}%
    \put(0.73684482,1.01197871){\color[rgb]{0,0,0}\makebox(0,0)[lb]{\smash{electrons}}}%
    \put(0.75683044,0.84332028){\makebox(0,0)[lb]{\smash{$\Theta_{xx}$}}}%
    \put(0.75683044,0.83048061){\makebox(0,0)[lb]{\smash{$\Theta_{yy}$}}}%
    \put(0.75683044,0.81764107){\makebox(0,0)[lb]{\smash{$\Theta_{zz}$}}}%
    \put(0.80176852,0.84340957){\makebox(0,0)[lb]{\smash{$\Theta_{xy}$}}}%
    \put(0.80176852,0.83056996){\makebox(0,0)[lb]{\smash{$\Theta_{xz}$}}}%
    \put(0.80176852,0.81773036){\makebox(0,0)[lb]{\smash{$\Theta_{zy}$}}}%
    \put(0.75683044,0.80497315){\makebox(0,0)[lb]{\smash{$j_y$}}}%
    \put(0.71060949,0.84286135){\color[rgb]{0,0,0}\makebox(0,0)[lb]{\smash{ions}}}%
    \put(0.41046667,0.84278644){\makebox(0,0)[lb]{\smash{$\Theta_{xx}$}}}%
    \put(0.41046667,0.82994676){\makebox(0,0)[lb]{\smash{$\Theta_{yy}$}}}%
    \put(0.41046667,0.81710723){\makebox(0,0)[lb]{\smash{$\Theta_{zz}$}}}%
    \put(0.45540474,0.84287572){\makebox(0,0)[lb]{\smash{$\Theta_{xy}$}}}%
    \put(0.45540474,0.83003612){\makebox(0,0)[lb]{\smash{$\Theta_{xz}$}}}%
    \put(0.45540474,0.81719651){\makebox(0,0)[lb]{\smash{$\Theta_{zy}$}}}%
    \put(0.41046667,0.80443931){\makebox(0,0)[lb]{\smash{$j_y$}}}%
    \put(0.36424572,0.8423275){\color[rgb]{0,0,0}\makebox(0,0)[lb]{\smash{ions}}}%
    \put(0.75711438,0.67104955){\makebox(0,0)[lb]{\smash{electron density}}}%
    \put(0.75711438,0.65957057){\makebox(0,0)[lb]{\smash{ion density}}}%
    \put(0.75711438,0.64809159){\makebox(0,0)[lb]{\smash{current $j_y$}}}%
    \put(0.57502567,0.62416054){\color[rgb]{0,0,0}\makebox(0,0)[lb]{\smash{island}}}%
    \put(0.91639295,0.62535414){\color[rgb]{0,0,0}\makebox(0,0)[lb]{\smash{island}}}%
    \put(0.65086105,0.00184148){\makebox(0,0)[lb]{\smash{electron inertial lengths $c/\omega_\mathrm{pe}(x=0,t=0)$}}}%
  \end{picture}%
\endgroup%

%% file: reconnection_rates_no_guide_field.pdf_tex
%% Creator: Inkscape inkscape 0.48.4, www.inkscape.org
%% Accompanies image file 'reconnection_rates_no_guide_field.pdf' (pdf, eps, ps)
\begingroup%
  \makeatletter%
  \providecommand\color[2][]{%
    \errmessage{(Inkscape) Color is used for the text in Inkscape, but the package 'color.sty' is not loaded}%
    \renewcommand\color[2][]{}%
  }%
  \providecommand\transparent[1]{%
    \errmessage{(Inkscape) Transparency is used (non-zero) for the text in Inkscape, but the package 'transparent.sty' is not loaded}%
    \renewcommand\transparent[1]{}%
  }%
  \providecommand\rotatebox[2]{#2}%
  \ifx\svgwidth\undefined%
    \setlength{\unitlength}{1476.41943359bp}%
    \ifx\svgscale\undefined%
      \relax%
    \else%
      \setlength{\unitlength}{\unitlength * \real{\svgscale}}%
    \fi%
  \else%
    \setlength{\unitlength}{\svgwidth}%
  \fi%
  \global\let\svgwidth\undefined%
  \global\let\svgscale\undefined%
  \makeatother%
  \begin{picture}(1,0.69753968)%
    \put(0,0){\includegraphics[width=\unitlength]{reconnection_rates_no_guide_field.pdf}}%
    \put(0.07257999,0.03357351){\makebox(0,0)[lb]{\smash{0}}}%
    \put(0.20920018,0.03357351){\makebox(0,0)[lb]{\smash{20}}}%
    \put(0.35587233,0.03357351){\makebox(0,0)[lb]{\smash{40}}}%
    \put(0.50272423,0.03357351){\makebox(0,0)[lb]{\smash{60}}}%
    \put(0.64948093,0.03357351){\makebox(0,0)[lb]{\smash{80}}}%
    \put(0.78299304,0.03357351){\makebox(0,0)[lb]{\smash{100}}}%
    \put(0.92976032,0.03357351){\makebox(0,0)[lb]{\smash{120}}}%
    \put(0.49301207,0.00191811){\makebox(0,0)[lb]{\smash{$t\omega_\mathrm{ci}$}}}%
    \put(0.55933355,0.60375458){\makebox(0,0)[lb]{\smash{~~~~~~~1,~~~~~~ $T_\mathrm{bg}=1.5\cdot10^7$\,K}}}%
    \put(0.55869203,0.56286632){\makebox(0,0)[lb]{\smash{~~~~~~~3,~~~~~~ $T_\mathrm{bg}=1.5\cdot10^7$\,K}}}%
    \put(0.55838182,0.52239811){\makebox(0,0)[lb]{\smash{~~~~~~~3,~~~~~~ $T_\mathrm{bg}=2\cdot10^8$\,K}}}%
    \put(0.55743678,0.48197129){\makebox(0,0)[lb]{\smash{~~~~~~~3,~~~~~~ $T_\mathrm{e,bg}=3\cdot10^9$\,K}}}%
    \put(0.55813039,0.44069282){\makebox(0,0)[lb]{\smash{~~~~~~~3,~~~~~~ $n_\mathrm{bg}=0.3n_\mathrm{cs}(0)$}}}%
    \put(0.55743678,0.39998848){\makebox(0,0)[lb]{\smash{~~~~~~~6,~~~~~~ $T_\mathrm{bg}=8\cdot10^8$\,K}}}%
    \put(0.55975249,0.65638555){\makebox(0,0)[lb]{\smash{$\omega_\mathrm{ce}/\omega_\mathrm{pe}$}}}%
    \put(-0.00033324,0.14283904){\makebox(0,0)[lb]{\smash{0.00}}}%
    \put(-0.00016391,0.21870599){\makebox(0,0)[lb]{\smash{0.05}}}%
    \put(-0.00033324,0.29457292){\makebox(0,0)[lb]{\smash{0.10}}}%
    \put(-0.00016391,0.37043987){\makebox(0,0)[lb]{\smash{0.15}}}%
    \put(-0.00033324,0.44630681){\makebox(0,0)[lb]{\smash{0.20}}}%
    \put(-0.00016391,0.52217374){\makebox(0,0)[lb]{\smash{0.25}}}%
    \put(-0.00033324,0.59804071){\makebox(0,0)[lb]{\smash{0.30}}}%
    \put(0.09988134,0.65407243){\color[rgb]{0,0,0}\makebox(0,0)[lb]{\smash{$E_y/(B_0V_\mathrm{A,in}^\mathrm{R})$}}}%
    \put(0.55845492,0.35879386){\makebox(0,0)[lb]{\smash{~~~~~~~3,~~~~~~ $m_\mathrm{i}=m_\mathrm{e}$}}}%
  \end{picture}%
\endgroup%

%% file: fig_particle_trajectory_3D_n=3601021854_zoomed_modif.pdf_tex
%% Creator: Inkscape inkscape 0.48.4, www.inkscape.org
%% Accompanies image file 'fig_particle_trajectory_3D_n=3601021854_zoomed_modif.pdf' (pdf, eps, ps)
\begingroup%
  \makeatletter%
  \providecommand\color[2][]{%
    \errmessage{(Inkscape) Color is used for the text in Inkscape, but the package 'color.sty' is not loaded}%
    \renewcommand\color[2][]{}%
  }%
  \providecommand\transparent[1]{%
    \errmessage{(Inkscape) Transparency is used (non-zero) for the text in Inkscape, but the package 'transparent.sty' is not loaded}%
    \renewcommand\transparent[1]{}%
  }%
  \providecommand\rotatebox[2]{#2}%
  \ifx\svgwidth\undefined%
    \setlength{\unitlength}{656.07705078bp}%
    \ifx\svgscale\undefined%
      \relax%
    \else%
      \setlength{\unitlength}{\unitlength * \real{\svgscale}}%
    \fi%
  \else%
    \setlength{\unitlength}{\svgwidth}%
  \fi%
  \global\let\svgwidth\undefined%
  \global\let\svgscale\undefined%
  \makeatother%
  \begin{picture}(1,0.91688286)%
    \put(0,0){\includegraphics[width=\unitlength]{fig_particle_trajectory_3D_n=3601021854_zoomed_modif.pdf}}%
    \put(0.84988469,0.17471798){\makebox(0,0)[lb]{\smash{y}}}%
    \put(0.955682,0.20058395){\makebox(0,0)[lb]{\smash{1500}}}%
    \put(0.91093185,0.14866098){\makebox(0,0)[lb]{\smash{2000}}}%
    \put(0.86424785,0.09408761){\makebox(0,0)[lb]{\smash{2500}}}%
    \put(0.81515416,0.03665554){\makebox(0,0)[lb]{\smash{3000}}}%
    \put(0.42042719,0){\makebox(0,0)[lb]{\smash{z}}}%
    \put(0.12886852,0.08378349){\makebox(0,0)[lb]{\smash{4500}}}%
    \put(0.30022112,0.06521254){\makebox(0,0)[lb]{\smash{5000}}}%
    \put(0.47423276,0.04632584){\makebox(0,0)[lb]{\smash{5500}}}%
    \put(0.65116879,0.02711527){\makebox(0,0)[lb]{\smash{6000}}}%
    \put(0.00486686,0.52694238){\makebox(0,0)[lb]{\smash{x}}}%
    \put(0.04307174,0.28059519){\makebox(0,0)[lb]{\smash{1500}}}%
    \put(0.03885666,0.44605718){\makebox(0,0)[lb]{\smash{2000}}}%
    \put(0.03491263,0.61417808){\makebox(0,0)[lb]{\smash{2500}}}%
    \put(0.68057557,0.70798053){\color[rgb]{0,0,0}\rotatebox{-4.91935428}{\makebox(0,0)[lb]{\smash{island}}}}%
    \put(0.80848912,0.79594852){\color[rgb]{0,0,0}\rotatebox{-5.10259677}{\makebox(0,0)[lb]{\smash{X-point}}}}%
    \put(0.39146008,0.73228154){\color[rgb]{0,0,0}\rotatebox{-5.10259677}{\makebox(0,0)[lb]{\smash{current sheet}}}}%
    \put(0.45077255,0.81882352){\color[rgb]{0,0,0}\makebox(0,0)[lb]{\smash{$B_\mathrm{rec}$}}}%
    \put(0.41083786,0.60902647){\color[rgb]{0,0,0}\makebox(0,0)[lb]{\smash{$B_\mathrm{rec}$}}}%
    \put(0.27416744,0.77446967){\color[rgb]{0,0,0}\makebox(0,0)[lb]{\smash{$B_\mathrm{G}$}}}%
  \end{picture}%
\endgroup%

%% file: reconnection_rates_guide_field.pdf_tex
%% Creator: Inkscape inkscape 0.48.4, www.inkscape.org
%% Accompanies image file 'reconnection_rates_guide_field.pdf' (pdf, eps, ps)
\begingroup%
  \makeatletter%
  \providecommand\color[2][]{%
    \errmessage{(Inkscape) Color is used for the text in Inkscape, but the package 'color.sty' is not loaded}%
    \renewcommand\color[2][]{}%
  }%
  \providecommand\transparent[1]{%
    \errmessage{(Inkscape) Transparency is used (non-zero) for the text in Inkscape, but the package 'transparent.sty' is not loaded}%
    \renewcommand\transparent[1]{}%
  }%
  \providecommand\rotatebox[2]{#2}%
  \ifx\svgwidth\undefined%
    \setlength{\unitlength}{1410.83408203bp}%
    \ifx\svgscale\undefined%
      \relax%
    \else%
      \setlength{\unitlength}{\unitlength * \real{\svgscale}}%
    \fi%
  \else%
    \setlength{\unitlength}{\svgwidth}%
  \fi%
  \global\let\svgwidth\undefined%
  \global\let\svgscale\undefined%
  \makeatother%
  \begin{picture}(1,0.63165116)%
    \put(0,0){\includegraphics[width=\unitlength]{reconnection_rates_guide_field.pdf}}%
    \put(0.07595403,0.03513423){\makebox(0,0)[lb]{\smash{0}}}%
    \put(0.21892521,0.03513423){\makebox(0,0)[lb]{\smash{20}}}%
    \put(0.37241575,0.03513423){\makebox(0,0)[lb]{\smash{40}}}%
    \put(0.5260943,0.03513423){\makebox(0,0)[lb]{\smash{60}}}%
    \put(0.67967327,0.03513423){\makebox(0,0)[lb]{\smash{80}}}%
    \put(0.81939195,0.03513423){\makebox(0,0)[lb]{\smash{100}}}%
    \put(0.97298199,0.03513423){\makebox(0,0)[lb]{\smash{120}}}%
    \put(0.51593065,0.00200732){\makebox(0,0)[lb]{\smash{$t\omega_\mathrm{ci}$}}}%
    \put(0.73697992,0.52686027){\makebox(0,0)[lb]{\smash{$B_\mathrm{G}=0$}}}%
    \put(0.73665528,0.48564488){\makebox(0,0)[lb]{\smash{$B_\mathrm{G}=0.5B_0$}}}%
    \put(0.73566632,0.4444728){\makebox(0,0)[lb]{\smash{$B_\mathrm{G}=1$}}}%
    \put(0.73289981,0.57819133){\makebox(0,0)[lb]{\smash{$\omega_\mathrm{ce}/\omega_\mathrm{pe}=3$}}}%
    \put(-0.00034872,0.14947921){\makebox(0,0)[lb]{\smash{0.00}}}%
    \put(-0.00017152,0.22887301){\makebox(0,0)[lb]{\smash{0.05}}}%
    \put(-0.00034872,0.30826673){\makebox(0,0)[lb]{\smash{0.10}}}%
    \put(-0.00017152,0.38766053){\makebox(0,0)[lb]{\smash{0.15}}}%
    \put(-0.00034872,0.46705425){\makebox(0,0)[lb]{\smash{0.20}}}%
    \put(-0.00017152,0.54644801){\makebox(0,0)[lb]{\smash{0.25}}}%
    \put(-0.00034872,0.62584182){\makebox(0,0)[lb]{\smash{0.30}}}%
    \put(0.10452452,0.58328554){\color[rgb]{0,0,0}\makebox(0,0)[lb]{\smash{$E_y/(B_0V_\mathrm{A,in}^\mathrm{R}\cos\theta)$}}}%
  \end{picture}%
\endgroup%

%% file: kappa_32.pdf_tex
%% Creator: Inkscape inkscape 0.48.4, www.inkscape.org
%% Accompanies image file 'kappa_32.pdf' (pdf, eps, ps)
\begingroup%
  \makeatletter%
  \providecommand\color[2][]{%
    \errmessage{(Inkscape) Color is used for the text in Inkscape, but the package 'color.sty' is not loaded}%
    \renewcommand\color[2][]{}%
  }%
  \providecommand\transparent[1]{%
    \errmessage{(Inkscape) Transparency is used (non-zero) for the text in Inkscape, but the package 'transparent.sty' is not loaded}%
    \renewcommand\transparent[1]{}%
  }%
  \providecommand\rotatebox[2]{#2}%
  \ifx\svgwidth\undefined%
    \setlength{\unitlength}{463.66201172bp}%
    \ifx\svgscale\undefined%
      \relax%
    \else%
      \setlength{\unitlength}{\unitlength * \real{\svgscale}}%
    \fi%
  \else%
    \setlength{\unitlength}{\svgwidth}%
  \fi%
  \global\let\svgwidth\undefined%
  \global\let\svgscale\undefined%
  \makeatother%
  \begin{picture}(1,0.83376889)%
    \put(0,0){\includegraphics[width=\unitlength]{kappa_32.pdf}}%
    \put(0.00224099,0.01290277){\makebox(0,0)[lb]{\smash{0.0}}}%
    \put(0.23909734,0.01290277){\makebox(0,0)[lb]{\smash{0.5}}}%
    \put(0.47597053,0.01290277){\makebox(0,0)[lb]{\smash{1.0}}}%
    \put(0.71282688,0.01323976){\makebox(0,0)[lb]{\smash{1.5}}}%
    \put(0.94867225,0.01290277){\makebox(0,0)[lb]{\smash{2.0}}}%
    \put(-0.02021578,0.19936428){\makebox(0,0)[lb]{\smash{2}}}%
    \put(-0.02196814,0.36523969){\makebox(0,0)[lb]{\smash{4}}}%
    \put(-0.02122676,0.53096345){\makebox(0,0)[lb]{\smash{6}}}%
    \put(-0.02115936,0.69667036){\makebox(0,0)[lb]{\smash{8}}}%
    \put(0.12472113,0.67081193){\makebox(0,0)[lb]{\smash{$4T/mc^2$}}}%
    \put(0.4713419,-0.04450726){\color[rgb]{0,0,0}\makebox(0,0)[lb]{\smash{$T/mc^2$}}}%
    \put(0.13266748,0.75042281){\color[rgb]{0,0,0}\makebox(0,0)[lb]{\smash{$h(T) = \frac{K_3(mc^2/T)}{K_2(mc^2/T)} = \kappa_{32}(mc^2/T)$}}}%
    \put(-0.01838471,0.11342098){\makebox(0,0)[lb]{\smash{1}}}%
  \end{picture}%
\endgroup%